\pdfoutput=1
\documentclass[aip,jcp,preprint]{revtex4-2}
\usepackage{adjustbox}
\usepackage{amsmath,amssymb}
\usepackage{upgreek}
\usepackage[ruled,vlined]{algorithm2e}

\SetKwFor{Loop}{loop}{}{end loop}   
\SetKw{Break}{break loop}

\usepackage{xcolor}

\newif\ifshowrevisions
\showrevisionstrue   

\ifshowrevisions

\else

\fi

\newcommand{\satoms}{_{\mathrm{a}}}
\newcommand{\sorbs}{_{\mathrm{orb}}}
\newcommand{\sconfs}{_\mathrm{conf}}

\newcommand{\appref}[1]{Appendix~\ref{#1}}
\newcommand{\mol}[1]{\mathrm{#1}}
\newcommand{\mat}[1]{\mathbf{#1}}
\newcommand{\Ntrain}{N_{\mathrm{train}}}
\newcommand{\UMat}[1]{\mat{I}_{#1}}
\newcommand{\kfunc}{k}
\newcommand{\Natoms}{N\satoms}
\newcommand{\ltwo}{l^{2}}
\newcommand{\Nfeatures}{N_{\mathrm{f}}}
\newcommand{\Ninit}{N_{\mathrm{init}}}
\newcommand{\Nstacks}{N_{\mathrm{stack}}}
\newcommand{\Ntransforms}{N_{\mathrm{tr}}}
\newcommand{\transpose}{\mathbf{T}}

\newcommand{\nc}{n}

\newcommand{\tK}{\tilde{\mat{K}}}

\newcommand{\locorb}{\psi}
\newcommand{\Norbs}{N\sorbs}
\newcommand{\orbweight}{w}
\newcommand{\slpair}[3]{\mol{#1}_{#2\rightarrow#3}}

\newcommand{\Eox}{E_{\mathrm{ox}}}
\newcommand{\Ehyd}{E_{\mathrm{hyd}}}

\newcommand{\randsigndiagtens}{\mathrm{D}}
\newcommand{\randsigndiag}{\mat{D}}
\newcommand{\randomphasestens}{\upphi}
\newcommand{\randomphases}{\vec{\phi}}

\newcommand{\sigmaglobal}{\sigma_{\mathrm{global}}}

\newcommand{\rcutorb}{\rcut^{\mathrm{orb}}}

\newcommand{\kglobal}{\kfunc_\mathrm{G}}
\newcommand{\klocal}{\kfunc_\mathrm{aG}}
\newcommand{\kdlocal}{\kfunc_\mathrm{eaG}}
\newcommand{\kFJK}{\kfunc_\mathrm{orb}}
\newcommand{\kSORF}{\kfunc_\mathrm{SORF}}
\newcommand{\korblin}{\kFJK^{\mathrm{lin}}}
\newcommand{\korbgauss}{\kFJK^{\mathrm{Gauss}}}

\newcommand{\rcut}{\rho_{\mathrm{cut}}}
\newcommand{\Nmorfconf}{N\sconfs^{\mathrm{Morfeus}}}
\newcommand{\Nconf}{N\sconfs}
\newcommand{\boltzweight}{w^{\mathrm{b}}}
\newcommand{\boltzweightvec}{\vec{w}^{\mathrm{b}}}
\newcommand{\cutboltzweight}{w^{\mathrm{cb}}}
\newcommand{\cutboltzweightvec}{\vec{w}^{\mathrm{cb}}}
\newcommand{\confid}{c}

\newcommand{\SORF}{\mathcal{F}}
\newcommand{\NORM}{\mathcal{N}}
\newcommand{\SUM}{\mathcal{S}}
\newcommand{\WEIGHTEDSUM}{\mathcal{W}}
\newcommand{\SWITCH}{\mathcal{T}}
\newcommand{\MIXEDEXTENSIVE}{\mathcal{E}}
\newcommand{\mixsigma}{\sigma_{\mathrm{mix}}}
\newcommand{\CONCATENATE}{\mathcal{C}}

\newcommand{\lmax}{l_{\mathrm{max}}}
\newcommand{\coeffvec}{\vec{c}}

\newcommand{\aoproj}{\mat{1}}

\newcommand{\pypackname}[1]{\mbox{\protect\path{#1}}}
\newcommand{\keyword}[1]{\protect\path{#1}}
\newcommand{\libname}[1]{\textsc{#1}}

\newcommand{\SciPy}{\libname{SciPy}}

\newcommand{\leaveoneoutadd}{_{\mathrm{LOO}}}
\newcommand{\dleaveoneout}{\Delta\leaveoneoutadd}
\newcommand{\leaveoneouterr}{\dleaveoneout q}
\newcommand{\leaveoneoutvecerr}{\dleaveoneout\vec{q}}
\newcommand{\reproducedquant}{q^{r}}
\newcommand{\reproducedvecquant}{\vec{q}^{r}}

\newcommand{\Lhyp}{L_{\mathrm{hyp}}}
\newcommand{\Lopthyp}{\tilde{L}_{\mathrm{hyp}}}

\newcommand{\ZU}{\mat{U}_{\mat{Z}}}
\newcommand{\ZD}{\mat{D}_{\mat{Z}}}
\newcommand{\ZV}{\mat{V}_{\mat{Z}}}
\newcommand{\zTR}{^{\mathrm{tr}}}
\newcommand{\LNSTEP}{S}
\newcommand{\allsigmas}{\vec{\sigma}_{\mathrm{all}}}
\newcommand{\leaveonegroupoutadd}{_{\mathrm{LOGO}}}
\newcommand{\dlogo}{\Delta\leaveonegroupoutadd}

\newcommand{\logovecerr}{\dlogo\vec{q}}
\newcommand{\cvvecerr}{\Delta \vec{q}}

\newcommand{\posconst}{C}

\newcommand{\LSCLC}{L_{\mathrm{SCLC}}}

\newcommand{\repokey}[1]{\protect\path{#1}}

\usepackage{url}
\usepackage{amsthm}
\usepackage{graphicx} 
\usepackage{amsmath}
\usepackage{bm}
\usepackage{siunitx,makecell}
\usepackage{tcolorbox,tabularx,booktabs}
\usepackage{url,hyperref}
\usepackage{multirow}

\begin{document}

\title[Ensemble-applied SORF]{Kernel Ridge Regression for conformer ensembles made easy with Structured Orthogonal Random Features}

\author{Konstantin Karandashev}

\affiliation{Chembricks Inc., Menlo Park, California, USA}
\affiliation{University of Vienna, Faculty of Physics, Kolingasse 14-16, AT-1090 Wien, Austria}
\email{kvkarandashev@gmail.com}
\vspace{10pt}

\begin{abstract}

A computationally efficient protocol for machine learning in chemical space using Boltzmann ensembles of conformers as input is proposed; the method is based on rewriting Kernel Ridge Regression expressions in terms of Structured Orthogonal Random Features, yielding physics-motivated trigonometric neural networks. To evaluate the method's utility for materials discovery, we test it on experimental datasets of two quantities related to battery electrolyte design, namely oxidation potentials in acetonitrile and hydration energies, using several geometry- and orbital-based molecular representations to demonstrate the method's flexibility. Despite only using computationally cheap force field calculations for conformer generation and generating orbital representations with GFN2-xTB calculations, we observe systematic decrease of prediction errors with increased training set size in all cases, with experimental accuracy reached after training on hundreds of molecules.

\end{abstract}

\maketitle

\section{Introduction\label{sec:introduction}}

Optimization in chemical compound space is an important part of chemical industry, with machine learning (ML) models emerging at the forefront of accelerating it~\cite{Saal_Meredig:2020,Malica_Roche:2025}. In this context a commonly posed ML problem is prediction of experimental outcomes based on limited experimental data; the main difference between approaches proposed for this problem is how prior physics knowledge enters the model. Quantitative structure--activity relationship (QSAR) methods\cite{Cherkasov2014QSARMW,Phan_Stadler:2026} rely on combining descriptors hand-crafted to cover specific quantities that are most often queried by the wider chemistry community. Approaches based on graph- and geometry-based deep learning\cite{Scarselli_Monfardini:2009,Gilmer_Dahl:2017,Schutt_Gastegger:2021} derive some of the knowledge from the neural network (NN) architecture and proceed to expand it with a large amount of chemical data by either deriving chemistry-aware features used in further ML\cite{Zang_Tang:2023,Qiao_Wei:2025} or augmenting experimental data of interest with results of \emph{ab initio} calculations\cite{Zhang_Zhang:2022,Xia_Zhang:2023,Kirschbaum_Bande:2024}, e.g. via pre-training and fine-tuning\cite{Zhuang_He:2021,Zhang_Zhang:2022}. The philosophy this work contributes to instead uses geometry-\cite{De_Ceriotti:2016,Huang_Lilienfeld:2020,Darby_Csanyi:2023} or orbital-based\cite{Welborn_Miller:2018,Fabrizio_Corminboeuf:2022,Llenga_Grynova:2023} descriptors hand-designed and benchmarked on predicting diverse \emph{ab initio} quantum mechanical properties of individual conformers\cite{Faber_Lilienfeld:2017} and applies them, unchanged, to experimental properties\cite{Lee_Lilienfeld:2024}; while this can be done with different regressors, the starting point of this work is Kernel Ridge Regression (KRR)\cite{Vapnik:1998}, a reliable approach in the low-data regime. The aim was to create a framework that would be able to seamlessly and efficiently combine two previously proposed ideas.

The first idea, which was also the main motivation for this work, is that since a molecule is more faithfully described by a Boltzmann ensemble rather than an individual conformer, its representation should be aggregated over multiple conformers too, e.g. by using representation of an averaged conformer\cite{Weinreich_Lilienfeld:2022} or representations averaged over several conformers\cite{Rauer_Bereau:2020,Weinreich_Lilienfeld:2021}. One drawback of the latter two approaches is the information loss implicit to averaging; augmenting representation averages with other information about their distributions (e.g. medians and higher distribution moments) as is done in Molecular Dynamics Fingerprints approach\cite{Riniker:2017} is possible, but systematically including higher-order moments beyond dispersion is hampered by statistical errors. Defining a KRR protocol that would both avoid information loss and lack statistical errors is straightforward, but it would be highly computationally inefficient (see Subsection~\ref{subsec:ensemble_app}). The second idea is that a local orbital-based representation suitable for predicting conformer energies can be easily repurposed to accurately predict energy differences associated with a change in orbital structure by combining the kernel elements of the initial and final states of that change\cite{Karandashev_Lilienfeld:2022}. However, the cost of evaluating those kernel elements scales poorly with molecular size (see Subsection~\ref{subsec:classic_KRR}).

Because both bottlenecks are at root problems of efficient kernel evaluation, it is natural to turn to efficient kernel approximation schemes~\cite{Rahimi_Recht:2007,Yu_Kumar:2016,Liu_Johan:2022,Reid_Weller:2023,Li_Mak:2024}, with this work focusing on Structured Orthogonal Random Feature (SORF) formalism~\cite{Yu_Kumar:2016}, which was previously used to speed up KRR calculations of molecular forces and energies~\cite{Browning_Lilienfeld:2022}. Rewriting a possible KRR protocol for ML based on Boltzmann ensembles in terms of SORF yields numerically efficient NNs characterized by using trigonometric activation functions and fast Walsh-Hadamard transform~\cite{Fino_Algazi:1976}. NN architectures sharing such characteristics also arise~\cite{Cutajar_Filippone:2017,Xie_Huang:2019,Rossi_Filippone:2020,Lu_Shafto:2022} when SORF or related feature expressions are combined with Deep Gaussian Processes~\cite{Damianou_Lawrence:2013} or Multilayer Kernel Machines~\cite{Cho_Saul:2009}. Since our NNs end up with several SORF layers corresponding to different levels of molecular structure (\emph{e.g.} conformer ensemble, individual conformer, conformer's atoms), the proposed method will be from now on referred to as Multilevel SORF (MSORF). As will be clarified in Section~\ref{sec:theory}, the crux of MSORF is encoding complicated chemical objects into feature vectors with physically meaningful dot products. While in this work we demonstrated their utility for approximate KRR, which is in this case ridge regression performed on these vectors, there are no fundamental obstacles to using them with other regressors.

MSORF was tested on experimental data pertaining to two compound properties fundamentally important to battery electrolyte design. The first one is electrolyte component's electrochemical stability window, which fundamentally affects battery's operating voltage and energy capacity~\cite{Borodin_Knap:2015}; this work used the collection of experimental oxidation potentials of molecules in acetonitrile $\Eox$ compiled in Ref.~\onlinecite{Lee_Lilienfeld:2024} from peer-reviewed scientific literature using Large Language Models. Since the authors did not provide an abbreviated name for their dataset, we will refer to it as ``literature electrochemical stability'' (LES) dataset for short. The second property of interest is compound's solvation energy, which is useful for approximating trends in solubility~\cite{Cheng_Curtiss:2015}, which in turn is fundamentally important for how usable a compound is as an electrolyte additive. We therefore also tested MSORF on experimental hydration energies $\Ehyd$ collected in the FreeSolv dataset~\cite{Mobley_Guthrie:2014,Matos_Mobley:2017,FreeSolv_Zenodo}. Importantly, $\Eox$ and $\Ehyd$ are examples of an intensive and an extensive property, allowing us to explore how this could be reflected in MSORF architectures. Additionally, since $\Eox$ is closely related to ionization potential, \emph{i.e.} energy change associated with removing an electron, the quantity provides an experimental testing ground for the ``orbital change'' trick of Ref.~\onlinecite{Karandashev_Lilienfeld:2022}. We also demonstrate MSORF's flexibility by presenting results with a diverse set of molecular representations, which will be discussed in Subsection~\ref{subsec:classic_KRR}.

The rest of the paper is structured as follows. The theory behind MSORF is explained in Section~\ref{sec:theory}, numerical details of computational experiments are presented in Section~\ref{sec:numerical_details}, with the latter's results discussed in Section~\ref{sec:results}. Summary of the paper's results along with potential further research directions are discussed in Section~\ref{sec:conclusions}. \appref{app:fjd_formalism} discusses a modification of the original FJK representation\cite{Karandashev_Lilienfeld:2022} which we combined with GFN2-xTB to generate orbital representations of molecules; \appref{app:hyperparameter_optimization} documents the hyperparameter optimization protocol used.

\section{Theory\label{sec:theory}}

\subsection{Kernel Ridge Regression in space of molecular conformers\label{subsec:classic_KRR}}

Given a training set of molecules $\mol{A}_{i}$ ($i=1,\ldots,\Ntrain$), which we for now assume to be represented by atomic coordinates and nuclear charges, and the corresponding quantities of interest $q_{i}$, the KRR estimate of the quantity of interest for a query molecule $\mol{B}$ is
\begin{equation}
q_{\mathrm{KRR}}(\mol{B})=\sum_{i=1}^{\Ntrain}\alpha_{i}^{\mathrm{KRR}}\kfunc(\mol{A}_{i},\mol{B}),
\label{eq:KRR_definition}
\end{equation}
where $\kfunc$ is the kernel function and $\vec{\alpha}^{\mathrm{KRR}}$ are regression coefficients solving the equation
\begin{equation}
(\mat{K}+\lambda \UMat{\Ntrain})\vec{\alpha}^{\mathrm{KRR}}=\vec{q},
\label{eq:KRR_alpha_coeffs}
\end{equation}
where $\mat{K}$ is a matrix such that $K_{ij}=\kfunc(\mol{A}_{i},\mol{A}_{j})$, $\UMat{\Ntrain}$ is the $\Ntrain\times\Ntrain$ unity matrix, $\lambda>0$ is a hyperparameter introduced for regularization. Equations~\eqref{eq:KRR_definition} and~\eqref{eq:KRR_alpha_coeffs} can be viewed as linear interpolation of quantities $q_{i}$ in a vector space of infinite dimensionality, with $\kfunc(\mol{A},\mol{B})$ being the dot product of vectors in that space where $\mol{A}$ and $\mol{B}$ are mapped \cite{Micchelli_Zhang:2006}.

We will now outline three strategies for representing molecular conformers and defining the kernel function $k$ in terms of those representations that were used in this work. The first strategy is to represent each molecule $\mol{A}$ with a vector $\vec{X}^{(\mol{A})}$ (referred to as the molecule's \emph{global representation}) and then calculate the kernel function $ k(\mol{A},\mol{B})$ using the \emph{Gaussian kernel function}
\begin{equation}
\kglobal(\mol{A},\mol{B}):=\exp\left(-\frac{|\vec{X}^{(\mol{A})}-\vec{X}^{(\mol{B})}|^{2}}{2\sigma^{2}}\right),
\label{eq:global_Gaussian}
\end{equation}
where $|\ldots|$ is a vector's $\ltwo$ norm and $\sigma$ is a hyperparameter. Examples of such molecular representations considered in this work are Coulomb Matrix\cite{Rupp_Lilienfeld:2012,Ramakrishnan_Lilienfeld:2015_CM} (CM) and Spectrum
of London and Axilrod-Teller-Muto potential\cite{Huang_Lilienfeld:2020} (SLATM).

The second strategy is to represent each atom in molecule $\mol{A}$ with a vector, the resulting molecular representation being a matrix $\mat{X}^{(\mol{A})}$ (referred to as the molecule's \emph{local representation}). We consider two popular choices for defining the corresponding kernel function, the first one being the \emph{atomic-sum Gaussian kernel function}\cite{Bartok_Csanyi:2013}
\begin{equation}
\klocal(\mol{A},\mol{B}):=\sum_{j^{\prime}=1}^{\Natoms^{(\mol{A})}}\sum_{j^{\prime\prime}=1}^{\Natoms^{(\mol{B})}}\kglobal(\vec{X}_{j^{\prime}}^{(\mol{A})},\vec{X}_{j^{\prime\prime}}^{(\mol{B})}),
\label{eq:local_Gaussian}
\end{equation}
where $\vec{X}_{j^{\prime}}^{(\mol{A})}$ and $\vec{X}_{j^{\prime\prime}}^{(\mol{B})}$ are the atomic representation vectors of atoms with indices $j^{\prime}$ and $j^{\prime\prime}$ in molecules $\mol{A}$ and $\mol{B}$, while $\Natoms^{(\mol{A})}$ and $\Natoms^{(\mol{B})}$ are numbers of atoms in molecules $\mol{A}$ and $\mol{B}$.  $\klocal$ is particularly useful for ML of extensive molecular quantities (\emph{e.g.} total potential energy); an example of a representation that can be used with $\klocal$ is Smooth Overlap of Atomic Orbitals (SOAP) descriptors\cite{Bartok_Csanyi:2013,De_Ceriotti:2016}. We note in passing that in the context of Gaussian Process Regression (GPR) $\kglobal$~\eqref{eq:global_Gaussian} is the squared-exponential kernel, while $\klocal$~\eqref{eq:local_Gaussian} is an additive kernel, which is a consequence of the connection between GPR and KRR discussed in Ref.~\onlinecite{Deringer_Csanyi:2021}. A popular version of $\klocal$~\eqref{eq:local_Gaussian} is the \emph{element-screened atomic-sum Gaussian kernel}\cite{Christensen_Lilienfeld:2020}
\begin{equation}
\kdlocal(\mol{A},\mol{B}):=\sum_{j^{\prime}=1}^{\Natoms^{(\mol{A})}}\sum_{j^{\prime\prime}=1}^{\Natoms^{(\mol{B})}}\delta(\nc_{j^{\prime}}^{(\mol{A})}-\nc_{j^{\prime\prime}}^{(\mol{B})})\kglobal(\vec{X}_{j^{\prime}}^{(\mol{A})},\vec{X}_{j^{\prime\prime}}^{(\mol{B})}),
\label{eq:dlocal_Gaussian}
\end{equation}
where $n_{j^{\prime}}^{(A)}$ and $n_{j^{\prime\prime}}^{(B)}$ are nuclear charges of atoms $j^{\prime}$ and $j^{\prime\prime}$ in molecules $\mol{A}$ and $\mol{B}$, and $\delta$ is the Kroneker delta function. Examples of representations used with $\kdlocal$ considered in this work are revised version of Faber-Christensen-Huang-Lilienfeld representation (FCHL19)~\cite{Faber_Lilienfeld:2018,Christensen_Lilienfeld:2020}, atomic version of SLATM (aSLATM)~\cite{Huang_Lilienfeld:2020}, and convolutional version of Many Body Distribution Functional (cMBDF)~\cite{Khan_Lilienfeld:2023,Khan_Lilienfeld:2024}.

The third strategy we will consider was originally introduced for the FJK representation~\cite{Karandashev_Lilienfeld:2022} (named so due to Fock $\mat{F}$, Coulomb $\mat{J}$, and exchange $\mat{K}$ used to generate it). The representation of molecule $\mol{A}$ takes results of cheap Hartree-Fock (HF) calculations, among them properties of localized orbitals $\locorb_{o}^{(\mol{A})}$ ($o=1,\ldots,\Norbs^{(\mol{A})}$, where $\Norbs^{(\mol{A})}$ is the number of occupied orbitals in molecule $\mol{A}$), then assigns a weight $\orbweight_{j}(\locorb_{o}^{(\mol{A})})$ and vector representation $\vec{v}_{j}(\locorb_{o}^{(\mol{A})})$ to each contribution to the localized orbital from atomic orbitals centered on an atom with index $j$ ($j\in 1,\ldots,\Natoms^{(\mol{A})}$). The corresponding expression for kernel function between molecules $\mol{A}$ and $\mol{B}$ can be written as
\begin{align}
\kFJK(\mol{A},\mol{B})&:=\sum_{o^{\prime}=1}^{\Norbs^{(\mol{A})}}\sum_{o^{\prime\prime}=1}^{\Norbs^{(\mol{B})}}\korbgauss(\locorb_{o^{\prime}}^{(\mol{A})},\locorb_{o^{\prime\prime}}^{(\mol{B})}),
\label{eq:FJK_kernel}\\
\korbgauss(\locorb_{o^{\prime}}^{(\mol{A})},\locorb_{o^{\prime\prime}}^{(\mol{B})})&:=\exp\left\{-\frac{1}{\sigmaglobal^{2}}\left[1-\frac{\korblin(\locorb_{o^{\prime}}^{(\mol{A})},\locorb_{o^{\prime\prime}}^{(\mol{B})})}{\sqrt{\korblin(\locorb_{o^{\prime}}^{(\mol{A})},\locorb_{o^{\prime}}^{(\mol{A})})\korblin(\locorb_{o^{\prime\prime}}^{(\mol{B})},\locorb_{o^{\prime\prime}}^{(\mol{B})})}}\right]\right\},\label{eq:FJK_gauss_orb_kernel}\\
\korblin(\locorb_{o^{\prime}}^{(\mol{A})},\locorb_{o^{\prime\prime}}^{(\mol{B})})&:=\sum_{j^{\prime}=1}^{\Natoms^{(\mol{A})}}\sum_{j^{\prime\prime}=1}^{\Natoms^{(\mol{B})}}\orbweight_{j^{\prime}}(\locorb_{o^{\prime}}^{(\mol{A})})\orbweight_{j^{\prime\prime}}(\locorb_{o^{\prime\prime}}^{(\mol{B})})\exp\left[\frac{|\vec{v}_{j^{\prime}}(\locorb_{o^{\prime}}^{(\mol{A})})-\vec{v}_{j^{\prime\prime}}(\locorb_{o^{\prime\prime}}^{(\mol{B})})|^{2}}{2\vec{\sigma}\odot\vec{\sigma}}\right]\label{eq:FJK_lin_orb_kernel}
\end{align}
where $\korblin$ and $\korbgauss$ are linear and Gaussian kernel functions for localized orbitals, $\sigma_{\mathrm{global}}$ is a hyperparameter, $\vec{\sigma}$ is a vector of hyperparameters of the same dimensionality as $\vec{v}_{j}(\locorb_{o}^{(\mol{A})})$ (the implied division by this vector is element-wise), and $\odot$ stands for element-wise multiplication.\footnote{The definition of $\korblin$~\eqref{eq:FJK_lin_orb_kernel} differs from the original one in Ref.~\onlinecite{Karandashev_Lilienfeld:2022} by rescaling $\vec{\sigma}$  by a factor of $\sqrt{2}$.} The number of weights $\orbweight_{j^{\prime}}(\locorb_{o^{\prime}}^{(\mol{A})})$ and $\orbweight_{j^{\prime\prime}}(\locorb_{o^{\prime\prime}}^{(\mol{B})})$ that are not exactly zero does not grow with system size, meaning the same can be said about the computational cost of evaluating $\korblin$~\eqref{eq:FJK_lin_orb_kernel}.

While $\kFJK$ was used for ML of extensive quantities, Ref.~\onlinecite{Karandashev_Lilienfeld:2022} additionally considered mapping properties associated with changes of energy due to changes of electronic structure (such as HOMO energies, associated to energy changes due to change of HOMO occupation number via Koopman's theorem) onto pairs of Slater determinants, each given a separate FJK representation. The kernel function for such Slater determinant pairs $\slpair{A}{1}{2}$ and $\slpair{B}{1}{2}$ reads
\begin{equation}
\kFJK^{\mathrm{pair}}(\slpair{A}{1}{2},\slpair{B}{1}{2})=\kFJK(\mol{A}_{1},\mol{B}_{1})+\kFJK(\mol{A}_{2},\mol{B}_{2})-\kFJK(\mol{A}_{1},\mol{B}_{2})-\kFJK(\mol{A}_{2},\mol{B}_{1}),
\label{eq:FJK_pair_kernel}
\end{equation}
where $\mol{A}_{1}/\mol{A}_{2}$ and $\mol{B}_{1}/\mol{B}_{2}$ are the Slater determinants constituting $\slpair{A}{1}{2}$ and $\slpair{B}{1}{2}$. The expression features cancellation between localized orbitals that are not affected by the change of electronic structure represented by the Slater determinant pair, which allowed FJK models to ``cut off'' parts of the molecule not relevant to the intensive quantity of interest while staying conceptually simple.

Lastly, as will be discussed in Section~\ref{sec:numerical_details} we encountered difficulties generating cheap HF calculations for systems relevant for this work, prompting us to switch towards GFN2-xTB extended tight-binding method\cite{Bannwarth_Grimme:2019} and develop a modification of FJK that could be used together with it. The latter was called ``FJD'' since the most prominent change is replacing terms derived from the $\mat{K}$ matrix with those derived from dipole operator. While the details behind the derivation are left for~\appref{app:fjd_formalism}, here we note that FJK and FJD share completely identical KRR and MSORF machinery.

\subsection{Rewriting kernels in terms of Structured Orthogonal Random Features\label{subsec:SORF_definitions}}

SORF is based on Random Fourier Features (RFF) formalism~\cite{Rahimi_Recht:2007}, which approximates the infinite dimensional vectors on which molecules are mapped in KRR with finite dimensional \emph{feature vectors} $\vec{z}$ of dimensionality $\Nfeatures$, resulting in the kernel function
\begin{equation}
\kSORF(\mol{A},\mol{B})=\vec{z}(\mol{A})\cdot\vec{z}(\mol{B}),
\label{eq:SORF_kernel}
\end{equation}
where ``$\cdot$'' is dot product. Rewriting Equations~\eqref{eq:KRR_definition} and~\eqref{eq:KRR_alpha_coeffs} in terms of $\kSORF$ results in the following expression for the SORF estimate of a quantity
\begin{equation}
q_{\mathrm{SORF}}(\mol{B})=\vec{\alpha}\cdot\vec{z}(\mol{B})
\label{eq:SORF_definition}
\end{equation}
with the regression coefficients $\vec{\alpha}$ found by solving the equation
\begin{equation}
\left(\mat{Z}^{\transpose}\mat{Z}+\lambda\UMat{\Nfeatures}\right)\vec{\alpha}=\mat{Z}^{\transpose}\vec{q},
\label{eq:SORF_alpha_coeffs}
\end{equation}
where $\mat{Z}$ is a matrix such that $\vec{Z}_{i}=\vec{z}(\mol{A}_{i})$. Though the expression implies finding $\vec{\alpha}$ by solving a $\Nfeatures\times\Nfeatures$ system of linear equations, the actual scaling is $\mathcal{O}[\Ntrain\Nfeatures\min(\Ntrain, \Nfeatures)]$ (which can be seen by rewriting Equation~\eqref{eq:SORF_alpha_coeffs} in terms of Singular Value Decomposition (SVD)~\cite{Petersen_Pedersen:2008} of the $\mat{Z}$ matrix), which made for a significant difference for our applications where $\Ntrain$ was always much smaller than $\Nfeatures$. There are two significant advantages of Equations~\eqref{eq:SORF_definition} and~\eqref{eq:SORF_alpha_coeffs} over their KRR versions~\eqref{eq:KRR_definition} and~\eqref{eq:KRR_alpha_coeffs}. The first one is that for larger datasets it is possible to choose $\Nfeatures$ smaller than $\Ntrain$, decreasing the computational time cost and memory requirements of finding $\vec{\alpha}$; it is important to note however that there are limits to how much $\Nfeatures$ can be decreased compared to $\Ntrain$ before SORF accuracy decreases compared to exact KRR, as observed in  Ref.~\onlinecite{Browning_Lilienfeld:2022} where learning curves started to flatten at some point after $\Ntrain$ surpassed $\Nfeatures$. The second one is most relevant for kernel functions whose evaluation scales bilinearly w.r.t. $\Natoms^{(\mol{A})}$ and $\Natoms^{(\mol{B})}$, such as $\klocal$~\eqref{eq:local_Gaussian}, $\kdlocal$~\eqref{eq:dlocal_Gaussian}, and $\kFJK$~\eqref{eq:FJK_kernel}; as will be demonstrated later evaluating the corresponding feature vectors $\vec{z}$ is linear w.r.t. the size of molecule of interest, eliminating computational cost scaling w.r.t. training set system size for calculating ML predictions and improving said scaling for finding the regression coefficients.

If the system of interest is represented by a single vector of size $\Ninit$ (which is a power of $2$) and the global Gaussian kernel~\eqref{eq:global_Gaussian} is used, the vectors $\vec{z}$ are generated with the standard SORF expression
\begin{equation}
\vec{\SORF}(\vec{X}^{(\mol{A})}, \randsigndiagtens,\randomphasestens,\sigma):=\sqrt{\frac{2}{\Nfeatures}}\left\{\cos\left[\frac{\sqrt{\Ninit}}{\sigma}\left(\prod_{t=1}^{\Ntransforms}\mat{H}_{\Ninit}\randsigndiag_{st}\right)\vec{X}^{(\mol{A})}+\randomphases_{s}\right]\right\}_{s=1}^{\Nstacks}.
\label{eq:base_SORF_features}
\end{equation}
Here $\mat{H}_{\Ninit}$ is $ \Ninit\times\Ninit$ Hadamard matrix~\cite{Fino_Algazi:1976}, $\randsigndiagtens$ is an $\Nstacks\times\Ntransforms$ array of $\Ninit\times\Ninit$ diagonal matrices $\mathbf{D}_{st}$ ($s=1,\ldots,\Nstacks$, $t=1,\ldots,\Ntransforms$), each diagonal element randomly chosen between $1$ and $-1$, $\randomphasestens$ is an $\Nstacks$ array of $\Ninit$ long vectors $\randomphases_{s}$, each value randomly chosen from the interval $[0,2\pi)$, $\{\ldots\}_{s=1}^{\Nstacks}$ stands for concatenating the $\Nstacks$ vectors into a single vector; the implied calculation of the cosine function is element-wise. Products of the Hadamard matrix with a vector are evaluated with fast Walsh-Hadamard transform (hence the name of the $\Ntransforms$ parameter) whose computational cost is $\mathcal{O}(\Ninit\ln\Ninit)$. Note that $\Nfeatures=\Ninit\Nstacks$. It can be demonstrated~\cite{Yu_Kumar:2016} that
\begin{equation}
\vec{\SORF}(\vec{X}^{(\mol{A})}, \randsigndiagtens,\randomphasestens,\sigma)\cdot\vec{\SORF}(\vec{X}^{(\mol{B})},\randsigndiagtens,\randomphasestens,\sigma)\approx\kglobal(\mol{A}, \mol{B}),
\label{eq:SORF_product_approximation}
\end{equation}
with the approximation introducing a statistical error which can be decreased by increasing $ \Nstacks$ and a systematic bias which can be decreased by increasing $\Ninit$~\cite{Yu_Kumar:2016} (\emph{e.g.} by padding the input vector with zeros). 
We note at this point that all the applications discussed in this work in the context of SORF can be analogously performed with other procedures~\cite{Reid_Weller:2023,Li_Mak:2024} for generating feature vectors that approximate $\kglobal$ analogously to Equation~\eqref{eq:SORF_product_approximation}. However, in this work we used SORF due to its balance of good numerical performance and algorithmic simplicity.

Using SORF can be considered~\cite{Lu_Shafto:2022} using a single-layer NN (the corresponding layer will be called $\SORF$ from now on), and we will now add layers to it to accommodate more complex kernel expressions. We first introduce the \emph{sum layer} $\SUM$, with $\SUM\satoms$ referring to sum over atoms. Combination of the latter with $\SORF$, which we will refer to as $\SUM\satoms\SORF$, is defined as
\begin{equation}
\SUM\satoms\SORF(\mat{X}^{(\mol{A})},\randsigndiagtens,\randomphasestens,\sigma):=\sum_{j=1}^{\Natoms^{(\mol{A})}}\SORF(\vec{X}_{j}^{(\mol{A})},\randsigndiagtens,\randomphasestens,\sigma).
\end{equation}
It is evident that dot product of $\SUM\satoms\SORF$ features reproduces $\klocal$~\eqref{eq:local_Gaussian}. To reproduce $\kdlocal$~\eqref{eq:dlocal_Gaussian} we additionally introduce the \emph{switch layer} $\SWITCH$, applicable if a vector $\vec{X}$ corresponding to an atom is augmented with the atom's nuclear charge $n$. Writing $\SWITCH\SORF$ means a separate $\SORF$ layer is initialized for each possible nuclear charge, with $\SUM\satoms\SWITCH\SORF$ written as
\begin{equation}
\SUM\satoms\SWITCH\SORF(\mat{X}^{(\mol{A})},\sigma):=\sum_{j}^{\Natoms^{(\mol{A})}}\SORF[\vec{X}_{j}^{(\mol{A})},\randsigndiagtens(n_{j}^{(\mol{A})}),\randomphasestens(n_{j}^{(\mol{A})}),\sigma],
\end{equation}
where $\randsigndiagtens(n_{j}^{(\mol{A})})$ and $\randomphasestens(n_{j}^{(\mol{A})})$ are tensors analogous to $\randsigndiagtens$ and $\randomphasestens$ in Equation~\eqref{eq:base_SORF_features} that have been initialized specifically for atoms with nuclear charge $n_{j}^{(\mol{A})}$. $\SUM\satoms\SWITCH\SORF$ features reproduce $\kdlocal$~\eqref{eq:dlocal_Gaussian} as their dot product and were previously used in Ref.~\onlinecite{Browning_Lilienfeld:2022}.

To reproduce the nested exponential expressions appearing in $\kFJK$~\eqref{eq:FJK_kernel}-\eqref{eq:FJK_lin_orb_kernel} we use $\SORF$~\eqref{eq:base_SORF_features} to generate vectors whose dot product reproduces $\korblin$~\eqref{eq:FJK_lin_orb_kernel} and note that applying $\kglobal$~\eqref{eq:global_Gaussian} to a pair of such vectors (after normalization) approximates $\korbgauss$~\eqref{eq:FJK_gauss_orb_kernel}. We introduce \emph{weighted sum layer} $\WEIGHTEDSUM$, with $\WEIGHTEDSUM\satoms$ corresponding to summation over atoms, that allows us to write features corresponding to $\korblin$~\eqref{eq:FJK_lin_orb_kernel} as
\begin{equation}
\WEIGHTEDSUM\satoms\SORF(\locorb^{\mol{A}}_{o}):=\sum_{j=1}^{\Natoms^{(\mol{A})}}\orbweight_{j}(\locorb_{o}^{(\mol{A})})\SORF\left[\vec{v}_{j}(\locorb_{o}^{(\mol{A})})/\vec{\sigma},\randsigndiagtens^{\mathrm{orb}},\randomphasestens^{\mathrm{orb}},1,\right].
\label{eq:linorb_features}
\end{equation}
Additionally defining \emph{normalization layer} $\NORM$ that divides a vector by its $l^{2}$ norm allows us to reproduce $\kFJK$~\eqref{eq:FJK_kernel} with features generated by $\SUM\sorbs\SORF\NORM\WEIGHTEDSUM\satoms\SORF$, formally written as
\begin{align}
\SUM\sorbs\SORF\NORM\WEIGHTEDSUM\satoms\SORF(\mol{A}):=&\sum_{o=1}^{\Norbs^{(\mol{A})}}\SORF\NORM\WEIGHTEDSUM\satoms\SORF(\locorb_{o}^{(\mol{A})})\\
\SORF\NORM\WEIGHTEDSUM\satoms\SORF(\locorb_{o}^{(\mol{A})}):=&\SORF\left[\frac{\WEIGHTEDSUM\satoms\SORF(\locorb^{(\mol{A})}_{o})}{\left|\WEIGHTEDSUM\satoms\SORF(\locorb^{(\mol{A})}_{o})\right|},\randsigndiagtens^{\mathrm{mol}},\randomphasestens^{\mathrm{mol}},\sigmaglobal\right]
\label{eq:gaussorb_features}
\end{align}

$\SUM\sorbs\SORF\NORM\WEIGHTEDSUM\satoms\SORF$ features allow to reproduce $\kFJK^{\mathrm{pair}}$~\eqref{eq:FJK_pair_kernel} as
\begin{equation}
\begin{split}
\kFJK^{\mathrm{pair}}(\slpair{A}{1}{2},\slpair{B}{1}{2})\approx & [\SUM\sorbs\SORF\NORM\WEIGHTEDSUM\satoms\SORF(\mol{A}_{2})-\SUM\sorbs\SORF\NORM\WEIGHTEDSUM\satoms\SORF(\mol{A}_{1})]\\
& \cdot[\SUM\sorbs\SORF\NORM\WEIGHTEDSUM\satoms\SORF(\mol{B}_{2})-\SUM\sorbs\SORF\NORM\WEIGHTEDSUM\satoms\SORF(\mol{B}_{1})].
\label{eq:FJK_pair_kernel_approx}
\end{split}
\end{equation}
We calculated $\SUM\sorbs\SORF\NORM\WEIGHTEDSUM\satoms\SORF(\mol{A}_{2})-\SUM\sorbs\SORF\NORM\WEIGHTEDSUM\satoms\SORF(\mol{A}_{1})$ by assigning representations of localized orbitals in $\mol{A}_{1}$ and $\mol{A}_{2}$ weights of $-1$ and $1$ and applying $\WEIGHTEDSUM\sorbs\SORF\NORM\WEIGHTEDSUM\satoms\SORF$ to them. Note that contributions to the feature vectors from localized orbitals that do not change between $\mol{A}_{1}$ and $\mol{A}_{2}$ cancel out, as is the case for $\kFJK^{\mathrm{pair}}(\slpair{A}{1}{2},\slpair{B}{1}{2})$~\eqref{eq:FJK_pair_kernel}. To simplify discussion, we will assume that for the single determinant case $\WEIGHTEDSUM\sorbs\SORF\NORM\WEIGHTEDSUM\satoms\SORF$ is used as well, with each localized orbital given a weight of $1$ (which is also the way FJD representations were processed in our code implementation).

Lastly, we note that $\Ehyd$ is an example of a quantity for which pairs $\slpair{A}{1}{2}$ can also be defined (in this case molecule in vacuum vs. solvated molecule) and the ML model could potentially benefit from information on both members of the pair, yet discarding contribution from parts of the molecule that do not change between $\mol{A}_{1}$ and $\mol{A}_{2}$ via the cancellation exhibited by $\kFJK^{\mathrm{pair}}$~\eqref{eq:FJK_pair_kernel} would be incorrect; in the specific case of solvation that happens because parts of the molecule not interacting with the solvent still contribute to $\Ehyd$ via solvent-excluded volume\cite{ehlert2021alpb}. We therefore also introduced \emph{concatenation layer} $\CONCATENATE$ that when acting on a pair $\mol{A}_{1}$ and $\mol{A}_{2}$ concatenates feature vectors generated for them by previous layers into a single feature vector. The corresponding kernel function reads
\begin{equation}
\kFJK^{\mathrm{pair,conc}}(\slpair{A}{1}{2},\slpair{B}{1}{2})=\kFJK(\mol{A}_{1},\mol{B}_{1})+\kFJK(\mol{A}_{2},\mol{B}_{2}).
\label{eq:k_pair_sum}
\end{equation}

\subsection{Applications to ensembles of conformers\label{subsec:ensemble_app}}

Consider doing ML of an extensive quantity of a molecule $\mol{A}$ about which we know conformers with indices $\confid=1,\ldots,\Nconf^{(\mol{A})}$, each assigned a matrix of atomic representations $\mat{X}_{c}^{(\mol{A)}}$ and a nonnegative Boltzmann weight $w_{c}^{\mathrm{b}(\mol{A})}$. A natural way to extend $\klocal$~\eqref{eq:local_Gaussian} to ensembles would be with the kernel function
\begin{equation}
\klocal^{\mathrm{ens}}=\sum_{\confid^{\prime}=1}^{\Nconf^{(\mol{A})}}\sum_{\confid^{\prime\prime}=1}^{\Nconf^{(\mol{B})}}w_{\confid^{\prime}}^{\mathrm{b}(\mol{A})}w_{\confid^{\prime\prime}}^{\mathrm{b}(\mol{B})}\klocal(\mat{X}_{\confid^{\prime}}^{(\mol{A})},\mat{X}_{\confid^{\prime\prime}}^{(\mol{B})}).
\label{eq:simple_ensemble_kernel}
\end{equation}
The cost of evaluating this kernel function is $\mathcal{O}(\Nconf^{(\mol{A})}\Nconf^{(\mol{B})}\Natoms^{(\mol{A})}\Natoms^{(\mol{B})})$, making the motivation to instead use SORF even stronger than for $\klocal$; in terms of Subsection~\ref{subsec:SORF_definitions}, the combination of layers corresponding to $\klocal^{\mathrm{ens}}$ is $\WEIGHTEDSUM\sconfs\SUM\satoms\SORF$ (with $\WEIGHTEDSUM\sconfs$ corresponding to summation over conformers). Using $\WEIGHTEDSUM\sconfs\SUM\satoms\SORF$ though leaves the ML algorithm unable to correct for $\boltzweightvec$ potentially inaccurately representing relative importance of different conformers. This motivated introduction of an alternative expression, $\MIXEDEXTENSIVE\WEIGHTEDSUM\sconfs\MIXEDEXTENSIVE\SUM\satoms\SORF$, where the \emph{mixed-extensive} layer $\MIXEDEXTENSIVE$ is defined as
\begin{equation}
\vec{\MIXEDEXTENSIVE}(\vec{X},\randsigndiagtens,\randomphasestens,\sigma,\mixsigma):=\frac{1}{\sqrt{\Nstacks(1+\mixsigma^{-1})}}\left\{\vec{X}\right\}_{s=1}^{\Nstacks}+\frac{|\vec{X}|}{\sqrt{1+\mixsigma}}\vec{F}\left(\frac{\vec{X}}{|\vec{X}|},\randsigndiagtens,\randomphases,\sigma\right),
\label{eq:mixedextensive_def}
\end{equation}
where $\mixsigma$ is a nonnegative hyperparameter. $\mixsigma$ controls switching between $\klocal^{\mathrm{ens}}$~\eqref{eq:simple_ensemble_kernel} and a different kernel that is related to $\klocal^{\mathrm{ens}}$ in a way similar to how $\korbgauss$~\eqref{eq:FJK_gauss_orb_kernel} is related to $\korblin$~\eqref{eq:FJK_lin_orb_kernel}. Note that for $\posconst>0$
\begin{equation}
\MIXEDEXTENSIVE(\posconst\vec{X}):=\posconst\MIXEDEXTENSIVE(\vec{X}),
\end{equation}
meaning that while quantity predictions of $\MIXEDEXTENSIVE\WEIGHTEDSUM\sconfs\MIXEDEXTENSIVE\SUM\satoms\SORF$ are not truly extensive, they demonstrate correct scaling with molecule's size. The case of representations using $\kdlocal$~\eqref{eq:dlocal_Gaussian} is completely analogous and yields $\WEIGHTEDSUM\sconfs\SUM\satoms\SWITCH\SORF$ and $\MIXEDEXTENSIVE\WEIGHTEDSUM\sconfs\MIXEDEXTENSIVE\SUM\satoms\SWITCH\SORF$ as possible layer combinations.

Since global representations do not allow to implicitly account for a quantity's extensivity during model construction, for them we simply used $\SORF\NORM\WEIGHTEDSUM\sconfs\SORF$, normalization layer added to remove unphysical dependence of the result on how the importance weights are normalized. For FJD models only using one Slater determinant (referred to from now on as \emph{single-determinant FJD}), considerations analogous to the case of local representations led us to consider $\WEIGHTEDSUM\sconfs\WEIGHTEDSUM\sorbs\SORF\NORM\WEIGHTEDSUM\satoms\SORF$ and $\MIXEDEXTENSIVE\WEIGHTEDSUM\sconfs\MIXEDEXTENSIVE\WEIGHTEDSUM\sorbs\SORF\NORM\WEIGHTEDSUM\satoms\SORF$ for calculating extensive properties. As mentioned in Subsec.~\ref{subsec:SORF_definitions}, we also tried generating FJD representations from orbitals obtained both for the molecule in solvent and in vacuum (from now on referred to as \emph{pair-determinant FJD}, or just FJD). Addition of the corresponding concatenation layers $\CONCATENATE$ yielded $\WEIGHTEDSUM\sconfs\CONCATENATE\WEIGHTEDSUM\sorbs\SORF\NORM\WEIGHTEDSUM\satoms\SORF$ and $\MIXEDEXTENSIVE\WEIGHTEDSUM\sconfs\MIXEDEXTENSIVE\CONCATENATE\WEIGHTEDSUM\sorbs\SORF\NORM\WEIGHTEDSUM\satoms\SORF$.

For intensive properties, we aim to create models whose prediction is invariant w.r.t. ``molecule multiplication'', \emph{i.e.} creating several copies of a molecule far away from each other. For global representations the requirement again cannot be satisfied implicitly, thus simple $\SORF\NORM\WEIGHTEDSUM\sconfs\SORF$ was used. For local representations $\NORM\SUM\satoms\SORF$ or $\NORM\SUM\satoms\SWITCH\SORF$ generates a feature vector for a conformer invariant w.r.t. molecule multiplication; treating it as a global representation vector yields $\SORF\NORM\WEIGHTEDSUM\sconfs\SORF\NORM\SUM\satoms\SORF$ and $\SORF\NORM\WEIGHTEDSUM\sconfs\SORF\NORM\SUM\satoms\SWITCH\SORF$ as the final layer combinations.

Treatment of intensive quantities with FJD depends on whether the molecule is represented by a single determinant or a pair of Slater determinants (as is the case for FreeSolv, we will drop ``pair-determinant'' for brevity when calling this FJD variation). For single Slater determinant case, analogously to the case of local representations, we obtain $\SORF\NORM\WEIGHTEDSUM\sconfs\SORF\NORM\WEIGHTEDSUM\sorbs\SORF\NORM\WEIGHTEDSUM\satoms\SORF$. For pairs of Slater determinants, we note that the differences of features appearing in $\kFJK^{\mathrm{pair}}$ are already invariant w.r.t. ``molecule multiplication,'' making normalization of conformer features excessive. The final expression is therefore $\SORF\NORM\WEIGHTEDSUM\sconfs\SORF\WEIGHTEDSUM\sorbs\SORF\NORM\WEIGHTEDSUM\satoms\SORF$. All layer combinations considered in this work for different representations and quantities are summarized in Table~\ref{tab:layer_cheatsheet}.

\begin{table}
\begin{tabular}{lccc}
\toprule
\multirow{2}{*}{rep. type} & \multicolumn{3}{c}{quantity} \\
\cline{2-4}& intensive & extensive & extensive (m.-ext.)\\ 
\midrule
global & \(\displaystyle\SORF\NORM\WEIGHTEDSUM\sconfs\SORF\) & 
(same as for intensive) & (same as for intensive)\\
local (w. $\klocal$)&
\(\displaystyle \SORF\NORM\WEIGHTEDSUM\sconfs\SORF\NORM\SUM\satoms\SORF\) &
\(\displaystyle \WEIGHTEDSUM\sconfs\SUM\satoms\SORF\) &
\(\displaystyle \MIXEDEXTENSIVE\WEIGHTEDSUM\sconfs\MIXEDEXTENSIVE\SUM\satoms\SORF\) \\
local (w. $\kdlocal$)&
\(\displaystyle \SORF\NORM\WEIGHTEDSUM\sconfs\SORF\NORM\SUM\satoms\SWITCH\SORF\) &
\(\displaystyle \WEIGHTEDSUM\sconfs\SUM\satoms\SWITCH\SORF\) &
\(\displaystyle \MIXEDEXTENSIVE\WEIGHTEDSUM\sconfs\MIXEDEXTENSIVE\SUM\satoms\SWITCH\SORF\) \\
FJD &
\(\displaystyle\SORF\NORM\WEIGHTEDSUM\sconfs\SORF\WEIGHTEDSUM\sorbs\SORF\NORM\WEIGHTEDSUM\satoms\SORF\)&
 \(\displaystyle\WEIGHTEDSUM\sconfs\CONCATENATE\WEIGHTEDSUM\sorbs\SORF\NORM\WEIGHTEDSUM\satoms\SORF\)&
 \(\displaystyle\MIXEDEXTENSIVE\WEIGHTEDSUM\sconfs\MIXEDEXTENSIVE\CONCATENATE\WEIGHTEDSUM\sorbs\SORF
 \NORM\WEIGHTEDSUM\satoms\SORF\)\\
FJD (s.-det.) & \(\displaystyle\SORF\NORM\WEIGHTEDSUM\sconfs\SORF\NORM\WEIGHTEDSUM\sorbs\SORF\NORM\WEIGHTEDSUM\satoms\SORF\) &  \(\displaystyle\WEIGHTEDSUM\sconfs\WEIGHTEDSUM\sorbs\SORF\NORM\WEIGHTEDSUM\satoms\SORF\) &  \(\displaystyle\MIXEDEXTENSIVE\WEIGHTEDSUM\sconfs\MIXEDEXTENSIVE\WEIGHTEDSUM\sorbs\SORF
 \NORM\WEIGHTEDSUM\satoms\SORF\)
\\
\bottomrule
\end{tabular}
\caption{Combinations of layers used in this work for different representations and quantities. ``Extensive (m.-ext.)'' means using the ``mixed-extensive'' layers $\MIXEDEXTENSIVE$~\eqref{eq:mixedextensive_def} to calculate an extensive property, ``FJD (s.-det.)'' means single-determinant FJD as opposed to the pair-determinant version. Representations considered in this work for each type are SLATM and CM for ``global'', SOAP for ``local (w. $\klocal$)'', FCHL19, cMBDF, and aSLATM for ``local (w. $\kdlocal$)''.\label{tab:layer_cheatsheet}}
\end{table}

We make two closing notes regarding how ensemble conformers were processed by our workflow. Firstly, in all examples considered in this work instead of Boltzmann weights we used their ``cut'' versions $\cutboltzweightvec$ defined as
\begin{equation}
    \cutboltzweight_{j}:=\frac{\mathrm{max}(\boltzweight_{j}-\rho_{-}, 0)}{\sum_{j^{\prime}=1}^{\Nconf}\mathrm{max}(\boltzweight_{j^{\prime}}-\rho_{-}, 0)},
    \label{eq:cut_weights_def}
\end{equation}
with $\rho_{-}$ such that
\begin{equation}
    \frac{\sum_{j=1}^{\Nconf}\mathrm{max}(\boltzweight_{j}-\rho_{-}, 0)}{\sum_{j=1}^{\Nconf}\boltzweight_{j}}=1-\rcut,
    \label{eq:cut_weights_companion_def}
\end{equation}
where $\rcut$ is a parameter chosen by the user; note that $\cutboltzweightvec$ changes continuously with $\boltzweight$. The procedure was introduced to avoid calculating features of conformers whose contribution is irrelevant. Secondly, our algorithm's implementation allows performing conformer generation not once, but several times to potentially account for statistical noise. Though we did not use the latter option and conformers were generated only once for each molecule, code-wise we were still performing ``summation over a single conformer generation.'' As a result in the actual scripts used in this work each $ \WEIGHTEDSUM\sconfs$ layer was preceded by an additional $\SUM$ layer corresponding to summation over conformer generations.

\section{Numerical details\label{sec:numerical_details}}

The MSORF code used in this work was implemented as part of QML2 package~\cite{QML2}, with the implementation based on NumPy~\cite{harris2020array}, Numba~\cite{lam2015numba}, and \SciPy~\cite{2020SciPy-NMeth}. By default we ran calculations with $\Nfeatures=32768$ at all $\SORF$ layers, since the value had performed well in Ref.~\onlinecite{Browning_Lilienfeld:2022}, and chose $\Ninit=\Nfeatures$ to minimize the non-statistical bias error introduced in the SORF product approximation~\eqref{eq:SORF_product_approximation} (the ``extra'' input vector values were padded with zeros). This could not be done straightforwardly for SLATM, aSLATM, and (in the case of LES) SOAP since their representation vectors corresponded to $\Ninit$ larger than $32768$. For SLATM, the issue was resolved by projecting all representation vectors onto linear space spanned by representation vectors of conformers in the training set, and then appending to the resulting vectors the $\ltwo$ distance between the projected and non-projected representation vector; note that using thus defined new representation vectors would not affect training and predictions of a KRR model. The procedure decreased $\Ninit$ to $4096$ for SLATM, but could not be used for aSLATM and SOAP due to large number of atomic environments in the training set. Ref.~\onlinecite{Browning_Lilienfeld:2022} used Principal Component Analysis to fit atomic representations into a small enough $\Ninit$, but we wanted to avoid that in order to accurately observe prediction errors of aSLATM and SOAP relative to other methods. Therefore, for aSLATM we used $\Ninit=131072$, for SOAP we used $\Ninit=65536$ for LES and $\Ninit=32768$ for FreeSolv.

We generated molecular conformers with the algorithm from Ref.~\onlinecite{Ebejer_Deane:2012}, as implemented in the Morfeus package~\cite{software:Morfeus}, at $T=298.15\,\mathrm{K}$ with the Merck molecular force field (MMFF94)~\cite{Halgren:1996_I,Halgren:1996_II,Halgren:1996_III,Halgren_Nachbar:1996_IV,Halgren:1996_V,Halgren:1999_VI,Halgren:1999_VII} as implemented~\cite{Tosco_Landrum:2014} in RDKit~\cite{software:RDKit}; unimportant conformers were cut off with $\rcut=0.05$. Morfeus calculations require a parameter $\Nmorfconf$ determining the number of conformers the package attempts to generate; we checked convergence of the $\mat{Z}\mat{Z}^{\transpose}$ matrix generated for one of the 4 training sets chosen for the learning curves (see Subsection~\ref{subsec:dataset_analysis_eval_metrics}) with cMBDF representations by scanning $\Nmorfconf=8,\ldots,1024$ and observed that the values were converged at $\Nmorfconf=32$ for both LES and FreeSolv, hence $\Nmorfconf=32$ was used throughout this work. In both cases, the $\sigma$ and $\vec{\sigma}$ hyperparameters used in $\SORF$ layers were taken to be the initial guess defined in~\appref{app:hyperparameter_optimization}. We used the same convergence criterion for optimizing $\Ntransforms$ by calculating a training set's $\mat{Z}\mat{Z}^{\transpose}$ $\Ntransforms=1,2,\ldots,6,7$ for each representation and combination of MSORF layers considered. We settled on using $\Ntransforms=3$ throughout the work (which is also consistent with observations in Refs.~\onlinecite{Yu_Kumar:2016} and~\onlinecite{Browning_Lilienfeld:2022}).

Almost all representation functions used in this work were calculated using the QML2 package, the exception being SOAP which was calculated with \libname{DScribe} package~\cite{dscribe,dscribe2}. While aSLATM, SLATM, FCHL19, and cMBDF were generated using default parameters present in QML2, SOAP was generated using $r_{\mathrm{cut}}^{\mathrm{SOAP}}=9.53$ \r{A}, $n_{\mathrm{max}}^{\mathrm{SOAP}}=8$, and $l_{\mathrm{max}}^{\mathrm{SOAP}}=8$, which were the parameters found to be optimal in Ref.~\onlinecite{Kirschbaum_Bande:2024} for FreeSolv computational hydration energies' prediction with SOAP descriptors and KRR. We note that optimizing beyond these last parameters was beyond the scope of this work.

FJD representations were generated from GFN2-xTB or, in case of charged ions, its spin-polarized version (sp)GFN2-xTB\cite{Neugebauer_Grimme:2023} calculations performed with the \libname{tblite}\cite{tblite} package, with orbitals localized using Pipek-Mezey localization\cite{pipek1989,lehtola2014pm} as implemented in the PySCF package~\cite{Sun:2015,Sun:2018,Sun:2020}; when needed presence of solvent was simulated with the xTB-parametrized\cite{ehlert2021alpb} analytical linearized Poisson--Boltzmann (ALPB) implicit solvation model. We initially considered generating FJK representations from minimal-basis HF calculations and later from Huckel calculations with a larger basis, but the former often failed to converge for charged species needed to generate determinant pairs for $\Eox$ predictions and representations generated from the latter failed to reach reasonable accuracies; the numerical results are available in Supplementary Materials for future reference. When referring to ``single Slater determinant'' FJD we will mean representations calculated from ground state GFN2-xTB calculations for conformers in solvent (acetonitrile for LES and water for FreeSolv); for ``pair Slater determinant'' FJD for LES we took ground states of the molecule at charges 0 and 1 [both solvated in acetonitrile, the charged state was calculated with (sp)GFN2-xTB\cite{Neugebauer_Grimme:2023}], while for FreeSolv we took ground states of the molecule in vacuum and solvated in water. For FJD we also used $l_{\mathrm{max}}=2$ and applied $\rcutorb=0.05$ to atomic contributions to localized orbitals (see Ref.~\onlinecite{Karandashev_Lilienfeld:2022} and~\appref{app:fjd_formalism}).

While the full procedure used to optimize hyperparameters is described in~\appref{app:hyperparameter_optimization}, here we note that apart from optimizing $\lambda$ value and different $\sigma$ hyperparameters, we also considered optimizing within the same framework quantity of interest by shifting it by a constant if it is intensive or by a stoichiometry-dependent shift if it is extensive; the latter procedure was inspired by and is a more sophisticated variation of the ``dressed'' atom approach~\cite{Hansen_Tkatchenko:2015,Huang_Lilienfeld:2021}. The main motivation for developing the procedure is given by Equations~\ref{eq:FJK_pair_kernel} and~\ref{eq:FJK_pair_kernel_approx}. When a KRR or MSORF procedure uses them, value 0 of the quantity of interest becomes physically meaningful as the value for a molecule $\mol{A}$ whose electronic structure does not change between the $\mol{A}_{1}/\mol{A}_{2}$ pair. While the procedure could prove useful for larger datasets, it did not improve prediction accuracy for the smaller datasets used in this work, except for one case that is discussed in the Supplementary Material. Therefore all data presented in Section~\ref{sec:results} have been obtained with standard dressed atom approach for FreeSolv and shifting $\Eox$ by its mean for LES, while results obtained while properly optimizing quantity shifts as hyperparameters are left for Supplementary Material.

Lastly, we tested some of our MSORF methods with $\Delta$-Machine Learning\cite{Ramakrishnan_Lilienfeld:2015_Delta} ($\Delta$-ML), for which the reference values were evaluated with the same calculations as used to generate FJD Slater determinant pairs, e.g. GFN2-xTB calculations on MMFF94-generated conformers. Namely, for LES the reference values were thus-computed and ensemble-weighted ionization energies (cation energy with positive sign, neutral energy with negative sign) calculated in acetonitrile; for FreeSolv - hydration energies. The GFN2-xTB baseline whose accuracy we will be comparing other methods against is defined for LES as the reference GFN2-xTB values shifted by an optimized intercept, and defined for FreeSolv as the reference GFN2-xTB values with a dressed atom stoichiometry-dependent model added. We also considered multiplying the GFN2-xTB values by an optimized slope coefficient, but as demonstrated in Supplementary Material its impact on baselines' prediction accuracy is negligible.

\section{Results and discussion\label{sec:results}}

\subsection{Dataset analysis and evaluation metrics}
\label{subsec:dataset_analysis_eval_metrics}

Both LES and FreeSolv are relatively small datasets, numbering 592 and 642 molecules respectively, making it particularly important to use learning metrics less affected by potential bias inside these datasets and more indicative of how adequately the models generalize from the datasets to the rest of chemical space. One of the most common tools for that, which has also been applied to FreeSolv\cite{Zang_Tang:2023,Qiao_Wei:2025}, is the frequency-ordered scaffold split\cite{Wu_Pande:2018}. The idea is that a dataset is divided into groups based on Bemis--Murcko (BM) scaffold\cite{Bemis_Murcko:1996} of a molecule and the training/test split is done in such a way that larger groups are picked first for the training set; thus groups of similar molecules are never split between train and test sets and the model is forced to extrapolate from as few such groups to as many such groups as possible for a given train/test ratio. Our main concern was that the most commonly used (including in Refs.~\onlinecite{Zang_Tang:2023,Qiao_Wei:2025}) definition of BM scaffolds puts non-cyclic molecules into a single group (\emph{``pool'' acyclic policy}); since both LES and FreeSolv have relatively large numbers of acyclic molecules, 67 and 320, we investigated a less strict alternative, where acyclic molecules are split into groups according to their heavy atom skeleton (\emph{``skeleton'' acyclic policy}). Another issue is that the default implementation of the frequency-ordered algorithm decides ties between groups of the same size considered for addition to the training set in a way which is not invariant with respect to random shuffling of the input data; we instead always had such ties decided randomly.

Table~\ref{tab:separation_analysis} summarizes comparative analysis of the two scaffold-based variants of defining the training and test set with 80/20 split (yielding training sets of 474 molecules for LES, and 514 molecules for FreeSolv), both to each other and to random split; the analysis was performed over 4 random number generator seeds and centered on finding for each test set molecule the training set molecule closest in terms of Tanimoto similarity w.r.t. ECFP4 representation\cite{Rogers_Hahn:2010} (Extended-Connectivity Fingerprint, diameter 4) and $\ltwo$ distance between SLATM representation vectors of MMFF94 coordinates generated by RDKit; full histogram plots can be found in Supplementary Material. We see that the difference between ``pool'' and ``skeleton'' acyclic policies does not affect train/test separation in LES much (at most ${\sim}15\%$ nearest neighbor value change), though both provide a much greater challenge than random split. For FreeSolv however, though scaffold split with ``skeleton'' policy is significantly better than random split for train/test separation, it is also significantly less strict than its pool counterpart. At the same time, the large acyclic group created by the ``pool'' policy drastically reduces the number of groups available to the training set, and with it the number of possible train/test partitions -- 5 for LES and only 1 for FreeSolv. To retain a split procedure with well-defined statistics w.r.t. shuffling we mostly used ``skeleton'' acyclic policy in the rest of the work, only switching to ``pool'' to have a fair comparison with prediction accuracies reported for FreeSolv using scaffold split with ``pool'' policy. From now on by ``scaffold split'' we will always mean ``frequency-ordered random-tie BM scaffold split with ``skeleton'' acyclic policy'' unless specified otherwise.

\begin{table}
\begin{tabular}{llcccccccccc}
\toprule
 \multirow{2}{*}{dataset} & \multirow{2}{*}{split} & \multicolumn{2}{c}{ECFP4 Tanimoto} & \phantom{.} & \multicolumn{2}{c}{SLATM distance} & \phantom{.} & \multicolumn{2}{c}{num. scaffolds} & \phantom{.} & distinct \\
\cline{3-4}\cline{6-7}\cline{9-10}  &  & median & mean &  & median & mean &  & total & train &  & splits \\
\midrule \multirow{3}{*}{FreeSolv} & random & $0.520$ & $0.542$ &  & $\phantom{0}4.20$ & $\phantom{0}6.00$ &  & \_ & \_ &  & \_ \\
  & scaffold (pool) & $0.250$ & $0.275$ &  & $12.83$ & $14.11$ &  & $\phantom{0}63$ & $\phantom{0}5$ &  & $1$ \\
  & scaffold (skeleton) & $0.389$ & $0.420$ &  & $10.46$ & $13.23$ &  & $136$ & $40$ &  & $2.1\times10^{5}$ \\
\midrule \multirow{3}{*}{LES} & random & $0.566$ & $0.562$ &  & $\phantom{0}7.12$ & $\phantom{0}7.51$ &  & \_ & \_ &  & \_ \\
  & scaffold (pool) & $0.357$ & $0.395$ &  & $10.94$ & $14.38$ &  & $112$ & $21$ &  & $5$ \\
  & scaffold (skeleton) & $0.385$ & $0.423$ &  & $10.46$ & $12.14$ &  & $139$ & $39$ &  & $1.1\times10^{7}$ \\
\bottomrule
\end{tabular}
\caption{Summary of training and test set separation analysis for FreeSolv and LES sets for 80/20 train/test split using random split (``random''), random-tie frequency-ordered Bemis--Murcko scaffold with acyclic molecules grouped by skeleton [``scaffold (skeleton)''] and with acyclic molecules pooled into a single group [``scaffold (pool)'']. Columns report, for each test molecule's nearest training-set neighbour, the median and mean ECFP4 Tanimoto similarity and SLATM $\ltwo$ distance, the total number of BM scaffold groups and the number falling in the training set, and the number of distinct train/test partitions reachable under random tie-breaking. Separation statistics are averaged over 4 random seeds; random-split rows have no scaffold count.}
\label{tab:separation_analysis}
\end{table}

Besides absolute prediction accuracies at 80/20 splits we also wanted to verify that all MSORF models tested in this work exhibit downwards learning curves. However, defining a learning curve with a scaffold split is problematic for both datasets even with ``skeleton'' acyclic policy due to FreeSolv having 152 and LES having 217 members belonging to the scaffold group of benzene. Therefore all learning curves were calculated with random train/test split. To construct a learning curve we split a dataset into a training set and a test set, then for each training size we picked a random subset of the training set, used it to optimize the model's hyperparameters, trained the model using these hyperparameters, and then calculated Mean Absolute Error (MAE) of model's predictions for the test set molecules. Since both datasets are compiled from inhomogeneous experimental reports, we reported learning curve accuracies as MAEs rather than Root Mean Square Errors (RMSEs), which are more sensitive to outliers; RMSEs at the largest training sizes are given for reference. Each learning curve construction was repeated 4 times with different random generator seeds, as was each scaffold split prediction accuracy evaluation.

\subsection{Computational time comparison\label{subsec:CPU_time}}

To illustrate our motivation to evaluate complicated ensemble-based kernel matrices via MSORF rather than exact expressions such as~(\ref{eq:simple_ensemble_kernel}), we first performed computational time measurements for different stages of KRR and SORF predictions for FreeSolv and LES datasets\footnote{``initial guess'' $\sigma$ values as defined in~\appref{app:hyperparameter_optimization} were used, along with $10^{-9}$ ratio of $\lambda$ parameter and average diagonal element of the training kernel} over the same 80/20 random splits between training and test set as used in the rest of the work; each molecule was represented with the lowest energy MMFF94 conformer. The results for LES using several representations are presented in Table~\ref{tab:CPU_LES_main} (see the Supplementary Material for FreeSolv results and LES results for other representations). The table demonstrates that while for the simple Gaussian kernel $\kglobal$ MSORF might not be helpful in this setup, MSORF significantly accelerated evaluation of more complicated kernel expressions with worse scaling w.r.t. number of atoms in the train and test sets. The counterpart of this table for FreeSolv, however, which is displayed in Supplementary Material, also demonstrates a situation where generating FJD training kernel costs less than the corresponding feature vectors; this reflects that for small enough datasets Hadamard transforms required to generate the feature vectors can be more costly then evaluating kernel elements using a compact enough representation (note representation sizes listed in Table~\ref{tab:CPU_LES_main}). At the same time, for our applications solving the regression equations~(\ref{eq:KRR_alpha_coeffs}) and~(\ref{eq:SORF_alpha_coeffs}) tended to take less computational time than calculating the training kernel, mirroring the timings observed in Ref.~\onlinecite{Khan_Lilienfeld:2023} for smaller datasets. Also note that computational time spent on training kernel matrices or feature vectors grows even more if model training includes thorough optimization of hyperparameters, requiring calculation of several kernel matrices or feature vector sets from the same representation vectors at different values of $\sigma$, though approaches have been proposed to address such tasks more efficiently~\cite{Grimblat_Rudorff:2026}.

\begin{table}
\begin{adjustbox}{center}
\begin{tabular}{lcccc}
\toprule
  & CM & FCHL19 & SOAP & FJD \\
\midrule training conformers & $\phantom{(}4.84\pm0.18\phantom{)}\phantom{{ \times } 10^{0\phantom{-}}}$\phantom{$^{b}$} & $\phantom{(}4.82\pm0.14\phantom{)}\phantom{{ \times } 10^{0\phantom{-}}}$ & $\phantom{(}4.80\pm0.12\phantom{)}\phantom{{ \times } 10^{0\phantom{-}}}$ & $\phantom{(}4.93\pm0.10\phantom{)}\phantom{{ \times } 10^{0\phantom{-}}}$ \\
 training GFN2-xTB & \_$^{a}$ & \_$^{a}$ & \_$^{a}$ & $(2.97\pm0.05){ \times } 10^{2\phantom{-}}$ \\
 training representations & $(3.25\pm0.43){ \times } 10^{-2}$$^{b}$ & $\phantom{(}5.91\pm0.12\phantom{)}\phantom{{ \times } 10^{0\phantom{-}}}$ & $\phantom{(}4.00\pm0.04\phantom{)}\phantom{{ \times } 10^{0\phantom{-}}}$ & $(3.70\pm0.02){ \times } 10^{1\phantom{-}}$ \\
 training $\mathbf{Z}$ & $(8.60\pm0.27){ \times } 10^{-1}$\phantom{$^{b}$} & $(1.54\pm0.10){ \times } 10^{1\phantom{-}}$ & $(3.07\pm0.02){ \times } 10^{1\phantom{-}}$ & $(2.97\pm0.03){ \times } 10^{2\phantom{-}}$ \\
 $\vec{\alpha}$ & $\phantom{(}1.86\pm0.14\phantom{)}\phantom{{ \times } 10^{0\phantom{-}}}$\phantom{$^{b}$} & $(7.80\pm0.16){ \times } 10^{-1}$ & $\phantom{(}1.40\pm0.01\phantom{)}\phantom{{ \times } 10^{0\phantom{-}}}$ & $\phantom{(}2.20\pm0.53\phantom{)}\phantom{{ \times } 10^{0\phantom{-}}}$ \\
 training $\mathbf{K}$ (KRR) & $(5.00\pm0.17){ \times } 10^{-1}$\phantom{$^{b}$} & $(8.62\pm0.39){ \times } 10^{1\phantom{-}}$ & $(2.80\pm0.04){ \times } 10^{3\phantom{-}}$ & $(1.08\pm0.03){ \times } 10^{3\phantom{-}}$ \\
 $\vec{\alpha}^{\mathrm{KRR}}$ & $(7.50\pm4.33){ \times } 10^{-3}$\phantom{$^{b}$} & $(1.50\pm0.87){ \times } 10^{-2}$ & $(1.25\pm0.43){ \times } 10^{-2}$ & $(2.87\pm4.81){ \times } 10^{-1}$ \\
 test conformers & $\phantom{(}1.19\pm0.12\phantom{)}\phantom{{ \times } 10^{0\phantom{-}}}$\phantom{$^{b}$} & $\phantom{(}1.18\pm0.12\phantom{)}\phantom{{ \times } 10^{0\phantom{-}}}$ & $\phantom{(}1.17\pm0.12\phantom{)}\phantom{{ \times } 10^{0\phantom{-}}}$ & $\phantom{(}1.47\pm0.18\phantom{)}\phantom{{ \times } 10^{0\phantom{-}}}$ \\
 test GFN2-xTB & \_$^{a}$ & \_$^{a}$ & \_$^{a}$ & $(7.64\pm0.57){ \times } 10^{1\phantom{-}}$ \\
 test representations & $(2.50\pm4.33){ \times } 10^{-3}$$^{b}$ & $\phantom{(}1.19\pm0.14\phantom{)}\phantom{{ \times } 10^{0\phantom{-}}}$ & $\phantom{(}1.01\pm0.04\phantom{)}\phantom{{ \times } 10^{0\phantom{-}}}$ & $\phantom{(}9.26\pm0.21\phantom{)}\phantom{{ \times } 10^{0\phantom{-}}}$ \\
 \begin{tabular}[t]{@{}l@{}}query feature vectors\\and final SORF result\end{tabular} & $(5.83\pm0.63){ \times } 10^{-1}$\phantom{$^{b}$} & $\phantom{(}4.67\pm0.13\phantom{)}\phantom{{ \times } 10^{0\phantom{-}}}$ & $\phantom{(}8.49\pm0.30\phantom{)}\phantom{{ \times } 10^{0\phantom{-}}}$ & $(7.58\pm0.32){ \times } 10^{1\phantom{-}}$ \\
 \begin{tabular}[t]{@{}l@{}}query kernel elements\\and final KRR result\end{tabular} & $(3.90\pm0.17){ \times } 10^{-1}$\phantom{$^{b}$} & $(4.45\pm0.07){ \times } 10^{1\phantom{-}}$ & $(1.42\pm0.05){ \times } 10^{3\phantom{-}}$ & $(6.58\pm0.12){ \times } 10^{2\phantom{-}}$ \\
\midrule representation size & $\phantom{0}3741$ & $\phantom{0}3408$ & $41904$ & $\phantom{0}\phantom{0}\phantom{0}56$ \\
\bottomrule
 \multicolumn{5}{l}{$^{a}$ No electronic-structure calculation is involved.}\\
 \multicolumn{5}{l}{$^{b}$ smaller than the time measurement error}\\
\end{tabular}
\end{adjustbox}
\caption{Computational time (in seconds) spent on different stages of training KRR or SORF models with different representations on $80\%$ of the LES dataset and predicting with them $\Eox$ of the remaining $20\%$; lowest energy MMFF94 conformers were used. Sizes of representation vectors (molecular for CM, atomic for FCHL19 and SOAP, and corresponding to orbital-atomic contributions in the case of FJD) are displayed for reference.}
\label{tab:CPU_LES_main}
\end{table}

\subsection{Oxidation potentials in acetonitrile}
\label{subsec:LES}

\begin{figure*}
\center
\includegraphics[width=1.\textwidth]{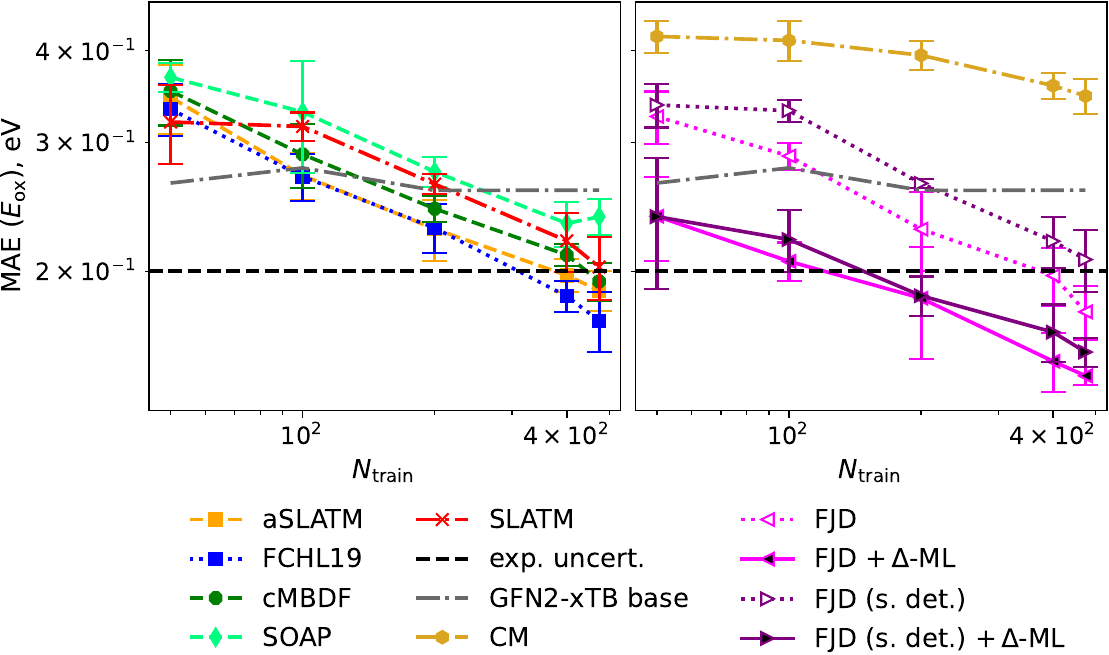}
    \caption{Mean Absolute Errors (MAE) of predictions of oxidation potentials $\Eox$ in acetonitrile from Ref.~\onlinecite{Lee_Lilienfeld:2024} obtained by combining MSORF with several molecular representations for different training set sizes $\Ntrain$  and random train/test split; plotted for comparison are assumed experimental uncertainty (``exp. uncert.'') and GFN2-xTB baseline (``GFN2-xTB base'') as defined in Section~\ref{sec:numerical_details}. ``FJD (s. det.)'' indicates FJD was used with a single Slater determinant.}
    \label{fig:LES_learning_curve} 
\end{figure*}

Figure~\ref{fig:LES_learning_curve} presents MAEs of $\Eox$ predictions on LES as a function of random split training set size for all representations considered in this work; for comparison we also show assumed experimental uncertainty, which the authors of Ref.~\onlinecite{Lee_Lilienfeld:2024} estimate to be of the order of 0.2 eV, and the learning curve of the $\Delta$-ML baseline as defined in Section~\ref{sec:numerical_details}. We can see that all learning curves exhibit a downward slope and the learning curves of all variations of FJD (but for single-determinant one with no $\Delta$-ML), as well as aSLATM, FCHL19, and cMBDF cross the experimental uncertainty line. Table~\ref{tab:LES_learning_summary} presents 80/20 train/test split prediction errors for random and scaffold splits. Among combinations that do not use $\Delta$-ML FJD is among the better-performing representations when tested on random split and becomes the best when tested on scaffold split; at the same time its single-determinant variant is outperformed by FCHL19 and cMBDF for both random and scaffold splits. This seems to support the point behind the ``pair'' expressions~\eqref{eq:FJK_pair_kernel} and~\eqref{eq:FJK_pair_kernel_approx} that allow to represent a molecule not with its orbitals, but with changes of its orbitals. However, in our setup the orbital changes in question, \emph{i.e.} caused by loss of an electron, can only be obtained as a byproduct to an estimate of $\Delta$-ML's reference value, and the latter proves to be of good enough quality that combining both variations of FJD with $\Delta$-ML seems to mostly mitigate the difference between the two approaches; we will come back to this observation in Subsection~\ref{subsec:delta_ml_vs_charge_aug}. Lastly, discussion of an alternative FJD expression that performs very similarly to the ``pair-determinant'' one presented here and thus does not affect our discussion is left for Supplementary Material.

\begin{table}
\begin{tabular}{lccccc}
\toprule
 \multirow{2}{*}{method} & \multicolumn{2}{c}{random} & \phantom{.} & \multicolumn{2}{c}{scaffold (skeleton)} \\
\cline{2-3}\cline{5-6}  & MAE & RMSE &  & MAE & RMSE \\
\midrule aSLATM & $0.188\pm0.012$ & $0.274\pm0.030$ &  & $0.356\pm0.008$ & $0.474\pm0.008$ \\
 FCHL19 & $0.171\pm0.016$ & $0.250\pm0.035$ &  & $0.308\pm0.009$ & $0.408\pm0.007$ \\
 cMBDF & $0.194\pm0.011$ & $0.285\pm0.032$ &  & $0.313\pm0.009$ & $0.434\pm0.013$ \\
 SOAP & $0.237\pm0.013$ & $0.345\pm0.040$ &  & $0.403\pm0.008$ & $0.517\pm0.010$ \\
 SLATM & $0.203\pm0.020$ & $0.299\pm0.057$ &  & $0.374\pm0.012$ & $0.587\pm0.026$ \\
 CM & $0.347\pm0.019$ & $0.458\pm0.024$ &  & $0.452\pm0.009$ & $0.581\pm0.014$ \\
\midrule FJD & $0.176\pm0.015$ & $0.246\pm0.026$ &  & $0.258\pm0.015$ & $0.368\pm0.022$ \\
 FJD $+\,\Delta$-ML & \underline{$0.144\pm0.004$} & \underline{$0.207\pm0.017$} &  & $0.212\pm0.010$ & $0.314\pm0.011$ \\
 FJD (s. det.) & $0.207\pm0.020$ & $0.295\pm0.030$ &  & $0.352\pm0.023$ & $0.461\pm0.035$ \\
 FJD (s. det.) $+\,\Delta$-ML & $0.155\pm0.007$ & $0.226\pm0.028$ &  & \underline{$0.193\pm0.004$} & \underline{$0.271\pm0.005$} \\
\midrule mean predictor & $0.425\pm0.030$ & $0.537\pm0.035$ &  & $0.498\pm0.017$ & $0.656\pm0.021$ \\
 GFN2-xTB base & $0.258\pm0.023$ & $0.363\pm0.062$ &  & $0.276\pm0.004$ & $0.359\pm0.005$ \\
\bottomrule
\end{tabular}
\caption{Mean Absolute Errors (MAEs) and Root Mean Square Errors (RMSEs) observed for machine learning $\Eox$ from Ref.~\onlinecite{Lee_Lilienfeld:2024} at $\Ntrain=474$ using two types of train/test splits: random splits and scaffold with random ties and ``skeleton'' policy for acyclic molecules. See Section~\ref{sec:numerical_details} for how the GFN2-xTB baseline (``GFN2-xTB base'') was defined. In each column the best observed value is underlined.}
\label{tab:LES_learning_summary}
\end{table}

\subsection{Hydration energies}
\label{subsec:FreeSolv}

\begin{figure*}
\center
\includegraphics[width=1.\textwidth]{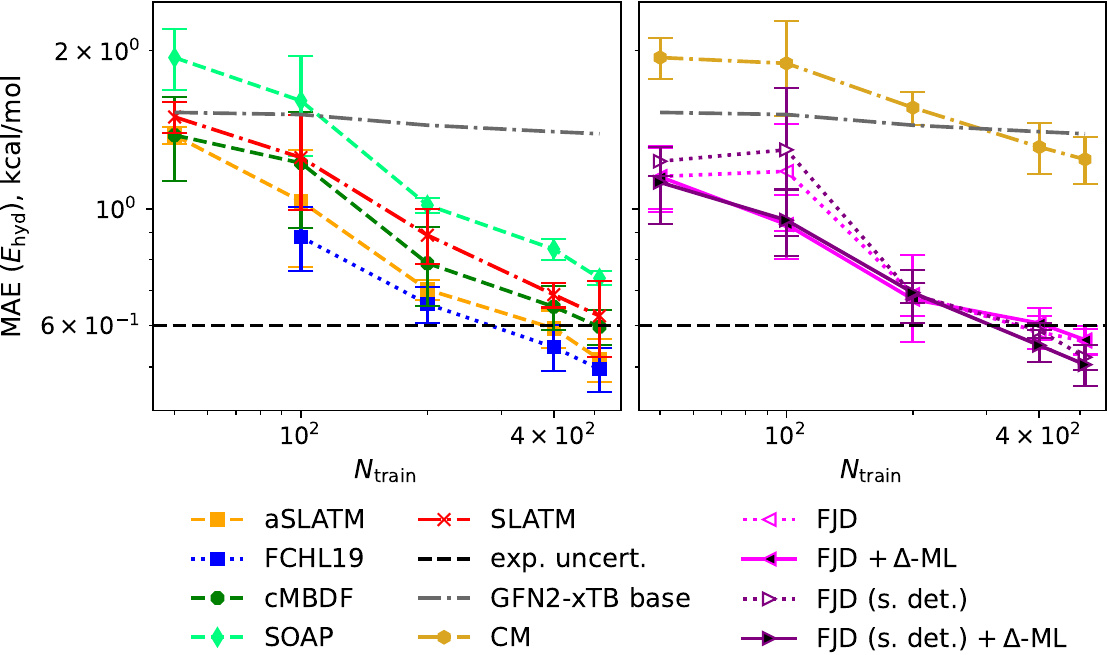}
    \caption{Mean Absolute Error (MAE) of predictions of $\Ehyd$ of the FreeSolv dataset obtained by combining MSORF with several molecular representations for different training set sizes $\Ntrain$ and random train/test split; plotted for comparison are assumed experimental uncertainty (``exp. uncert.'') and GFN2-xTB baseline (``GFN2-xTB base'') as defined in Section~\ref{sec:numerical_details}. ``S. det.'' next to FJD entries indicates single Slater determinant was used.}
    \label{fig:FreeSolv_learning_curves} 
\end{figure*}

Figure~\ref{fig:FreeSolv_learning_curves} presents MAEs as a function of training set size for all representations considered in this work; for comparison we also show assumed experimental uncertainty of 0.6 kcal/mol\footnote{See FreeSolv's ``text notes'' data fields stating this value as the ``default'' experimental error.} as well as the GFN2-xTB baseline from Section~\ref{sec:numerical_details}. As was the case for LES, we observe downward learning curves in all cases. Table~\ref{tab:FreeSolv_learning_summary} presents 80/20 train/test split prediction errors for random and scaffold splits; similarly to LES we see aSLATM and FCHL19 as the best-performing geometry-based representations, coming close to the best-performing combination of single-determinant FJD with $\Delta$-ML. We also note that introduction of the $\MIXEDEXTENSIVE$ layer~\eqref{eq:mixedextensive_def} only noticeably improved prediction errors in the case of cMBDF, netting same or worse results for all other representations. Comparing accuracies of single- and pair-determinant FJD variants (using representation of just the solvated molecule and the molecule in both solvent and vacuum) reveals that adding more information into the representation also decreased prediction accuracy, possibly exemplifying that for small enough datasets making a model more complex can damage learning rather than help it.

\begin{table}
\begin{tabular}{lccccc}
\toprule
 \multirow{2}{*}{method} & \multicolumn{2}{c}{random} & \phantom{.} & \multicolumn{2}{c}{scaffold (skeleton)} \\
\cline{2-3}\cline{5-6}  & MAE & RMSE &  & MAE & RMSE \\
\midrule aSLATM & $0.52\pm0.05$ & $0.90\pm0.14$ &  & $1.36\pm0.12$ & $2.32\pm0.30$ \\
 aSLATM (m.-ext.) & $0.51\pm0.05$ & $0.98\pm0.18$ &  & $1.77\pm0.20$ & $3.02\pm0.35$ \\
 FCHL19 & \underline{$0.50\pm0.05$} & $0.89\pm0.12$ &  & $1.25\pm0.16$ & $2.02\pm0.32$ \\
 FCHL19 (m.-ext.) & $0.51\pm0.04$ & $0.89\pm0.11$ &  & $1.40\pm0.17$ & $2.41\pm0.37$ \\
 cMBDF & $0.60\pm0.05$ & $1.14\pm0.20$ &  & $1.63\pm0.19$ & $2.72\pm0.52$ \\
 cMBDF (m.-ext.) & $0.53\pm0.02$ & $0.91\pm0.03$ &  & $1.57\pm0.07$ & $2.54\pm0.19$ \\
 SOAP & $0.74\pm0.02$ & $1.14\pm0.06$ &  & $1.73\pm0.18$ & $2.66\pm0.25$ \\
 SOAP (m.-ext.) & $0.73\pm0.03$ & $1.15\pm0.06$ &  & $1.79\pm0.10$ & $2.71\pm0.18$ \\
 SLATM & $0.63\pm0.10$ & $1.14\pm0.22$ &  & $1.35\pm0.06$ & $2.02\pm0.02$ \\
 CM & $1.24\pm0.13$ & $1.99\pm0.28$ &  & $2.44\pm0.21$ & $4.21\pm0.36$ \\
\midrule FJD & $0.56\pm0.03$ & $0.94\pm0.08$ &  & $1.23\pm0.06$ & $2.14\pm0.03$ \\
 FJD $+\,\Delta$-ML & $0.56\pm0.04$ & $0.95\pm0.14$ &  & $1.08\pm0.02$ & $1.85\pm0.03$ \\
 FJD (s. det.) & $0.52\pm0.03$ & $0.88\pm0.03$ &  & $1.21\pm0.08$ & $2.05\pm0.09$ \\
 FJD (s. det.) $+\,\Delta$-ML & $0.51\pm0.05$ & \underline{$0.82\pm0.07$} &  & \underline{$1.01\pm0.10$} & \underline{$1.71\pm0.24$} \\
 FJD (m.-ext.) & $0.57\pm0.03$ & $0.92\pm0.06$ &  & $1.20\pm0.08$ & $2.21\pm0.11$ \\
 FJD (m.-ext.) $+\,\Delta$-ML & $0.53\pm0.08$ & $0.86\pm0.14$ &  & $1.08\pm0.05$ & $1.89\pm0.08$ \\
 FJD (m.-ext., s. det.) & $0.59\pm0.02$ & $1.05\pm0.05$ &  & $1.10\pm0.08$ & $1.98\pm0.19$ \\
 FJD (m.-ext., s. det.) $+\,\Delta$-ML & $0.53\pm0.06$ & $0.87\pm0.11$ &  & $1.05\pm0.07$ & $1.84\pm0.11$ \\
\midrule dressed atom & $2.01\pm0.09$ & $2.84\pm0.29$ &  & $2.54\pm0.08$ & $3.75\pm0.14$ \\
 GFN2-xTB base & $1.39\pm0.11$ & $1.83\pm0.18$ &  & $1.90\pm0.02$ & $2.40\pm0.07$ \\
\bottomrule
\end{tabular}
\caption{Summary of results for machine learning $\Ehyd$ of the FreeSolv dataset labeled analogously to Table~\ref{tab:LES_learning_summary}; note that for FreeSolv the maximum training set size was 514 molecules.}
\label{tab:FreeSolv_learning_summary}
\end{table}

Unlike LES, FreeSolv has been used as a benchmark in numerous ML studies~\cite{Riniker:2017,Wang_Wei:2019,Rong_Huang:2020,Gao_Wang:2021,Fang_Chen:2022,Zhang_Zhang:2022,Low_Izgorodina:2022,Zeng_Cheng:2022,Fang_Wang:2022,Fang_Chen:2023,Wang_Farimani:2022,Xia_Zhang:2023,Kirschbaum_Bande:2024,Yadav_Bandyopadhyay:2025,Qiao_Wei:2025}, with Table~\ref{tab:FreeSolv_literature_summary} containing a representative slice of previously reported accuracy values as well as prediction accuracies of the better-performing methods proposed in this work evaluated on the scaffold split with ``pool'' acyclic policy commonly used in the literature, though due to the difference in train/test splits and their implementation reliability of the resulting accuracy comparisons is limited. Three of the methods presented in Table~\ref{tab:FreeSolv_literature_summary}, AIML, C-SGEN, and MDFP+, are philosophically similar to MSORF in learning properties of interest with largely handcrafted general descriptors (though in MDFP+ case the descriptors are leaning towards being hydration-centric); we see that MSORF approaches outperform all of them but for C-SGEN. KCL, HiMol, and SCAGE are GNNs that are pre-trained in self-supervised manner with graph (and in case of SCAGE also geometric) information from large molecular databases to learn general-purpose representations of chemical space; we can see that MSORF displayed accuracy comparable to or better than theirs. sPhysNet-MT-ens5 and A3D-PNAConv-FT augmented FreeSolv with DFT solvation energy calculations, achieving transfer learning from \emph{ab initio} calculations to experimental data via multitask learning or fine-tuning, respectively, reporting better accuracies than any other method on random train/test splits. ProQSAR\cite{Phan_Stadler:2026}, a QSAR method that chooses the best model architecture among combinations of implemented descriptors and regressors, proves to be overwhelmingly effective compared to other methods in the context of a small dataset and a well-studied quantity of interest by reporting significantly lower MAEs and RMSEs for scaffold train/test splits. Also, our prediction accuracies improve on those previously reported for two KRR-based methods of Ref.~\onlinecite{Weinreich_Lilienfeld:2022}, namely AIML and the separately reported MAE of 0.68 kcal/mol for FML~\cite{Weinreich_Lilienfeld:2021} approach combined with Bag-of-Bonds (BoB) representation~\cite{Hansen_Tkatchenko:2015} . Possible reasons for the improvement are discussed in Subsection~\ref{subsec:minimum_en_vs_ensemble}.

\begin{table}
\begin{tabular}{llcc}
\toprule
 method & split & MAE & RMSE \\
\midrule MDFP+~\cite{Riniker:2017} & random & $0.9\phantom{0}\phantom{0}\pm0.1\phantom{0}\phantom{0}$ & $1.2\phantom{0}\phantom{0}\pm0.1\phantom{0}\phantom{0}$ \\
 C-SGEN + fingerprint$^{a}$~\cite{Wang_Wei:2019} & random & \_ & $0.78\phantom{0}\pm0.02\phantom{0}$ \\
 KCL(KMPNN)$^{a}$~\cite{Fang_Chen:2022} & random & \_ & $0.795\phantom{{}\pm0.000}$ \\
 AIML~\cite{Weinreich_Lilienfeld:2022} & random & $0.82\phantom{0}\phantom{{}\pm0.000}$ & $1.35\phantom{0}\phantom{{}\pm0.000}$ \\
 sPhysNet-MT-ens5$^{a,b}$~\cite{Xia_Zhang:2023} & random & $0.359\pm0.063$ & $0.620\pm0.140$ \\
 A3D-PNAConv-FT$^{a,b,c}$~\cite{Zhang_Zhang:2022} & random & $0.417\pm0.066$ & $0.719\pm0.168$ \\
\midrule HiMol (LARGE)$^{b,d}$~\cite{Zang_Tang:2023} & scaffold (pool) & \_ & $2.283\phantom{{}\pm0.000}$ \\
 SCAGE$^{b,d}$~\cite{Qiao_Wei:2025} & scaffold (pool) & \_ & $1.688\pm0.080$ \\
 ProQSAR$^{d,e}$~\cite{Phan_Stadler:2026} & scaffold (pool) & $0.359\phantom{{}\pm0.000}$ & $0.494\phantom{{}\pm0.000}$ \\
 FCHL19$^{f}$ & scaffold (pool) & $1.23\phantom{0}\phantom{{}\pm0.000}$ & $1.71\phantom{0}\phantom{{}\pm0.000}$ \\
 aSLATM$^{f}$ & scaffold (pool) & $1.71\phantom{0}\phantom{{}\pm0.000}$ & $2.34\phantom{0}\phantom{{}\pm0.000}$ \\
 FJD $+\,\Delta$-ML$^{f}$ & scaffold (pool) & $1.52\phantom{0}\phantom{{}\pm0.000}$ & $2.12\phantom{0}\phantom{{}\pm0.000}$ \\
 FJD (s. det.) $+\,\Delta$-ML$^{f}$ & scaffold (pool) & $1.43\phantom{0}\phantom{{}\pm0.000}$ & $1.96\phantom{0}\phantom{{}\pm0.000}$ \\
 GFN2-xTB base$^{f}$ & scaffold (pool) & $2.07\phantom{0}\phantom{{}\pm0.000}$ & $2.69\phantom{0}\phantom{{}\pm0.000}$ \\
\bottomrule
 \multicolumn{4}{l}{$^{a}$ 80/10/10 training/validation/test split}\\
 \multicolumn{4}{l}{$^{b}$ uses transfer learning / large-scale pretraining}\\
 \multicolumn{4}{l}{$^{c}$ excluded iodine molecules}\\
 \multicolumn{4}{l}{$^{d}$ scaffold split; the scaffold-split implementation differs between papers and from this work, so scaffold values are not strictly comparable}\\
 \multicolumn{4}{l}{$^{e}$ descriptor-based classical ML; single external test split}\\
 \multicolumn{4}{l}{$^{f}$ This work; scaffold split with the acyclic molecules pooled.}\\
\end{tabular}
\caption{Accuracy of machine learning $\Ehyd$ of the FreeSolv dataset reported previously for training on randomly chosen $80\%$ of the dataset (unless explicitly mentioned otherwise) compared to prediction accuracies observed for our approaches for 80/20 train/test scaffold split with ``pool'' acyclic policy.}
\label{tab:FreeSolv_literature_summary}
\end{table}

\subsection{Comparison of machine learning with ensembles or minimum energy conformers\label{subsec:minimum_en_vs_ensemble}}

\begin{figure*}
\center
\includegraphics[width=1.\textwidth]{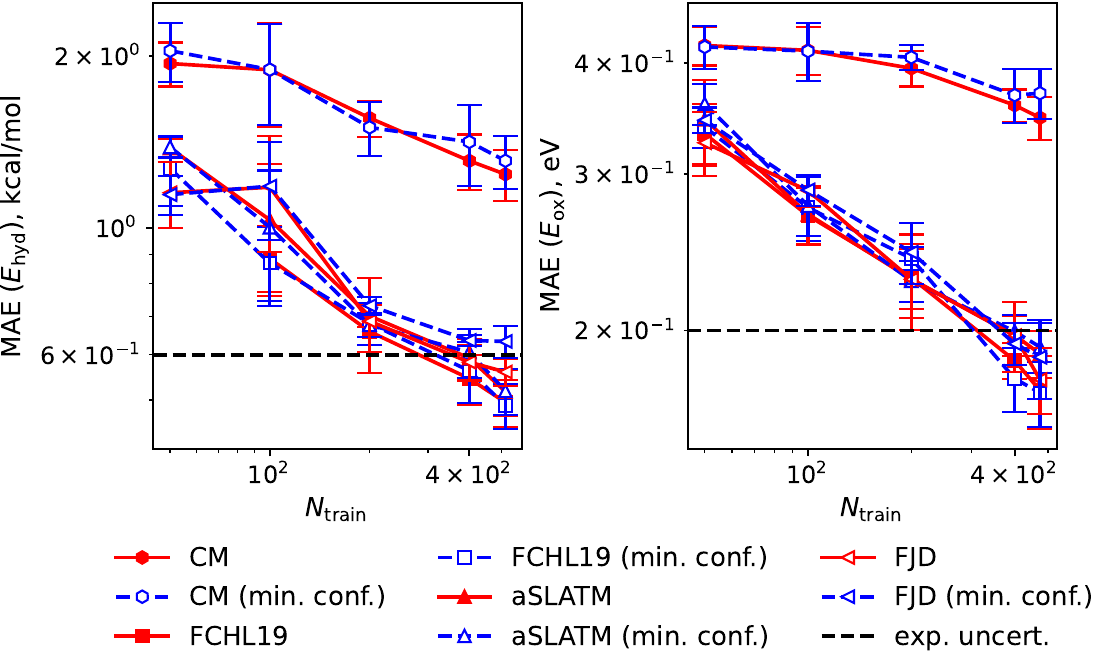}
    \caption{Comparison of random train/test split MAEs obtained by representing molecules either with a conformer ensemble or the lowest energy conformer (``min. conf.'') for different representations and training set sizes $\Ntrain$. $\Eox$ and $\Ehyd$ are obtained from Ref.~\onlinecite{Lee_Lilienfeld:2024} and the FreeSolv dataset; assumed experimental uncertainty (``exp. uncert.'') values are displayed for comparison.}
    \label{fig:minconf_learning_curves} 
\end{figure*}

We now evaluate how much representing a molecule with a conformer ensemble rather than a single conformer actually affected ML results presented in this work, using as examples combinations of MSORF representations that had demonstrated best and worst performance, namely aSLATM, FCHL19, FJD, and CM. For these representations we reran the same workflows as in the previous subsections, but for each ensemble we only considered the conformer with the lowest MMFF94 energy. For CM we used single $\SORF$ layer (\emph{i.e.} we used simple SORF), for aSLATM and FCHL19 we used $\SORF\NORM\SUM\satoms\SWITCH\SORF$ with LES and $\SUM\satoms\SWITCH\SORF$ with FreeSolv, for FJD we used $\SORF\WEIGHTEDSUM\sorbs\SORF\NORM\WEIGHTEDSUM\satoms\SORF$. The resulting learning curves are presented in Figure~\ref{fig:minconf_learning_curves}, with prediction accuracies observed for largest training sets gathered in Tables~\ref{tab:minconf_LES} and~\ref{tab:minconf_FreeSolv}. We observe that the difference between representing a molecule with an ensemble and one optimal conformer tends to be of the order of statistical error, and is significantly less pronounced than differences between MAEs observed in this work with different conformer representations. The only exceptions are prediction accuracies for FJD for the more challenging scaffold train/test split, yet even in this case the difference disappears once $\Delta$-ML protocol is applied to the labels, supporting the point from Subsection~\ref{subsec:LES} about $\Delta$-ML reference being more important in this setting then using its other results to generate a physics-motivated representation. This implies that, at least for the quantities considered in this work, choosing how to represent molecular geometry is more important than how many molecular geometries are used. One implication is that our reported improvement over MAEs obtained with FML~\cite{Weinreich_Lilienfeld:2021} and BoB (as reported in Ref.~\onlinecite{Weinreich_Lilienfeld:2022}) is likely caused by a combination of a different hyperparameter optimization procedure and statistical noise introduced to the ensemble representation in FML due to averaging over a molecular dynamics trajectory.

\begin{table}
\begin{tabular}{lccccc}
\toprule
 \multirow{2}{*}{method} & \multicolumn{2}{c}{random} & \phantom{.} & \multicolumn{2}{c}{scaffold (skeleton)} \\
\cline{2-3}\cline{5-6}  & MAE & RMSE &  & MAE & RMSE \\
\midrule CM (ensemble) & $0.347\pm0.019$ & $0.458\pm0.024$ &  & $0.452\pm0.009$ & $0.581\pm0.014$ \\
 CM (min. conf.) & $0.370\pm0.024$ & $0.476\pm0.019$ &  & $0.455\pm0.012$ & $0.583\pm0.015$ \\
\midrule FCHL19 (ensemble) & $0.171\pm0.016$ & $0.250\pm0.035$ &  & $0.308\pm0.009$ & $0.408\pm0.007$ \\
 FCHL19 (min. conf.) & $0.170\pm0.014$ & $0.249\pm0.031$ &  & $0.316\pm0.008$ & $0.426\pm0.018$ \\
\midrule aSLATM (ensemble) & $0.188\pm0.012$ & $0.274\pm0.030$ &  & $0.356\pm0.008$ & $0.474\pm0.008$ \\
 aSLATM (min. conf.) & $0.192\pm0.012$ & $0.276\pm0.032$ &  & $0.366\pm0.013$ & $0.487\pm0.018$ \\
\midrule FJD (ensemble) & $0.176\pm0.015$ & $0.246\pm0.026$ &  & $0.258\pm0.015$ & $0.368\pm0.022$ \\
 FJD (min. conf.) & $0.187\pm0.019$ & $0.258\pm0.034$ &  & $0.300\pm0.036$ & $0.427\pm0.044$ \\
\midrule FJD (ensemble) $+\,\Delta$-ML & $0.144\pm0.004$ & \underline{$0.207\pm0.017$} &  & $0.212\pm0.010$ & $0.314\pm0.011$ \\
 FJD (min. conf.) $+\,\Delta$-ML & \underline{$0.143\pm0.006$} & $0.208\pm0.020$ &  & $0.214\pm0.003$ & $0.313\pm0.003$ \\
\midrule FJD (s. det., ensemble) & $0.207\pm0.020$ & $0.295\pm0.030$ &  & $0.352\pm0.023$ & $0.461\pm0.035$ \\
 FJD (s. det., min. conf.) & $0.215\pm0.018$ & $0.304\pm0.035$ &  & $0.387\pm0.015$ & $0.506\pm0.036$ \\
\midrule FJD (s. det., ensemble) $+\,\Delta$-ML & $0.155\pm0.007$ & $0.226\pm0.028$ &  & \underline{$0.193\pm0.004$} & $0.271\pm0.005$ \\
 FJD (s. det., min. conf.) $+\,\Delta$-ML & $0.162\pm0.011$ & $0.236\pm0.033$ &  & $0.195\pm0.004$ & \underline{$0.269\pm0.005$} \\
\bottomrule
\end{tabular}
\caption{Comparison of prediction accuracies for machine learning $\Eox$ from Ref.~\onlinecite{Lee_Lilienfeld:2024} while representing molecules with minimum energy conformers or ensembles labeled analogously to Table~\ref{tab:LES_learning_summary}.}
\label{tab:minconf_LES}
\end{table}

\begin{table}
\begin{tabular}{lccccc}
\toprule
 \multirow{2}{*}{method} & \multicolumn{2}{c}{random} & \phantom{.} & \multicolumn{2}{c}{scaffold (skeleton)} \\
\cline{2-3}\cline{5-6}  & MAE & RMSE &  & MAE & RMSE \\
\midrule CM (ensemble) & $1.24\pm0.13$ & $1.99\pm0.28$ &  & $2.44\pm0.21$ & $4.21\pm0.36$ \\
 CM (min. conf.) & $1.31\pm0.14$ & $2.00\pm0.23$ &  & $2.40\pm0.18$ & $4.13\pm0.34$ \\
\midrule FCHL19 (ensemble) & $0.50\pm0.05$ & $0.89\pm0.12$ &  & $1.25\pm0.16$ & $2.02\pm0.32$ \\
 FCHL19 (min. conf.) & \underline{$0.49\pm0.04$} & $0.85\pm0.13$ &  & $1.36\pm0.20$ & $2.27\pm0.38$ \\
\midrule aSLATM (ensemble) & $0.52\pm0.05$ & $0.90\pm0.14$ &  & $1.36\pm0.12$ & $2.32\pm0.30$ \\
 aSLATM (min. conf.) & $0.52\pm0.05$ & $0.90\pm0.14$ &  & $1.34\pm0.12$ & $2.26\pm0.27$ \\
\midrule FJD (ensemble) & $0.56\pm0.03$ & $0.94\pm0.08$ &  & $1.23\pm0.06$ & $2.14\pm0.03$ \\
 FJD (min. conf.) & $0.63\pm0.04$ & $1.04\pm0.14$ &  & $1.31\pm0.07$ & $2.31\pm0.10$ \\
\midrule FJD (ensemble) $+\,\Delta$-ML & $0.56\pm0.04$ & $0.95\pm0.14$ &  & $1.08\pm0.02$ & $1.85\pm0.03$ \\
 FJD (min. conf.) $+\,\Delta$-ML & $0.59\pm0.05$ & $0.96\pm0.14$ &  & $1.15\pm0.02$ & $1.95\pm0.05$ \\
\midrule FJD (s. det., ensemble) & $0.52\pm0.03$ & $0.88\pm0.03$ &  & $1.21\pm0.08$ & $2.05\pm0.09$ \\
 FJD (s. det., min. conf.) & $0.56\pm0.04$ & $0.93\pm0.11$ &  & $1.14\pm0.03$ & $1.94\pm0.06$ \\
\midrule FJD (s. det., ensemble) $+\,\Delta$-ML & $0.51\pm0.05$ & \underline{$0.82\pm0.07$} &  & \underline{$1.01\pm0.10$} & \underline{$1.71\pm0.24$} \\
 FJD (s. det., min. conf.) $+\,\Delta$-ML & $0.51\pm0.04$ & \underline{$0.82\pm0.09$} &  & $1.05\pm0.04$ & $1.75\pm0.08$ \\
\bottomrule
\end{tabular}
\caption{Comparison of prediction accuracies for machine learning $\Ehyd$ of the FreeSolv dataset while representing molecules with minimum energy conformers or ensembles labeled analogously to Table~\ref{tab:LES_learning_summary}.}
\label{tab:minconf_FreeSolv}
\end{table}

\subsection{Comparing $\Delta$-ML with augmenting geometric representations with orbital information}
\label{subsec:delta_ml_vs_charge_aug}

Given how $\Delta$-ML largely mitigated the difference between different FJD variations for both LES and FreeSolv, we decided to test whether the observation would extend to other methods for including orbital information in molecular representation for ML. To that end we largely followed the philosophy of Ref.~\onlinecite{Caldas_Sandonas:2026}, who augmented geometric representations of molecules with descriptors extracted from density-functional tight-binding\cite{Spiegelman_Rapacioli:2020} calculations. In our case the geometric representations of choice were aSLATM and FCHL19 as they have performed well over both datasets; since those are local representations the only components of the descriptors of Ref.~\onlinecite{Caldas_Sandonas:2026} that could be straightforwardly appended to them were Mulliken population charges, with further possible expansions of the procedure being beyond the scope of this work. There are two possible ways to do that. The first one, referred to as ``xTB charges'', involves simply adding atomic charge to an atom's aSLATM or FCHL19 representation and assigning a separate $\sigma$ hyperparameter to it, but otherwise preserving the rest of MSORF protocol. The second one, referred to as ``xTB pair charges'', means generating the same pairs of representations as in the case of FJD and combining them analogously, \emph{i.e.} taking the difference between feature vectors for the charged and neutral conformer for LES, and concatenating feature vectors for the conformer in water and in vacuum for FreeSolv. As a result for ``xTB pair charges'' the used MSORF expressions were $\SORF\NORM\WEIGHTEDSUM\sconfs\SORF\WEIGHTEDSUM_{\mathrm{state}}\SUM\satoms\SWITCH\SORF$ for LES (with $\WEIGHTEDSUM_{\mathrm{state}}$ corresponding to taking the difference of charged and neutral states) and $\WEIGHTEDSUM\sconfs\CONCATENATE\SUM\satoms\SWITCH\SORF$ for FreeSolv.

While the learning curve MAEs are left for Supplementary Material, Tables~\ref{tab:LES_charge_augment} and~\ref{tab:FreeSolv_charge_augment} present prediction accuracies observed at 80/20 train/test splits. For FreeSolv augmenting geometric representation with charges did not produce statistically noticeable improvements for prediction accuracies, same can be said about LES for only adding ground state xTB charges. However, for LES adding ``xTB pair charges'' and thus transforming extensive aSLATM and FCHL19 representations into intensive feature vectors produced significant improvements when $\Delta$-ML was not used. Adding $\Delta$-ML to a large degree mitigates the difference between different representations; for LES it significantly reduces the gap between best prediction accuracies observed for pure geometric representations and their charge-augmented counterparts, eliminating it completely for scaffold splits, while for FreeSolv this gap is removed altogether.

\begin{table}
\begin{tabular}{lccccc}
\toprule
 \multirow{2}{*}{method} & \multicolumn{2}{c}{random} & \phantom{.} & \multicolumn{2}{c}{scaffold (skeleton)} \\
\cline{2-3}\cline{5-6}  & MAE & RMSE &  & MAE & RMSE \\
\midrule aSLATM $+\,\Delta$-ML & $0.149\pm0.015$ & $0.225\pm0.034$ &  & $0.210\pm0.013$ & $0.310\pm0.004$ \\
 aSLATM $+$ xTB pair charges $+\,\Delta$-ML & \underline{$0.128\pm0.013$} & \underline{$0.186\pm0.030$} &  & \underline{$0.204\pm0.002$} & $0.306\pm0.005$ \\
 aSLATM $+$ xTB charges $+\,\Delta$-ML & $0.148\pm0.015$ & $0.220\pm0.035$ &  & $0.214\pm0.011$ & \underline{$0.304\pm0.006$} \\
 FCHL19 $+\,\Delta$-ML & $0.151\pm0.016$ & $0.228\pm0.042$ &  & $0.243\pm0.007$ & $0.339\pm0.006$ \\
 FCHL19 $+$ xTB pair charges $+\,\Delta$-ML & $0.137\pm0.012$ & $0.204\pm0.034$ &  & $0.205\pm0.003$ & \underline{$0.304\pm0.005$} \\
 FCHL19 $+$ xTB charges $+\,\Delta$-ML & $0.145\pm0.013$ & $0.221\pm0.043$ &  & $0.238\pm0.022$ & $0.345\pm0.046$ \\
\midrule aSLATM & $0.188\pm0.012$ & $0.274\pm0.030$ &  & $0.356\pm0.008$ & $0.474\pm0.008$ \\
 aSLATM $+$ xTB pair charges & $0.143\pm0.015$ & $0.208\pm0.025$ &  & $0.257\pm0.007$ & $0.368\pm0.008$ \\
 aSLATM $+$ xTB charges & $0.189\pm0.015$ & $0.275\pm0.037$ &  & $0.363\pm0.014$ & $0.494\pm0.024$ \\
 FCHL19 & $0.171\pm0.016$ & $0.250\pm0.035$ &  & $0.308\pm0.009$ & $0.408\pm0.007$ \\
 FCHL19 $+$ xTB pair charges & $0.152\pm0.008$ & $0.223\pm0.020$ &  & $0.275\pm0.006$ & $0.374\pm0.010$ \\
 FCHL19 $+$ xTB charges & $0.172\pm0.010$ & $0.249\pm0.023$ &  & $0.315\pm0.006$ & $0.424\pm0.009$ \\
\bottomrule
\end{tabular}
\caption{Summary of results for machine learning $\Eox$ of the LES dataset using FCHL19 and aSLATM combined with $\Delta$-ML as well as augmenting these representations with GFN2-xTB Mulliken population charges as explained in Subsection~\ref{subsec:delta_ml_vs_charge_aug}; labeled analogously to Table~\ref{tab:LES_learning_summary}.}
\label{tab:LES_charge_augment}
\end{table}

\begin{table}
\begin{tabular}{lccccc}
\toprule
 \multirow{2}{*}{method} & \multicolumn{2}{c}{random} & \phantom{.} & \multicolumn{2}{c}{scaffold (skeleton)} \\
\cline{2-3}\cline{5-6}  & MAE & RMSE &  & MAE & RMSE \\
\midrule aSLATM $+\,\Delta$-ML & $0.55\pm0.05$ & $0.90\pm0.08$ &  & $1.21\pm0.16$ & $2.24\pm0.44$ \\
 aSLATM $+$ xTB pair charges $+\,\Delta$-ML & \underline{$0.45\pm0.04$} & $0.75\pm0.11$ &  & $1.20\pm0.06$ & $2.18\pm0.24$ \\
 aSLATM $+$ xTB charges $+\,\Delta$-ML & \underline{$0.45\pm0.04$} & \underline{$0.73\pm0.12$} &  & $1.17\pm0.18$ & $2.14\pm0.41$ \\
 FCHL19 $+\,\Delta$-ML & \underline{$0.45\pm0.05$} & $0.74\pm0.09$ &  & $1.09\pm0.12$ & \underline{$1.75\pm0.30$} \\
 FCHL19 $+$ xTB pair charges $+\,\Delta$-ML & \underline{$0.45\pm0.04$} & $0.77\pm0.07$ &  & \underline{$1.07\pm0.11$} & $1.79\pm0.28$ \\
 FCHL19 $+$ xTB charges $+\,\Delta$-ML & $0.46\pm0.04$ & $0.78\pm0.09$ &  & $1.12\pm0.06$ & $1.89\pm0.20$ \\
\midrule aSLATM & $0.52\pm0.05$ & $0.90\pm0.14$ &  & $1.36\pm0.12$ & $2.32\pm0.30$ \\
 aSLATM $+$ xTB pair charges & $0.46\pm0.06$ & $0.83\pm0.17$ &  & $1.31\pm0.22$ & $2.45\pm0.57$ \\
 aSLATM $+$ xTB charges & $0.47\pm0.06$ & $0.85\pm0.18$ &  & $1.37\pm0.18$ & $2.58\pm0.54$ \\
 FCHL19 & $0.50\pm0.05$ & $0.89\pm0.12$ &  & $1.25\pm0.16$ & $2.02\pm0.32$ \\
 FCHL19 $+$ xTB pair charges & $0.47\pm0.06$ & $0.83\pm0.09$ &  & $1.29\pm0.18$ & $2.45\pm0.51$ \\
 FCHL19 $+$ xTB charges & $0.47\pm0.04$ & $0.86\pm0.09$ &  & $1.27\pm0.15$ & $2.48\pm0.48$ \\
\bottomrule
\end{tabular}
\caption{Summary of results for machine learning $\Ehyd$ of the FreeSolv dataset using FCHL19 and aSLATM combined with $\Delta$-ML as well as augmenting these representations with GFN2-xTB Mulliken population charges as explained in Subsection~\ref{subsec:delta_ml_vs_charge_aug}; labeled analogously to Table~\ref{tab:LES_learning_summary}.}
\label{tab:FreeSolv_charge_augment}

\end{table}

\section{Conclusions and outlook\label{sec:conclusions}}

We have devised a general SORF-based procedure for generating physically motivated molecular representations from Boltzmann ensembles of conformers and demonstrated its utility for predicting experimental values of molecular properties, with Subsection~\ref{subsec:delta_ml_vs_charge_aug} showing how the same combination of geometry- and orbital-based descriptors can efficiently predict two fundamentally different properties by readjusting only the MSORF protocol applied to them, rather than the descriptors themselves. The method both demonstrates systematic decrease of prediction error with increased training set size and reaches experimental accuracy with hundreds of molecules as training set. At the same time, we tended to not observe a significant difference between representing a molecule with an ensemble of conformers or a single conformer, implying that, at least for the quantities considered in this work, how conformers are represented is a more important question than how many of them are used. We also observed that FJD descriptors generated from GFN2-xTB calculations, as well as some geometric descriptors augmented with GFN2-xTB's orbital charges, were among the better performers for the calculations considered in this work, especially for the case of $\Eox$ in acetonitrile, where the ``pair-determinant'' version allowed the model to capitalize on qualitative information from GFN2-xTB on which part of a molecule is most affected by subtraction of an electron. However, the semiempirical calculations used to generate orbitals for these representations also generated baselines of high enough quality that using them in a standard $\Delta$-ML protocol largely leveled out the difference between representations (see also Supplementary Material). The tendencies observed in this work for using ensemble-based and orbital-based ML are similar to those observed in Ref.~\onlinecite{Axelrod_Gomez-Bombarelli:2023} for ensemble-based ML of biomolecular properties and Ref.~\onlinecite{Caldas_Sandonas:2026} for augmenting geometric representations with orbital information.

There are several possible further directions for this work. Given impressive prediction accuracies observed on FreeSolv for approaches based on transfer learning\cite{Vermeire_Green:2021,Zhang_Zhang:2022,Li_Zeng:2022,Xia_Zhang:2023}, it would be lucrative to bridge the gap between MSORF and finetuning by moving away from hand-crafted molecular descriptors to those automatically generated and validated on large amounts of \emph{ab initio} data. Another possible direction could be going beyond the simple expressions for weighting contributions of different conformers, atoms, or localized orbitals used in this work to more complicated expressions gleaned from explainable artificial intelligence applied to \emph{ab initio} calculations~\cite{Esders_Muller:2025}. In both cases, it would be possible to capitalize on additional \emph{ab initio} data while keeping MSORF grounded in KRR formalism. Lastly, while we focused on combining representations of conformers into representations of molecular ensembles, MSORF could analogously combine representations of molecular ensembles into representations of mixtures of molecules. This could make MSORF models useful for exploring chemical spaces of mixtures, a problem relevant for several industrial applications that involve using organic liquids~\cite{Flores_Bhowmik:2022,Sose_Deshmukh:2023,Blasio_Bhowmik:2024}.

\section*{Supplementary Material}

See the Supplementary Material for additional dataset analysis, continuation of Table~\ref{tab:CPU_LES_main} for other combinations of representations and datasets considered in this work, learning curves analogous to the ones presented in Subsections~\ref{subsec:LES} and~\ref{subsec:FreeSolv}, but with quantity shifts optimized w.r.t. cross-validation errors, some additional learning curves expanding the discussions in Subsections~\ref{subsec:LES}-\ref{subsec:delta_ml_vs_charge_aug}, and some extra derivations behind~\appref{app:hyperparameter_optimization}.

\section*{Acknowledgments}

A significant portion of the computations was performed on hardware at the University of Vienna acquired with funding from the European Union's Horizon 2020 research and innovation programme under grant agreements No.~957189 (BIG-MAP) and No.~957213 (BATTERY 2030+).

\section*{Author Declarations}
\subsection*{Conflict of Interest}

The author has no conflicts to disclose.

\subsection*{Author Contributions}
\textbf{Konstantin Karandashev:} Conceptualization (lead); Methodology (lead);
Software (lead); Formal analysis (lead); Investigation (lead);
Data curation (lead); Writing -- original draft (lead);
Writing -- review \& editing (lead).

\section*{Data availability statement}

The data that support the findings of this study are openly available in Zenodo at \url{https://doi.org/10.5281/zenodo.23705912}, Ref.~\onlinecite{qml2_ensemble_data_Zenodo}; they comprise all prediction-accuracy data shown in the plots and tables of this work. The QML2 code and scripts used to generate them are openly available in Zenodo at \url{https://doi.org/10.5281/zenodo.23703754}, Ref.~\onlinecite{qml2_Zenodo}. The LES and FreeSolv datasets were obtained from Refs.~\onlinecite{lee_Lilienfeld:2023_Zenodo} and~\onlinecite{FreeSolv_Zenodo}, respectively.

\appendix

\section{Combining FJK with extended tight-binding calculations}
\label{app:fjd_formalism}

FJK was originally introduced with HF calculations in mind and before specifying how we adapted it for use with xTB calculations a recap of the original expressions is needed. To define atomic contribution weights $\orbweight_{j}(\locorb_{o}^{(\mol{A})})$ appearing in the FJK kernel expression~\eqref{eq:FJK_lin_orb_kernel} when considering atom with index $j$ of conformer $\mol{A}$ and one of the latter's localized orbitals $\locorb_{o}^{(\mol{A})}$, we first introduce their ``uncut'' versions as
\begin{equation}
 \orbweight_{j}^{\mathrm{u}}(\locorb_{o}^{(\mol{A})})=\left[\aoproj_{j}\coeffvec(\locorb_{o}^{(\mol{A})})\right]^{\transpose}\mat{S}^{(\mol{A})}\aoproj_{j}\coeffvec(\locorb_{o}^{(\mol{A})}),
\label{eq:fjk_atomic_weight}
\end{equation}
where $\vec{c}(\locorb_{o}^{(\mol{A})})$ are coefficients of the localized orbital $\locorb_{o}^{(\mol{A})}$, $\mat{1}_{j}$ is projector onto atomic orbital basis functions centered on atom $j$, and $\mat{S}^{(\mol{A})}$ is the overlap matrix. $\orbweight_{j}(\locorb_{o}^{(\mol{A})})$ are then found with
\begin{equation}
    \orbweight_{j}(\locorb_{o}^{(\mol{A})}):=\mathrm{max}[\orbweight_{j}^{\mathrm{u}}(\locorb_{o}^{(\mol{A})})-\rho_{-}^{\mathrm{orb}}, 0],
    \label{eq:orb_cut_weights_def}
\end{equation}
with $\rho_{-}^{\mathrm{orb}}$ such that
\begin{equation}
    \frac{\sum_{j=1}^{\Natoms^{(\mol{A})}}\mathrm{max}[\orbweight_{j}^{\mathrm{u}}(\locorb_{o}^{(\mol{A})})-\rho_{-}^{\mathrm{orb}}, 0]}{\sum_{j=1}^{\Natoms^{(\mol{A})}}\orbweight_{j}^{\mathrm{u}}(\locorb_{o}^{(\mol{A})})}=1-\rcutorb,
    \label{eq:orb_cut_weights_companion_def}
\end{equation}
where $\rcutorb$ is a user-defined parameter. The expressions are very similar to Equations~\eqref{eq:cut_weights_def} and~\eqref{eq:cut_weights_companion_def}, except normalization is absent since it is taken care of inside the Gaussian FJK kernel~\eqref{eq:FJK_gauss_orb_kernel}. The corresponding vector representation of the atomic contribution $\vec{v}_{j}(\locorb_{o}^{(\mol{A})})$ is composed from ``self-coupling'' and ``other-coupling'' functions that for a ``coupling'' matrix $\mat{M}$ defined over atomic orbitals and angular momentum values $l_{1}$ and $l_{2}$ take the form
\begin{align}
v^{(\mathrm{s}),\mat{M}}_{j,l_{1}l_{2}}(\locorb_{o}^{(\mol{A})})=&\frac{1}{\orbweight_{j}^{\mathrm{u}}(\locorb_{o}^{(\mol{A})})}\left[\aoproj_{j,l_{1}}\coeffvec(\locorb_{o}^{(\mol{A})})\right]^{\transpose}\mat{M}\aoproj_{j,l_{2}}\coeffvec(\locorb_{o}^{(\mol{A})}),\label{eq:fjk_self_coupling}\\
v^{(\mathrm{o}),\mat{M}}_{j,l_{1}l_{2}}(\locorb_{o}^{(\mol{A})})=&\frac{1}{\orbweight_{j}^{\mathrm{u}}(\locorb_{o}^{(\mol{A})})}\sum_{\substack{j^{\prime}=1 \\ j^{\prime}\neq j}}^{\Natoms^{(\mol{A})}}\left[\aoproj_{j,l_{1}}\coeffvec(\locorb_{o}^{(\mol{A})})\right]^{\transpose}\mat{M}\aoproj_{j^{\prime},l_{2}}\coeffvec(\locorb_{o}^{(\mol{A})}),\label{eq:fjk_other_coupling}
\end{align}
where $\aoproj_{j,l}$ is projector on atomic orbitals centered on atom $j$ and having angular momentum $l$. The maximum values of $l_{1}$ and $l_{2}$ to be included into the representation are regulated by the $\lmax$ parameter.

The original FJK paper used cheap HF calculations to generate the representation from intrinsic bond orbitals\cite{Knizia:2013} (IBOs), with coupling matrices $\mat{M}$ chosen as Fock $\mat{F}$, electronic repulsion $\mat{J}$, and electronic exchange $\mat{K}$ matrices; it additionally added the self-coupling terms derived from the overlap matrix $\mat{S}$ for the same angular momentum values, which corresponds to angular momentum distribution. To transfer the protocol to xTB calculations we first note that the non-zero ``electronic temperature'' parameter used to improve xTB's convergence results in smeared occupation numbers, making the choice of occupied orbitals to be transformed into localized orbitals non-unique. Our implementation assumed the $\Norbs$ canonical orbitals with the lowest energies to be occupied, and then generated Pipek-Mezey (PM) orbitals\cite{pipek1989,lehtola2014pm} from them, in a way assuming that GFN2-xTB's basis is well-behaved enough that it should not require the depolarization procedure which is the main difference between IBO and PM localization. Secondly, since HF's $\mat{K}$ lacks a close analogue in xTB formalism we replaced it with matrices derived from shifted dipole and bond-projected shifted dipole matrices aiming to generate multipole- and bond-focused descriptors philosophically similar to those of Refs.~\onlinecite{Sokalski_Poirier:1983,Briling_Corminboeuf:2024}. Given matrix $\mat{D}^{(A),\gamma}$ of dipole component on coordinate $\gamma\in\{x,y,z\}$ we introduce the shifted dipole matrix whose element for two atomic orbitals $\mu$ and $\nu$ is
\begin{equation}
\tilde{D}^{(\mol{A}),\gamma}_{\mu\nu}=D^{(\mol{A}),\gamma}_{\mu\nu}-R^{(\mol{A}),\gamma}_{j(\mu)}S^{(\mol{A})}_{\mu\nu},
\end{equation}
where $R^{(\mol{A}),\gamma}_{j}$ is position on coordinate $\gamma$ of atom $j$ and $j(\mu)$ is index of the atom on which atomic orbital $\mu$ is centered. Note that an expression very similar to $\tilde{D}^{(\mol{A}),\gamma}_{\mu\nu}$ is also used in calculating the anisotropic electrostatic energy term in GFN2-xTB. $\tilde{\mat{D}}^{(\mol{A}),\gamma}$ allows us to define ``dipole norm coupling functions'' as
\begin{align}
\tilde{v}^{(\mathrm{s}),\tilde{\mat{D}}}_{j,l_{1}l_{2}}(\locorb_{o}^{(\mol{A})})=&\sqrt{\sum_{\gamma=\{x,y,z\}}\left[v^{(\mathrm{s}),\tilde{\mat{D}}^{(\gamma)}}_{j,l_{1}l_{2}}(\locorb_{o}^{(\mol{A})})\right]^{2}},\label{eq:fjk_self_coupling_dipole_norm}\\
\tilde{v}^{(\mathrm{o}),\tilde{\mat{D}}}_{j,l_{1}l_{2}}(\locorb_{o}^{(\mol{A})})=&\sqrt{\sum_{\gamma=\{x,y,z\}}\left[v^{(\mathrm{o}),\tilde{\mat{D}}^{(\gamma)}}_{j,l_{1}l_{2}}(\locorb_{o}^{(\mol{A})})\right]^{2}},\label{eq:fjk_other_coupling_dipole_norm}
\end{align}
The bond-projected dipole matrix $\mat{B}$ is defined as
\begin{equation}
B^{(\mol{A})}_{\mu\nu}=\begin{cases}
\sum_{\gamma=\{x,y,z\}}\tilde{D}^{(\mol{A}),\gamma}_{\mu\nu}(\vec{R}^{(\mol{A}),\gamma}_{j(\nu)}-\vec{R}^{(\mol{A}),\gamma}_{j(\mu)})/|\vec{R}^{(\mol{A})}_{j(\nu)}-\vec{R}^{(\mol{A})}_{j(\mu)}|, j(\nu)\neq j(\mu),\\
0, j(\nu)=j(\mu),
\end{cases}
\label{eq:bond_proj_def}
\end{equation}
with the aim to generate coupling elements sharing the bond-centric character of SPAHM(b)\cite{Briling_Corminboeuf:2024} descriptors.

The finalized list of components that included into $\vec{v}_{j}(\locorb_{o}^{(\mol{A})})$ was thus: self-couplings for $\mat{S}$ corresponding to angular momentum distribution (same as original FJK); self- and other-couplings for the full Hamiltonian and the difference between the full Hamiltonian and xTB's one-electron Hamiltonian, chosen due to physical similarity of HF's $\mat{F}$ and $\mat{J}$ matrices; both self- and other- dipole norm couplings of Eqs.~\eqref{eq:fjk_self_coupling_dipole_norm} and~\eqref{eq:fjk_other_coupling_dipole_norm}; and other-couplings from matrix $\mat{B}$~\eqref{eq:bond_proj_def}. Since the finalized set of coupling methods is derived by replacing $\mat{K}$ matrix-derived representation components with dipole-derived ones the method is referred to as ``FJD'' throughout the paper. Lastly, during hyperparameter optimization we assigned separate $\sigma$ values to blocks of $\vec{v}_{j}(\locorb_{o}^{(\mol{A})})$ corresponding to different coupling matrices, \emph{i.e.} five in total.

\section{Hyperparameter optimization\label{app:hyperparameter_optimization}}

Our hyperparameter optimization protocol largely followed the commonly known philosophy documented in Ref.~\onlinecite{Rifkin_Lippert:2007} of relatively simply optimizing $\lambda$ for pre-calculated feature vectors, though as will be clarified later we additionally implemented an option for optimizing w.r.t. cross-validation errors quantity shifts by a constant or a stoichiometry-dependent term. To keep the hyperparameter optimization target consistent with the assumed train/test split we optimized different types of errors depending on whether random or scaffold-based split was used. For the random split we considered leave-one-out (LOO) cross-validation errors $\leaveoneoutvecerr$, \emph{i.e.} prediction errors for a point given a model had been trained on all other points in the dataset. They can be efficiently calculated\cite{Cawley_Talbot:2003} as
\begin{equation}
\leaveoneouterr_{i}=\frac{\reproducedquant_{i}-q_{i}}{1-\vec{Z}_{i}^{\transpose}\tK^{-1}\vec{Z}_{i}}
\label{eq:leaveoneouterrs}
\end{equation}
where $\reproducedvecquant$ are predictions for training points of the model trained on the entire training set, $\tK:=\mat{Z}^{\transpose}\mat{Z}+\lambda \UMat{\Nfeatures}$. Note that
\begin{equation}
\reproducedvecquant=\mat{Z}\tK^{-1}\mat{Z}^{\transpose}\vec{q}.
\label{eq:reproducedquants}
\end{equation}

To see how optimization of $\lambda$ and property shifts can be separated out consider SVD of $\mat{Z}$
\begin{equation}
\mat{Z}=\ZU \ZD \ZV^{\transpose},
\label{eq:SVD_decomposition}
\end{equation}
where $\ZU$ is the $\Ntrain\times\Ntrain$ matrix of left singular vectors, $\ZD$ is $\Ntrain\times\Ntrain$ diagonal matrix whose diagonal elements are the $\Ntrain$ singular values, $\ZV$ is the $\Nfeatures\times\Ntrain$ matrix of right singular vectors. Using Equations~\eqref{eq:reproducedquants} and~\eqref{eq:SVD_decomposition} to rewrite $\leaveoneoutvecerr$~\eqref{eq:leaveoneouterrs} yields
\begin{equation}
\leaveoneouterr_{i}=\frac{(\vec{Z}\zTR_{i})^{\transpose}[\ZD^{2}+\lambda\UMat{\Ntrain}]^{-1}(\mat{Z}\zTR)^{\transpose}\vec{q}-q_{i}}{1-(\vec{Z}_{i}\zTR)^{\transpose}[\ZD^{2}+\lambda\UMat{\Ntrain}]^{-1}\vec{Z}_{i}\zTR}
\label{eq:leaveoneouterrs_svd}
\end{equation}
where $\mat{Z}\zTR$ is the matrix of features transformed into the basis of right singular vectors
\begin{equation}
\mat{Z}\zTR:=\mat{Z}\ZV.
\end{equation}
Note how for fixed $\sigma$ and pre-calculated SVD of $\mat{Z}$ evaluating thus rewritten $\leaveoneoutvecerr$~\eqref{eq:leaveoneouterrs_svd} costs $\mathcal{O}[\Ntrain\min(\Nfeatures,\Ntrain)]$, while for fixed $\sigma$ and $\lambda$ $\leaveoneoutvecerr$ is a linear function of the training set quantities $\vec{q}$.

If a training set is divided into clusters of molecules, as was done in this work when evaluating prediction accuracies on scaffold-based splits, we would use leave-one-group-out (LOGO) cross-validation errors $\logovecerr$, \emph{i.e.} prediction errors for a point given a model had been trained on all groups but for the group containing the point of interest. An efficient way to evaluate such errors in the context of optimization of $\lambda$ for a fixed $\sigma$ was proposed in Ref.~\onlinecite{Grimblat_Rudorff:2026} for KRR and reused eigenvalue decomposition of the kernel for each $\lambda$ value; we used the general philosophy of that work, but applied it to reusing SVD of $\mat{Z}$~\eqref{eq:SVD_decomposition}. Consider group $g$ consisting of $N^{(g)}$ molecules in the training set and auxiliary matrices
\begin{align}
\mat{A}^{(g)}&:=\UMat{N^{(g)}}-\mat{W}^{(g)}\left[\ZU^{(g)}\right]^{\transpose},
\label{eq:logo_stat_matrix}\\
\mat{W}^{(g)}&:=\ZU^{(g)}\ZD^{2}\left[\ZD^{2}+\lambda\UMat{\Ntrain}\right]^{-1},
\end{align}
where $\ZU^{(g)}$ is matrix of rows of $\ZU$ corresponding to molecules in group $g$. The LOGO errors for group $g$, $\logovecerr^{(g)}$, are then evaluated via a generalization of~\eqref{eq:leaveoneouterrs_svd}
\begin{equation}
\begin{split}
\logovecerr^{(g)}&=\left[\mat{A}^{(g)}\right]^{-1}\left(\vec{q}^{r,(g)}-\vec{q}^{(g)}\right),\\
&=\left[\mat{A}^{(g)}\right]^{-1}\left(\mat{W}^{(g)}\ZU^{\transpose}\vec{q}-\vec{q}^{(g)}\right),
\label{eq:logo_errors}
\end{split}
\end{equation}
where $\vec{q}^{r,(g)}$ and $\vec{q}^{(g)}$ are quantity predictions of model trained on the entire training set and quantity values for molecules in group $g$. Evaluating~\eqref{eq:logo_errors} for a single group costs $\mathcal{O}[(N^{(g)})^{3}]+\mathcal{O}[(N^{(g)})^{2}\Ntrain]$.

We note that while several algorithms have been proposed for combining errors from non-equivalent cross-validation splits\cite{Arlot_Celisse:2010}, in this work we simply pooled together all LOGO-errors into a single $\logovecerr$ treated analogously to $\leaveoneoutvecerr$. We will therefore refer to the vector of optimized errors as just $\cvvecerr$ from now on. Since this work focuses on datasets that compile experimental data from different authors, we wanted to find hyperparameters that minimize MAE over $\cvvecerr$, given that MAE is less affected by outliers than RMSE. However, MAE's discontinuous derivatives would make it difficult to use the gradient during optimization. We instead use ``self-consistent LogCosh'' function $\LSCLC(\cvvecerr,\gamma)$, whose in-depth discussion is left for the Supplementary Material, but for now we note that it can be regarded as a modification of the Orlicz-norm estimators of Ref.~\onlinecite{Andoni2018Orlicz}, its square is convex and exhibits continuous derivatives, and the relative difference between the MAE at its minimum and the smallest possible MAE does not exceed $\gamma$. We will refer to our final hyperparameter optimization target as $\Lhyp$
\begin{equation}
\Lhyp:=\LSCLC^{2}(\cvvecerr,\gamma).
\label{eq:hyploss}
\end{equation}

Since $\cvvecerr$ is linear function of $\vec{q}$, optimizing coefficients for shifting $\vec{q}$ by a constant or by stoichiometry-dependent shifts of the same form as in the dressed atom approach can be done for fixed $\sigma$ values (collectively referred to as $\allsigmas$ from now on) and fixed $\lambda$ value with a quasi-Newton \cite{Boyd_Vandenberghe:2004} method, with this work using implementation of Limited-memory Broyden--Fletcher--Goldfarb--Shanno algorithm (L-BFGS) \cite{Fletcher:1987,Liu_Nocedal:1989} in {\SciPy}. For fixed $\allsigmas$ the optimization w.r.t. $\lambda$ consisted of two steps. Bayesian Optimization Structure Search (BOSS) package~\cite{boss} was used to approximately find the minimum of $-1/\Lhyp$, inversion done to give minimized function a well-defined limit should a value in $\cvvecerr$ go to infinity due to numerical instability. The minimum was refined with gradient optimization using "Sequential Least-Squares Quadratic Programming" (SLSQP)\footnote{The original publication does not clarify what ``SLSQP'' stands for, the meaning behind the abbreviation presented here is taken from current documentation of the NLopt package~\cite{NLopt}.} algorithm~\cite{Kraft:1994} as implemented in \SciPy. We thus define $\Lopthyp(\allsigmas,\gamma)$ as value of $\Lhyp$ optimal w.r.t. $\lambda$ and, optionally, quantity shifts. The general idea is to optimize it w.r.t. $\allsigmas$ via steepest descent, but setting $\gamma=0$ from the beginning will likely make the optimization loop stuck in a local minimum. Hence we used several gradient descent steps and $\gamma$ values to avoid the problem, as documented in Algorithm~\ref{alg:hyperparameter_optimization}.

\begin{algorithm}
\caption{Gradient optimization of $\Lopthyp$.}
\label{alg:hyperparameter_optimization}
\KwIn{Initial guess for hyperparameters $\allsigmas^{\mathrm{init}}$}
\For{$\gamma=\gamma_{1},\ldots,\gamma_{\mathrm{final}}$}{
  \For{$\LNSTEP=\LNSTEP_{1},\ldots,\LNSTEP_{\mathrm{final}}$}{
    \Loop{}{
      $\vec{G}\leftarrow\left.\mathrm{d}\Lopthyp(\allsigmas,\gamma)/\mathrm{d}(\ln\allsigmas)\right|_{\allsigmas=\allsigmas^{\mathrm{init}}}$\;
      $\ln\vec{\sigma}_{\mathrm{all}}^{\mathrm{next}}\leftarrow\ln\vec{\sigma}_{\mathrm{all}}^{\mathrm{init}}-\LNSTEP\vec{G}/|\vec{G}|$\;
      \eIf{$\Lopthyp(\allsigmas^{\mathrm{next}},\gamma) > \Lopthyp(\allsigmas^{\mathrm{init}},\gamma)$}{
        \Break\;
      }{
        $\allsigmas^{\mathrm{init}}\leftarrow\allsigmas^{\mathrm{next}}$\;
      }
    }
  }
}
\end{algorithm}

For initial guesses, we always set $\mixsigma$ from $\MIXEDEXTENSIVE$~\eqref{eq:mixedextensive_def} as unity. Initial guesses for $\sigma $ values appearing in each $\SORF$~\eqref{eq:base_SORF_features} were calculated as root mean square distance from training set vectors used as procedure's input to their average. For example, for FJD features that meant that first $\vec{\sigma}$ values are guessed, then used to generate the corresponding $\WEIGHTEDSUM\SORF$ features~\eqref{eq:linorb_features}, which are in turn used to estimate the initial guess for $\sigmaglobal$ used to generate the $\SORF\NORM\WEIGHTEDSUM\SORF$ features~\eqref{eq:gaussorb_features}. Also, as mentioned in Subsec.~\ref{subsec:classic_KRR}, FJD applies five different values of $\sigma$ to five different types of entries in the $\vec{v}(\locorb)$ vectors; each of these $\sigma$ values was calculated analogously to the case of single $\sigma$ hyperparameter, but with only entries rescaled by the corresponding $\sigma$ considered and the final result multiplied by $\sqrt{5}$.

Lastly, in this work hyperparameter optimization always ran with consecutive $\gamma$ values of $0.5$, $0.25$, and $0.125$ and gradient step values of $0.5$, $0.25$, and $0.125$. During L-BFGS and SLSQP algorithm calls we set gradient tolerance parameter as $10^{-9}$. While performing BOSS optimization the search bounds for $\ln\lambda$ were $-25.0$ and $0.0$ shifted by $\ln[\mathrm{Tr}(\ZD^{2})/\Ntrain]$, which is approximately the logarithm of average diagonal element of the training kernel matrix $\mat{K}$ in the SORF approximation. Each BOSS run started with putting 16 points on a uniform grid in the search bounds, then doing 16 more evaluations according to the Bayesian optimization algorithm.

%aipnum4-2.bst 2019-01-14 (MD) hand-edited version of apsrev4-1.bst

\end{document}

\title[SM: Ensemble-applied SORF]{Supplementary Material: Kernel Ridge Regression for conformer ensembles made easy with Structured Orthogonal Random Features}
\author{Konstantin Karandashev}

\affiliation{Chembricks Inc., Menlo Park, California, USA}
\affiliation{University of Vienna, Faculty of Physics, Kolingasse 14-16, AT-1090 Wien, Austria}
\email{kvkarandashev@gmail.com}
\vspace{10pt}

\maketitle

\section{Dataset analysis}

Tables~\ref{tab:xTB_calibration_LES} and~\ref{tab:xTB_calibration_FreeSolv} illustrate the statement from Section~III of the main text that additionally optimizing the slope with which GFN2-xTB reference value enters the final expression for GFN2-xTB baseline model, as opposed to only optimizing the intercept in case of LES or per-element shifts in case of FreeSolv, yielded negligible gains in accuracy. For FreeSolv the fitted slope does not change much from unity, while for LES it significantly differs from unity and is clearly overfitted.

\begin{table}
\begin{adjustbox}{center}
  \begin{tabular}{lcccccccc}
\toprule
 \multirow{2}{*}{quantity} & \multicolumn{2}{c}{random} & \phantom{.} & \multicolumn{2}{c}{scaffold (skeleton)} & \phantom{.} & \multicolumn{2}{c}{scaffold (pool)} \\
\cline{2-3}\cline{5-6}\cline{8-9}  & \makecell{GFN2-xTB\\base} & \makecell{xTB\\(fitted)} &  & \makecell{GFN2-xTB\\base} & \makecell{xTB\\(fitted)} &  & \makecell{GFN2-xTB\\base} & \makecell{xTB\\(fitted)} \\
\midrule MAE & $\phantom{-}0.258\pm0.023$ & $\phantom{-}0.251\pm0.014$ &  & $\phantom{-}0.276\pm0.004$ & $\phantom{-}0.266\pm0.002$ &  & $\phantom{-}0.258$ & $\phantom{-}0.242$ \\
 RMSE & $\phantom{-}0.363\pm0.062$ & $\phantom{-}0.347\pm0.042$ &  & $\phantom{-}0.359\pm0.005$ & $\phantom{-}0.359\pm0.003$ &  & $\phantom{-}0.326$ & $\phantom{-}0.305$ \\
 intercept & $-9.523\pm0.006$ & $-6.611\pm0.276$ &  & $-9.532\pm0.001$ & $-6.238\pm0.045$ &  & $-9.529$ & $-6.479$ \\
 slope & $\phantom{-}1\phantom{.}\phantom{0}\phantom{0}\phantom{0}\phantom{{}\pm0.000}$ & $\phantom{-}0.732\pm0.025$ &  & $\phantom{-}1\phantom{.}\phantom{0}\phantom{0}\phantom{0}\phantom{{}\pm0.000}$ & $\phantom{-}0.696\pm0.004$ &  & $\phantom{-}1\phantom{.}\phantom{0}\phantom{0}\phantom{0}$ & $\phantom{-}0.719$ \\
\bottomrule
\end{tabular}
\end{adjustbox}
  \caption{Prediction accuracies of GFN2-xTB baseline model (``GFN2-xTB base'') used for comparison throughout the main text and of a linear function of GFN2-xTB ionization potential with both slope and intercept fitted [``xTB (fitted)''], observed for train/test splits that were random (``random'') or scaffold, the latter tested with ``skeleton'' and ``pool'' acyclic policies [``scaffold (skeleton)'' and ``scaffold (pool)'', see Subsection~IV\,A of the main text for the definition]. For ``random'' and ``scaffold (skeleton)'' splits the error bars come from averaging over 4 splits.}
  \label{tab:xTB_calibration_LES}
\par\medskip{\footnotesize\textit{Alt text:} A table of baseline-model accuracies for LES: the GFN2-xTB baseline and a fitted linear function of the GFN2-xTB ionization potential (both slope and intercept fitted). Columns give the error metrics for random, scaffold-skeleton and scaffold-pool splits, with error bars from averaging over 4 splits for the random and scaffold-skeleton cases.}
\end{table}

\begin{table}
\begin{adjustbox}{center}
  \begin{tabular}{lcccccccc}
\toprule
 \multirow{2}{*}{quantity} & \multicolumn{2}{c}{random} & \phantom{.} & \multicolumn{2}{c}{scaffold (skeleton)} & \phantom{.} & \multicolumn{2}{c}{scaffold (pool)} \\
\cline{2-3}\cline{5-6}\cline{8-9}  & \makecell{GFN2-xTB\\base} & \makecell{dressed atom\\$+$ xTB (fitted)} &  & \makecell{GFN2-xTB\\base} & \makecell{dressed atom\\$+$ xTB (fitted)} &  & \makecell{GFN2-xTB\\base} & \makecell{dressed atom\\$+$ xTB (fitted)} \\
\midrule MAE & $1.39\pm0.11$ & $1.39\pm0.12$ &  & $1.90\pm0.02$ & $1.89\pm0.03$ &  & $2.07$ & $2.07$ \\
 RMSE & $1.83\pm0.18$ & $1.83\pm0.18$ &  & $2.40\pm0.07$ & $2.38\pm0.08$ &  & $2.69$ & $2.68$ \\
 slope & $1\phantom{.}\phantom{0}\phantom{0}\phantom{{}\pm0.00}$ & $1.08\pm0.01$ &  & $1\phantom{.}\phantom{0}\phantom{0}\phantom{{}\pm0.00}$ & $1.03\pm0.02$ &  & $1\phantom{.}\phantom{0}\phantom{0}$ & $1.03$ \\
\bottomrule
\end{tabular}
\end{adjustbox}
  \caption{Prediction accuracies of GFN2-xTB baseline model (``GFN2-xTB base'') used for comparison throughout the main text and of a linear function of GFN2-xTB $\Ehyd$ with both optimized ``dressed atom'' element-wise intercepts and optimized slope w.r.t. GFN2-xTB $\Ehyd$ [``dressed atom + xTB (fitted)''], observed for train/test splits that were random (``random'') or scaffold, the latter tested with ``skeleton'' and ``pool'' acyclic policies [``scaffold (skeleton)'' and ``scaffold (pool)'', see Subsection~IV\,A of the main text for the definition]. For ``random'' and ``scaffold (skeleton)'' splits the error bars come from averaging over 4 splits.}
  \label{tab:xTB_calibration_FreeSolv}
\par\medskip{\footnotesize\textit{Alt text:} A table of baseline-model accuracies for FreeSolv: the GFN2-xTB baseline and a fitted linear model using element-wise ``dressed atom'' intercepts together with a fitted slope. Columns give the error metrics for random, scaffold-skeleton and scaffold-pool splits, with error bars from averaging over 4 splits for the random and scaffold-skeleton cases.}
\end{table}

\begin{figure*}
\center
\includegraphics[width=1.\textwidth]{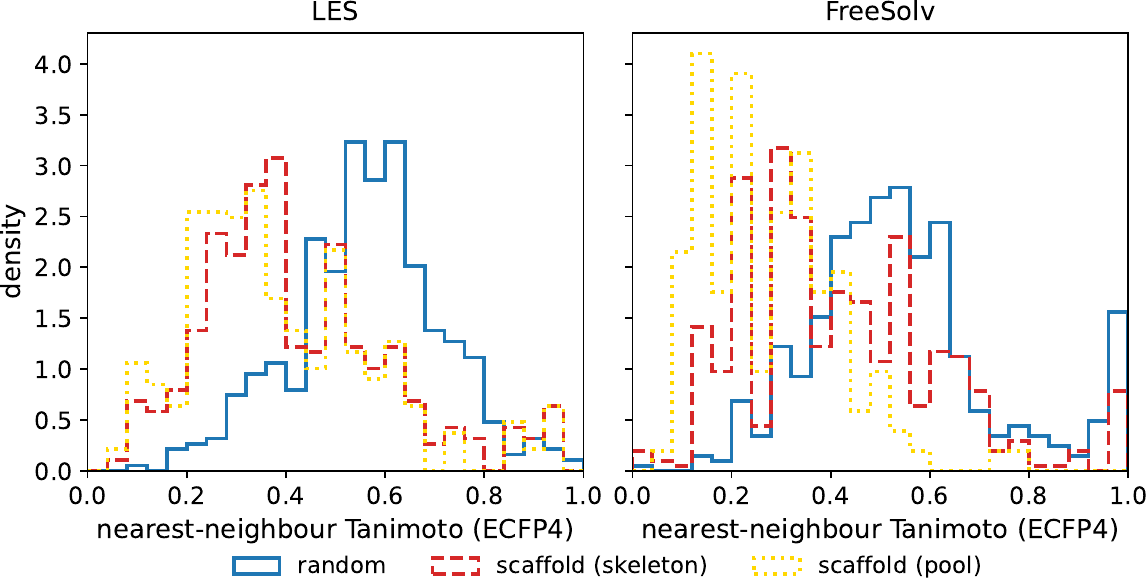}
    \caption{Distributions of the ECFP4\cite{Rogers_Hahn:2010} (Extended-Connectivity Fingerprint, diameter 4) Tanimoto similarity between each test-set molecule and its nearest neighbour in the training set, for 
FreeSolv and LES datasets at an 80/20 train/test split, with the latter done via the three procedures outlined in Subsection~IV\,A of the main text: random or via scaffold grouping, the latter done with ``skeleton'' or ``pool'' acyclic policies. Values are averaged over 4 random seeds and normalized to unit area.}
    \label{fig:nn_separation_ecfp4}
\par\medskip{\footnotesize\textit{Alt text:} A two-panel histogram (LES left, FreeSolv right) of the nearest-neighbour ECFP4 Tanimoto similarity between each test molecule and its closest training molecule (x-axis 0 to 1, y-axis density). Three overlaid step distributions are shown: random split (blue solid), scaffold-skeleton (red dashed), and scaffold-pool (yellow dotted). The random distributions peak at higher similarity ($\sim$0.5--0.65), while the scaffold distributions are shifted toward lower similarity -- scaffold-pool most strongly, peaking near 0.1--0.2 for FreeSolv -- showing scaffold splits place test molecules farther from the training set.}
\end{figure*}

\begin{figure*}
\center
\includegraphics[width=1.\textwidth]{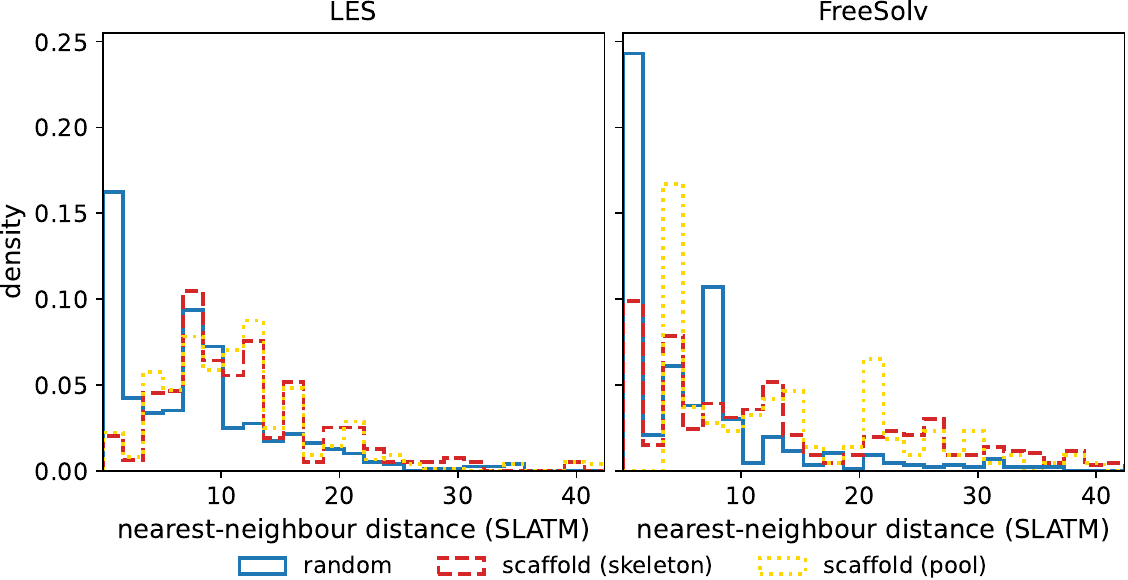}
    \caption{Distributions of the Spectrum
of London and Axilrod-Teller-Muto potential\cite{Huang_Lilienfeld:2020} (SLATM) $l^{2}$ distance between each test-set molecule and its nearest neighbour in the training set, for 
FreeSolv and LES datasets at an 80/20 train/test split, with the latter done via the three procedures outlined in Subsection~IV\,A of the main text: random or via scaffold grouping, the latter done with ``skeleton'' or ``pool'' acyclic policies. Values are averaged over4 random seeds and normalized to unit area.}
    \label{fig:nn_separation_slatm}
\par\medskip{\footnotesize\textit{Alt text:} A two-panel histogram (LES left, FreeSolv right) of the nearest-neighbour SLATM $l^{2}$ distance between each test molecule and its closest training molecule (x-axis 0 to $\sim$42, y-axis density). Three overlaid step distributions: random (blue solid), scaffold-skeleton (red dashed), scaffold-pool (yellow dotted). The random distributions have a tall spike at small distance, while the scaffold distributions are shifted toward larger distances, indicating scaffold test molecules are less similar to the training set.}
\end{figure*}

Figures~\ref{fig:nn_separation_ecfp4} and~\ref{fig:nn_separation_slatm} present histograms demonstrating in full the separation between train and test sets achieved by different scaffolds considered in this work as summarized in Table~II. Note how for LES both plotting ECFP4\cite{Rogers_Hahn:2010} (Extended-Connectivity Fingerprint, diameter 4) Tanimoto similarities and SLATM vector distances reveals that scaffold splits yield similar separations both with ``skeleton'' and ``pool'' acyclic policies, while for FreeSolv acyclic policy has a significant influence of train/test separation, which is appears more pronounced when Tanimoto similarities are considered.

\section{Additional computational time measurements}

Tables~\ref{tab:CPU_LES_additional},~\ref{tab:CPU_FreeSolv_main}, and~\ref{tab:CPU_FreeSolv_additional} continue Table~III in Subsection~IV\,B of the main text for all other combinations of representations and datasets. Note that the tendencies observed here are analogous to those discussed in Subsection~IV\,B, with the FJD-related exception discussed in Subsection~IV\,B.

\begin{table}
\begin{adjustbox}{center}
\begin{tabular}{lcccc}
\toprule
  & SLATM & SLATM (proj.) & cMBDF & aSLATM \\
\midrule training conformers & $\phantom{(}4.82\pm0.11\phantom{)}\phantom{{ \times } 10^{0\phantom{-}}}$ & $\phantom{(}4.83\pm0.14\phantom{)}\phantom{{ \times } 10^{0\phantom{-}}}$ & $\phantom{(}4.76\pm0.15\phantom{)}\phantom{{ \times } 10^{0\phantom{-}}}$ & $\phantom{(}4.75\pm0.13\phantom{)}\phantom{{ \times } 10^{0\phantom{-}}}$ \\
 training GFN2-xTB & \_$^{a}$ & \_$^{a}$ & \_$^{a}$ & \_$^{a}$ \\
 training representations & $\phantom{(}6.55\pm0.06\phantom{)}\phantom{{ \times } 10^{0\phantom{-}}}$ & $\phantom{(}6.56\pm0.12\phantom{)}\phantom{{ \times } 10^{0\phantom{-}}}$ & $\phantom{(}2.93\pm0.11\phantom{)}\phantom{{ \times } 10^{0\phantom{-}}}$ & $(1.27\pm0.01){ \times } 10^{1\phantom{-}}$ \\
 training $\mathbf{Z}$ & $\phantom{(}2.76\pm0.33\phantom{)}\phantom{{ \times } 10^{0\phantom{-}}}$ & $(6.03\pm0.11){ \times } 10^{-1}$ & $(2.03\pm0.09){ \times } 10^{1\phantom{-}}$ & $(6.43\pm0.07){ \times } 10^{1\phantom{-}}$ \\
 $\vec{\alpha}$ & $\phantom{(}2.27\pm0.42\phantom{)}\phantom{{ \times } 10^{0\phantom{-}}}$ & $(8.22\pm0.88){ \times } 10^{-1}$ & $\phantom{(}2.53\pm1.41\phantom{)}\phantom{{ \times } 10^{0\phantom{-}}}$ & $\phantom{(}2.02\pm0.02\phantom{)}\phantom{{ \times } 10^{0\phantom{-}}}$ \\
 training $\mathbf{K}$ (KRR) & $(1.41\pm0.03){ \times } 10^{1\phantom{-}}$ & $(7.50\pm0.50){ \times } 10^{-2}$ & $\phantom{(}3.86\pm0.20\phantom{)}\phantom{{ \times } 10^{0\phantom{-}}}$ & $(3.51\pm0.12){ \times } 10^{3\phantom{-}}$ \\
 $\vec{\alpha}^{\mathrm{KRR}}$ & $(2.00\pm0.71){ \times } 10^{-2}$ & $(2.25\pm0.83){ \times } 10^{-2}$ & $(3.08\pm5.10){ \times } 10^{-1}$ & $(1.00\pm0.00){ \times } 10^{-2}$ \\
 test conformers & $\phantom{(}1.18\pm0.12\phantom{)}\phantom{{ \times } 10^{0\phantom{-}}}$ & $\phantom{(}1.19\pm0.12\phantom{)}\phantom{{ \times } 10^{0\phantom{-}}}$ & $\phantom{(}1.19\pm0.12\phantom{)}\phantom{{ \times } 10^{0\phantom{-}}}$ & $\phantom{(}1.18\pm0.12\phantom{)}\phantom{{ \times } 10^{0\phantom{-}}}$ \\
 test GFN2-xTB & \_$^{a}$ & \_$^{a}$ & \_$^{a}$ & \_$^{a}$ \\
 test representations & $(9.48\pm0.53){ \times } 10^{-1}$ & $(9.40\pm0.51){ \times } 10^{-1}$ & $(3.33\pm0.84){ \times } 10^{-1}$ & $\phantom{(}2.21\pm0.10\phantom{)}\phantom{{ \times } 10^{0\phantom{-}}}$ \\
 \begin{tabular}[t]{@{}l@{}}query feature vectors\\and final SORF result\end{tabular} & $\phantom{(}1.44\pm0.06\phantom{)}\phantom{{ \times } 10^{0\phantom{-}}}$ & $(6.58\pm1.21){ \times } 10^{-1}$ & $\phantom{(}5.60\pm0.11\phantom{)}\phantom{{ \times } 10^{0\phantom{-}}}$ & $(1.70\pm0.06){ \times } 10^{1\phantom{-}}$ \\
 \begin{tabular}[t]{@{}l@{}}query kernel elements\\and final KRR result\end{tabular} & $\phantom{(}7.48\pm0.05\phantom{)}\phantom{{ \times } 10^{0\phantom{-}}}$ & $(1.60\pm0.07){ \times } 10^{-1}$ & $\phantom{(}2.58\pm0.08\phantom{)}\phantom{{ \times } 10^{0\phantom{-}}}$ & $(1.86\pm0.12){ \times } 10^{3\phantom{-}}$ \\
\midrule representation size & $123106$ & $\phantom{0}\phantom{0}\phantom{0}475$ & $\phantom{0}\phantom{0}\phantom{0}\phantom{0}40$ & $123106$ \\
\bottomrule
 \multicolumn{5}{l}{$^{a}$ No electronic-structure calculation is involved.}\\
\end{tabular}
\end{adjustbox}
\caption{Computational time (in seconds) measured for different stages of training KRR or SORF models using different representations on $80\%$ of LES dataset and predicting with them $\Eox$ of the remaining $20\%$; lowest energy MMFF94 conformers were used. Molecular and atomic representation sizes for different representation functions are added for reference. Continuation of Table~III of the main text.}
\label{tab:CPU_LES_additional}
\par\medskip{\footnotesize\textit{Alt text:} A continuation table of wall-clock timings (seconds) for the stages of KRR/SORF training and prediction on LES for additional representations, one row per representation, with columns for the timed stages and the representation vector sizes.}
\end{table}

\begin{table}
\begin{adjustbox}{center}
  \begin{tabular}{lcccc}
\toprule
  & CM & FCHL19 & SOAP & FJD \\
\midrule training conformers & $\phantom{(}2.34\pm0.03\phantom{)}\phantom{{ \times } 10^{0\phantom{-}}}$\phantom{$^{b}$} & $\phantom{(}2.33\pm0.04\phantom{)}\phantom{{ \times } 10^{0\phantom{-}}}$ & $\phantom{(}2.42\pm0.14\phantom{)}\phantom{{ \times } 10^{0\phantom{-}}}$ & $\phantom{(}2.52\pm0.02\phantom{)}\phantom{{ \times } 10^{0\phantom{-}}}$ \\
 training GFN2-xTB & \_$^{a}$ & \_$^{a}$ & \_$^{a}$ & $(1.32\pm0.03){ \times } 10^{2\phantom{-}}$ \\
 training representations & $(2.25\pm0.43){ \times } 10^{-2}$$^{b}$ & $\phantom{(}3.42\pm0.02\phantom{)}\phantom{{ \times } 10^{0\phantom{-}}}$ & $\phantom{(}2.40\pm0.01\phantom{)}\phantom{{ \times } 10^{0\phantom{-}}}$ & $(3.11\pm0.03){ \times } 10^{1\phantom{-}}$ \\
 training $\mathbf{Z}$ & $\phantom{(}1.02\pm0.06\phantom{)}\phantom{{ \times } 10^{0\phantom{-}}}$\phantom{$^{b}$} & $(1.13\pm0.00){ \times } 10^{1\phantom{-}}$ & $(1.16\pm0.00){ \times } 10^{1\phantom{-}}$ & $(1.38\pm0.02){ \times } 10^{2\phantom{-}}$ \\
 $\vec{\alpha}$ & $\phantom{(}2.21\pm0.44\phantom{)}\phantom{{ \times } 10^{0\phantom{-}}}$\phantom{$^{b}$} & $(9.02\pm0.67){ \times } 10^{-1}$ & $(8.62\pm0.63){ \times } 10^{-1}$ & $\phantom{(}2.30\pm0.06\phantom{)}\phantom{{ \times } 10^{0\phantom{-}}}$ \\
 training $\mathbf{K}$ (KRR) & $(1.98\pm0.15){ \times } 10^{-1}$\phantom{$^{b}$} & $(3.32\pm0.03){ \times } 10^{1\phantom{-}}$ & $(1.07\pm0.01){ \times } 10^{3\phantom{-}}$ & $(1.02\pm0.02){ \times } 10^{2\phantom{-}}$ \\
 $\vec{\alpha}^{\mathrm{KRR}}$ & $(8.60\pm8.95){ \times } 10^{-1}$\phantom{$^{b}$} & $(2.75\pm0.43){ \times } 10^{-2}$ & $(1.75\pm0.43){ \times } 10^{-2}$ & $(1.00\pm0.00){ \times } 10^{-2}$ \\
 test conformers & $(6.38\pm0.52){ \times } 10^{-1}$\phantom{$^{b}$} & $(5.67\pm0.13){ \times } 10^{-1}$ & $(6.23\pm0.55){ \times } 10^{-1}$ & $(6.27\pm0.08){ \times } 10^{-1}$ \\
 test GFN2-xTB & \_$^{a}$ & \_$^{a}$ & \_$^{a}$ & $(3.35\pm0.17){ \times } 10^{1\phantom{-}}$ \\
 test representations & $(7.50\pm4.33){ \times } 10^{-3}$$^{b}$ & $(4.68\pm0.20){ \times } 10^{-1}$ & $(6.00\pm0.07){ \times } 10^{-1}$ & $\phantom{(}7.91\pm0.16\phantom{)}\phantom{{ \times } 10^{0\phantom{-}}}$ \\
 \begin{tabular}[t]{@{}l@{}}query feature vectors\\and final SORF result\end{tabular} & $(6.07\pm1.03){ \times } 10^{-1}$\phantom{$^{b}$} & $\phantom{(}3.54\pm0.03\phantom{)}\phantom{{ \times } 10^{0\phantom{-}}}$ & $\phantom{(}3.60\pm0.03\phantom{)}\phantom{{ \times } 10^{0\phantom{-}}}$ & $(3.45\pm0.12){ \times } 10^{1\phantom{-}}$ \\
 \begin{tabular}[t]{@{}l@{}}query kernel elements\\and final KRR result\end{tabular} & $(2.17\pm0.08){ \times } 10^{-1}$\phantom{$^{b}$} & $(1.65\pm0.02){ \times } 10^{1\phantom{-}}$ & $(5.28\pm0.06){ \times } 10^{2\phantom{-}}$ & $(8.09\pm0.22){ \times } 10^{1\phantom{-}}$ \\
\midrule representation size & $\phantom{0}\phantom{0}990$ & $\phantom{0}2440$ & $29160$ & $\phantom{0}\phantom{0}\phantom{0}56$ \\
\bottomrule
 \multicolumn{5}{l}{$^{a}$ No electronic-structure calculation is involved.}\\
 \multicolumn{5}{l}{$^{b}$ smaller than the time measurement error}\\
\end{tabular}
\end{adjustbox}
  \caption{Computational time (in seconds) measured for different stages of training KRR or SORF models using representations considered in Table~III on $80\%$ of the FreeSolv dataset and predicting with them $\Ehyd$ of the remaining $20\%$; lowest energy MMFF94 conformers were used. Sizes of representation vectors (molecular, atomic, and corresponding to orbital-atomic contributions) are displayed for reference.}
  \label{tab:CPU_FreeSolv_main}
\par\medskip{\footnotesize\textit{Alt text:} A table of wall-clock timings (seconds) for the stages of KRR/SORF training and prediction on FreeSolv, one row per representation, with columns for the timed stages and the representation vector sizes (molecular, atomic, and orbital-atomic).}
\end{table}

\begin{table}
\begin{adjustbox}{center}
  \begin{tabular}{lcccc}
\toprule
  & SLATM & SLATM (proj.) & cMBDF & aSLATM \\
\midrule training conformers & $\phantom{(}2.35\pm0.01\phantom{)}\phantom{{ \times } 10^{0\phantom{-}}}$ & $\phantom{(}2.36\pm0.03\phantom{)}\phantom{{ \times } 10^{0\phantom{-}}}$ & $\phantom{(}2.38\pm0.05\phantom{)}\phantom{{ \times } 10^{0\phantom{-}}}$ & $\phantom{(}2.33\pm0.02\phantom{)}\phantom{{ \times } 10^{0\phantom{-}}}$ \\
 training GFN2-xTB & \_$^{a}$ & \_$^{a}$ & \_$^{a}$ & \_$^{a}$ \\
 training representations & $\phantom{(}5.00\pm0.03\phantom{)}\phantom{{ \times } 10^{0\phantom{-}}}$ & $\phantom{(}5.03\pm0.04\phantom{)}\phantom{{ \times } 10^{0\phantom{-}}}$ & $\phantom{(}1.93\pm0.03\phantom{)}\phantom{{ \times } 10^{0\phantom{-}}}$ & $\phantom{(}8.51\pm0.04\phantom{)}\phantom{{ \times } 10^{0\phantom{-}}}$ \\
 training $\mathbf{Z}$ & $\phantom{(}2.73\pm0.01\phantom{)}\phantom{{ \times } 10^{0\phantom{-}}}$ & $(6.57\pm0.19){ \times } 10^{-1}$ & $(1.61\pm0.06){ \times } 10^{1\phantom{-}}$ & $(4.80\pm0.01){ \times } 10^{1\phantom{-}}$ \\
 $\vec{\alpha}$ & $\phantom{(}2.12\pm0.02\phantom{)}\phantom{{ \times } 10^{0\phantom{-}}}$ & $(9.05\pm0.27){ \times } 10^{-1}$ & $\phantom{(}2.45\pm0.94\phantom{)}\phantom{{ \times } 10^{0\phantom{-}}}$ & $\phantom{(}2.13\pm0.01\phantom{)}\phantom{{ \times } 10^{0\phantom{-}}}$ \\
 training $\mathbf{K}$ (KRR) & $(1.01\pm0.01){ \times } 10^{1\phantom{-}}$ & $(8.25\pm0.43){ \times } 10^{-2}$ & $\phantom{(}2.20\pm0.14\phantom{)}\phantom{{ \times } 10^{0\phantom{-}}}$ & $(1.16\pm0.01){ \times } 10^{3\phantom{-}}$ \\
 $\vec{\alpha}^{\mathrm{KRR}}$ & $(1.00\pm0.00){ \times } 10^{-2}$ & $(2.25\pm0.83){ \times } 10^{-2}$ & $(5.15\pm8.57){ \times } 10^{-1}$ & $(1.25\pm0.43){ \times } 10^{-2}$ \\
 test conformers & $(5.70\pm0.10){ \times } 10^{-1}$ & $(5.70\pm0.12){ \times } 10^{-1}$ & $(5.77\pm0.04){ \times } 10^{-1}$ & $(5.65\pm0.11){ \times } 10^{-1}$ \\
 test GFN2-xTB & \_$^{a}$ & \_$^{a}$ & \_$^{a}$ & \_$^{a}$ \\
 test representations & $(5.45\pm0.15){ \times } 10^{-1}$ & $(5.37\pm0.15){ \times } 10^{-1}$ & $(8.25\pm0.83){ \times } 10^{-2}$ & $\phantom{(}1.15\pm0.04\phantom{)}\phantom{{ \times } 10^{0\phantom{-}}}$ \\
 \begin{tabular}[t]{@{}l@{}}query feature vectors\\and final SORF result\end{tabular} & $\phantom{(}1.47\pm0.03\phantom{)}\phantom{{ \times } 10^{0\phantom{-}}}$ & $(6.78\pm1.21){ \times } 10^{-1}$ & $\phantom{(}4.35\pm0.24\phantom{)}\phantom{{ \times } 10^{0\phantom{-}}}$ & $(1.26\pm0.01){ \times } 10^{1\phantom{-}}$ \\
 \begin{tabular}[t]{@{}l@{}}query kernel elements\\and final KRR result\end{tabular} & $\phantom{(}5.47\pm0.02\phantom{)}\phantom{{ \times } 10^{0\phantom{-}}}$ & $(1.62\pm0.04){ \times } 10^{-1}$ & $\phantom{(}1.45\pm0.03\phantom{)}\phantom{{ \times } 10^{0\phantom{-}}}$ & $(5.77\pm0.07){ \times } 10^{2\phantom{-}}$ \\
\midrule representation size & $78789$ & $\phantom{0}\phantom{0}515$ & $\phantom{0}\phantom{0}\phantom{0}40$ & $78789$ \\
\bottomrule
 \multicolumn{5}{l}{$^{a}$ No electronic-structure calculation is involved.}\\
\end{tabular}
\end{adjustbox}
  \caption{Computational time (in seconds) measured for different stages of training KRR or SORF models using representations considered in Table~\ref{tab:CPU_LES_additional} on $80\%$ of the FreeSolv dataset and predicting with them $\Ehyd$ of the remaining $20\%$; lowest energy MMFF94 conformers were used. Sizes of representation vectors (molecular, atomic, and corresponding to orbital-atomic contributions) are displayed for reference.}
  \label{tab:CPU_FreeSolv_additional}
\par\medskip{\footnotesize\textit{Alt text:} A table of wall-clock timings (seconds) for the stages of KRR/SORF training and prediction on FreeSolv for a further set of representations, one row per representation, with columns for the timed stages and the representation vector sizes.}
\end{table}

\section{Additional protocols considered for machine learning $\Eox$ and $\Ehyd$}
\label{sec:additional_protocols}

\begin{figure*}
\center
\includegraphics[width=1.\textwidth]{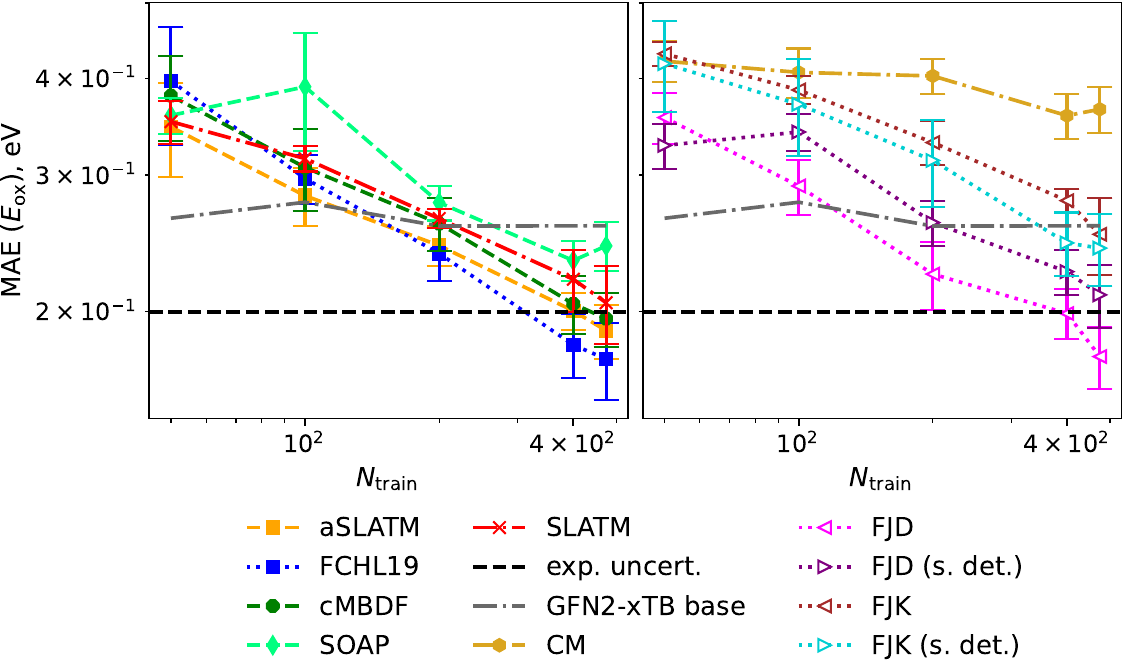}
    \caption{Mean Absolute Errors (MAE) of predictions of oxidation potentials $\Eox$ in acetonitrile from Ref.~\onlinecite{Lee_Lilienfeld:2024} obtained by combining MSORF with several molecular representations for different training set sizes $\Ntrain$. ``s. det.'' indicates FJD or FJK was used with a single Slater determinant; ``exp. uncert.'' is the assumed experimental uncertainty; ``GFN2-xTB base'' is the GFN2-xTB baseline as defined in Section~III of the main text. Unlike Figure~1 of the main text, the results were obtained while shifting $\Eox$ by a constant optimized w.r.t. cross-validation errors.}
    \label{fig:LES_noshifts_False}
\par\medskip{\footnotesize\textit{Alt text:} A two-panel line plot like main-text Figure~1 -- $\Eox$ MAE (eV, log scale) versus $\Ntrain$ (log scale) -- but with the constant $\Eox$ shift optimized against cross-validation errors, and with FJK variants added to the right panel. The left panel shows geometric representations, all declining except the near-flat Coulomb matrix; the right panel shows FJD and FJK variants (standard and single-determinant), declining toward $\sim$0.15 eV with FJD reaching lowest. A black dashed line marks experimental uncertainty at 0.2 eV and a grey dash-dot line the GFN2-xTB baseline; points carry error bars.}
\end{figure*}

\begin{figure*}
\center
\includegraphics[width=1.\textwidth]{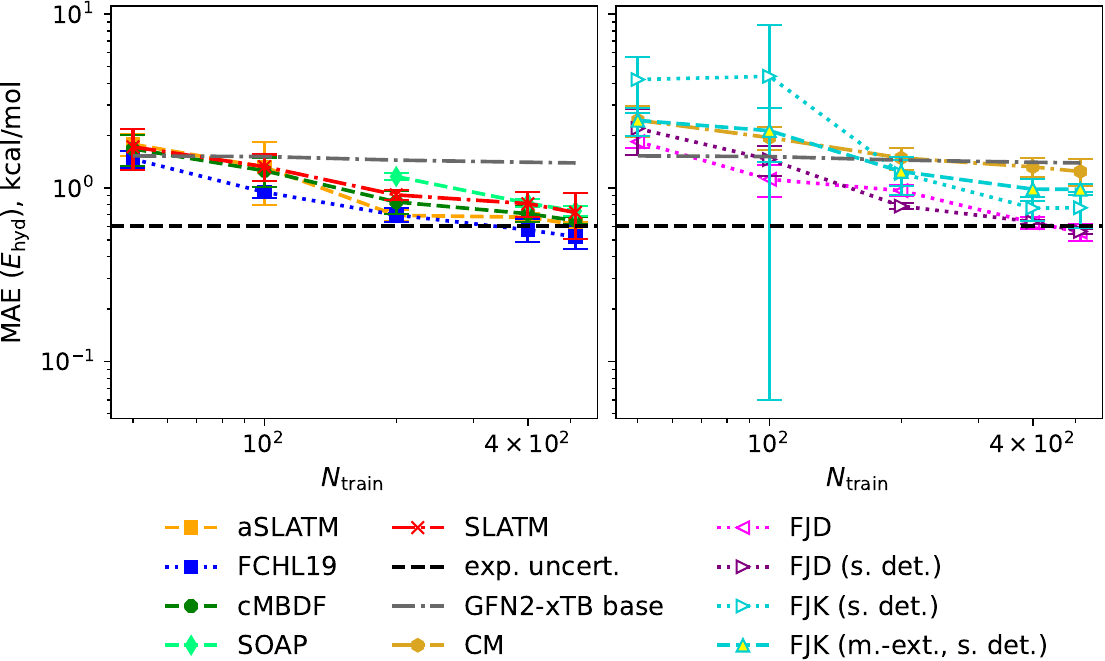}
    \caption{Mean Absolute Error (MAE) of predictions of $\Ehyd$ of the FreeSolv dataset obtained by combining MSORF with several molecular representations for different training set sizes $\Ntrain$ and random train/test split; plotted for comparison are assumed experimental uncertainty (``exp. uncert.'') and GFN2-xTB baseline (``GFN2-xTB base'') as defined in Section~III of the main text. ``S. det.'' next to FJD and FJK entries indicates single Slater determinant was used. Unlike Figure~2 of the main text, the results were obtained with shifting $\Ehyd$ by a stoichiometry-dependent term optimized w.r.t. cross-validation errors.}
    \label{fig:FreeSolv_noshifts_False} 
\par\medskip{\footnotesize\textit{Alt text:} A two-panel line plot of $\Ehyd$ MAE (kcal/mol, log scale, $\sim$0.1 to 10) versus $\Ntrain$ (log scale), with the stoichiometry-dependent shift optimized against cross-validation errors and FJK variants added. The left panel shows geometric representations declining to $\sim$0.55; the right panel shows FJD and FJK variants that largely overlap and converge near 0.6, with one FJK single-determinant point at $\Ntrain\sim100$ carrying a very large error bar. A black dashed line marks experimental uncertainty at 0.6 and a grey dash-dot line the GFN2-xTB baseline.}
\end{figure*}

We start with results analogous to the ones in Subsections~IV\,C and~IV\,D of the main text, but with optimizing as hyperparameters the constant by which LES oxidation potential is shifted and coefficients in FreeSolv hydration energies' ``dressed atom'' shifts. The results for LES are presented in Figure~\ref{fig:LES_noshifts_False} and Table~\ref{tab:FreeSolv_noshifts_False}, demonstrating that shifting the quantities did not affect the learning curves significantly. The results for FreeSolv are presented in Figure~\ref{fig:FreeSolv_noshifts_False} and Table~\ref{tab:FreeSolv_noshifts_False}; comparing the table values demonstrates that applying the stoichiometry-dependent shift often increased MAE and RMSE for different representations, and the large statistical errors occasionally appearing on the learning curves correspond to cases when the procedure proved numerically unstable.

\begin{table}
\begin{tabular}{lcc}
\toprule
 \multirow{2}{*}{method} & \multicolumn{2}{c}{random} \\
\cline{2-3}  & MAE & RMSE \\
\midrule aSLATM & $0.189\pm0.015$ & $0.275\pm0.039$ \\
 FCHL19 & \underline{$0.174\pm0.020$} & $0.254\pm0.044$ \\
 cMBDF & $0.196\pm0.016$ & $0.288\pm0.039$ \\
 SOAP & $0.243\pm0.018$ & $0.363\pm0.059$ \\
 SLATM & $0.205\pm0.024$ & $0.302\pm0.062$ \\
 CM & $0.365\pm0.025$ & $0.472\pm0.022$ \\
\midrule FJD & $0.175\pm0.016$ & \underline{$0.244\pm0.031$} \\
 FJD (s. det.) & $0.210\pm0.020$ & $0.294\pm0.030$ \\
\midrule FJK & $0.252\pm0.028$ & $0.352\pm0.039$ \\
 FJK (s. det.) & $0.241\pm0.026$ & $0.368\pm0.053$ \\
\bottomrule
\end{tabular}
\caption{Mean Absolute Errors (MAEs) and Root Mean Square Errors (RMSEs) observed for machine learning $\Eox$ from Ref.~\onlinecite{Lee_Lilienfeld:2024} at $\Ntrain=474$ using two types of train/test splits: random splits and scaffold with random ties and ``skeleton'' policy for acyclic molecules. See Section~III of main text for how the GFN2-xTB baseline (``GFN2-xTB base'') was defined. In each column the best observed value is underlined. Unlike Table~IV of the main text, the results were obtained with shifting $E_{\mathrm{ox}}$ by a constant optimized w.r.t. cross-validation errors.}
\label{tab:LES_noshifts_False}
\par\medskip{\footnotesize\textit{Alt text:} A table of LES prediction accuracies (MAE and RMSE) at $\Ntrain=474$ for random and scaffold (skeleton) splits, one row per representation, best value per column underlined; here the constant $\Eox$ shift is optimized against cross-validation errors.}
\end{table}

\begin{table}
\begin{tabular}{lcc}
\toprule
 \multirow{2}{*}{method} & \multicolumn{2}{c}{random} \\
\cline{2-3}  & MAE & RMSE \\
\midrule aSLATM & $0.61\pm0.02$ & $1.48\pm0.38$ \\
 aSLATM (m.-ext.) & $0.57\pm0.03$ & $1.21\pm0.10$ \\
 FCHL19 & $0.52\pm0.08$ & $0.92\pm0.18$ \\
 FCHL19 (m.-ext.) & \underline{$0.50\pm0.05$} & $0.91\pm0.14$ \\
 cMBDF & $0.65\pm0.03$ & $1.24\pm0.20$ \\
 cMBDF (m.-ext.) & $0.57\pm0.06$ & $0.96\pm0.17$ \\
 SOAP & $0.73\pm0.05$ & $1.14\pm0.10$ \\
 SOAP (m.-ext.) & $0.74\pm0.02$ & $1.14\pm0.06$ \\
 SLATM & $0.72\pm0.22$ & $1.98\pm1.58$ \\
 CM & $1.24\pm0.22$ & $1.88\pm0.42$ \\
\midrule FJD & $0.55\pm0.06$ & $0.90\pm0.14$ \\
 FJD (s. det.) & $0.56\pm0.02$ & $0.98\pm0.03$ \\
 FJD (m.-ext.) & $0.54\pm0.04$ & \underline{$0.86\pm0.09$} \\
 FJD (m.-ext., s. det.) & $0.62\pm0.01$ & $1.11\pm0.09$ \\
\midrule FJK (s. det.) & $0.77\pm0.18$ & $1.34\pm0.43$ \\
 FJK (m.-ext., s. det.) & $0.98\pm0.07$ & $1.80\pm0.09$ \\
\bottomrule
\end{tabular}
\caption{Summary of results for machine learning $E_{\mathrm{hyd}}$ of the FreeSolv dataset labeled analogously to Table~\ref{tab:LES_noshifts_False}; note that for FreeSolv the maximum training set size was 514 molecules. Unlike Table~V of the main text, the results were obtained with shifting $\Ehyd$ by a stoichiometry-dependent term optimized w.r.t. cross-validation errors.}
\label{tab:FreeSolv_noshifts_False}
\par\medskip{\footnotesize\textit{Alt text:} A table of FreeSolv prediction accuracies, one row per representation, with columns for random and scaffold splits; laid out like the corresponding main-text table but with the stoichiometry-dependent $\Ehyd$ shift optimized against cross-validation errors. The maximum training set is 514 molecules.}
\end{table}

Also note that Figures~\ref{fig:LES_noshifts_False} and~\ref{fig:FreeSolv_noshifts_False}, as well as Tables~\ref{tab:LES_noshifts_False} and~\ref{tab:FreeSolv_noshifts_False}, demonstrate results obtained with FJK representation with the same protocol as for FJD (in this case with optimized quantity shifts), but the orbital information was taken from Huckel calculations with 6-311G basis~\cite{Krishnan_Pople:1980,McLean_Chandler:1980,Glukhovtsev_Radom:1995,Curtiss_Radom:1995} and Boys localized orbitals~\cite{Foster_Boys:1960}, with solvent presence simulated with Solvation Model based on Density~\cite{Marenich_Truhlar:2009}. Originally, we considered to generate FJK descriptors with the same protocol as in Ref.~\onlinecite{Karandashev_Lilienfeld:2022}, i.e. with Hartree-Fock (HF) calculations with STO-3G basis set~\cite{Hehre_Pople:1969,Hehre_Pople:1970} and intrinsic bond orbitals~\cite{Knizia:2013} (IBOs), but found that sometimes the HF calculations failed to converge for charged species while IBOs were challenging to obtain for some elements; the turn to using results of Huckel calculations was motivated by decreasing calculation cost and Refs.~\onlinecite{Fabrizio_Corminboeuf:2022,Briling_Corminboeuf:2024}. The main consideration behind this experiment was that FJK and FJD representations are significantly more compact than aSLATM and FCHL19 and also conceptually simple, only requiring $\rho_{\mathrm{cut}}^{\mathrm{orb}}$ and $l_{\mathrm{max}}$ to be optimized or assumed from general considerations; thus an FJK/Huckel combination could be cited as a compact descriptor with significantly less hand-crafting than many geometry-based analogues. However, FJK/Huckel also demonstrated worse prediction accuracies than the better-performing geometry-based representations in this work, and switching to GFN2-xTB both drastically decreased the computational time needed to generate the orbital information (approximately by factors of 200 for ``pair determinants'' and 300-500 for ``single determinants'') and increased prediction accuracy to comparable to or better than aSLATM and FCHL19. Unlike FJK/Huckel, FJD/GFN2-xTB has significant amount of hand crafting invested in it, though on the side of GFN2-xTB semiempirical method rather than the descriptor generated from its results.

\begin{figure*}
\center
\includegraphics[width=1.\textwidth]{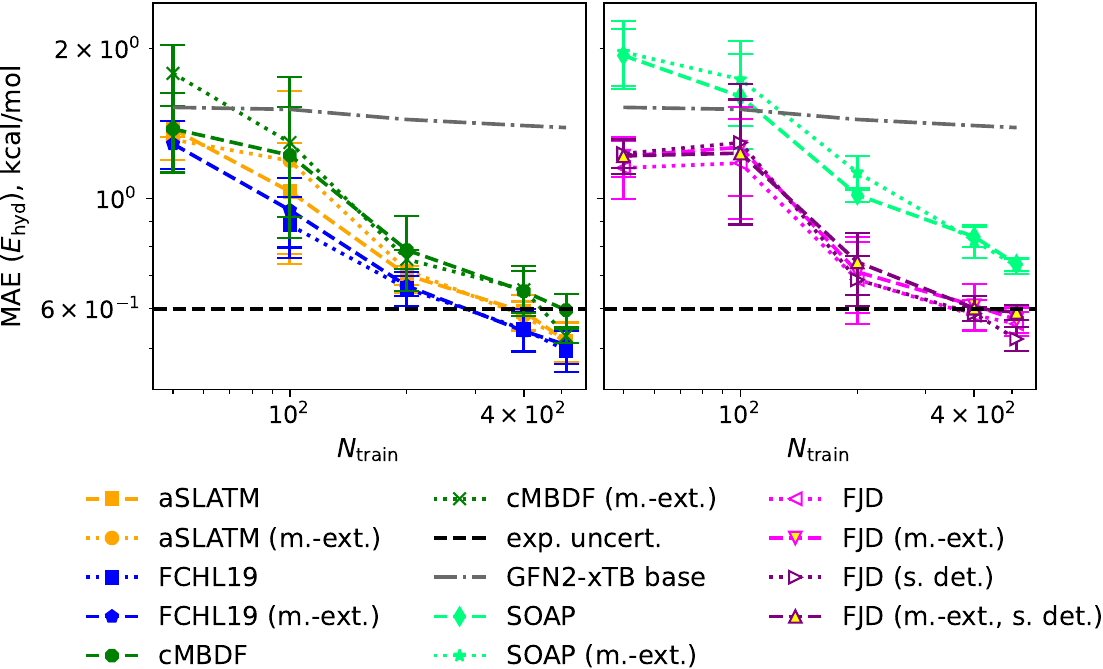}
    \caption{Comparison of Mean Absolute Error (MAE) of predictions of $\Ehyd$ of the FreeSolv dataset obtained with and without ``mixed-extensive'' ($\MIXEDEXTENSIVE$) layer, presence of the latter marked with ``m.-ext.''. Continuation of Figure~2 of the main text.}
    \label{fig:FreeSolv_mixed_sorf} 
\par\medskip{\footnotesize\textit{Alt text:} A two-panel line plot of $\Ehyd$ MAE (kcal/mol, log scale) versus $\Ntrain$ (log scale) comparing each representation with and without the mixed-extensive (``m.-ext.'') layer, drawn as paired curves. The left panel shows aSLATM, FCHL19 and cMBDF; the right panel SOAP and FJD variants. For most representations the m.-ext. and standard curves overlap closely, indicating little effect; FCHL19 reaches lowest on the left and the FJD variants lowest on the right, while SOAP and cMBDF (m.-ext.) sit higher. A black dashed line marks experimental uncertainty and a grey dash-dot line the GFN2-xTB baseline.}
\end{figure*}

\begin{figure*}
\center
\includegraphics[width=1.\textwidth]{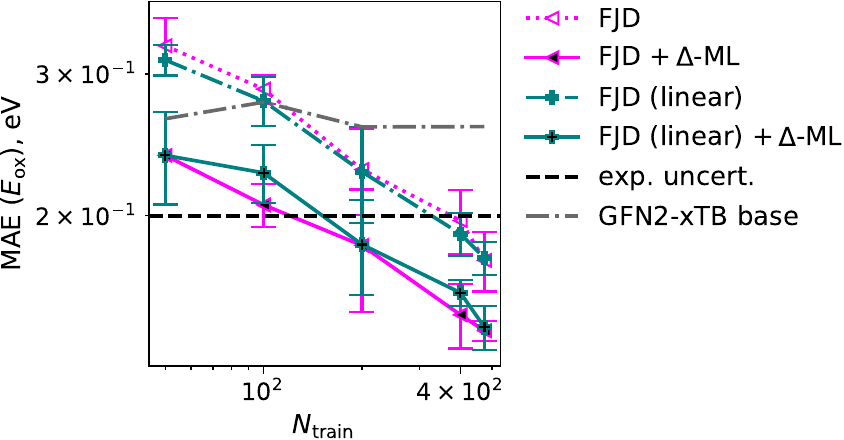}
    \caption{Comparison of Mean Absolute Errors (MAEs) of predictions of $\Eox$ using the ``linear'' version of FJD as defined in Section~\ref{sec:additional_protocols} and the pair-determinant version discussed in Subsection~IV\,C of the main text. Assumed experimental uncertainty (``exp. uncert.'') and GFN2-xTB baseline model (``GFN2-xTB base'') are plotted for comparison.}
    \label{fig:LES_FJD_linear} 
\par\medskip{\footnotesize\textit{Alt text:} A single-panel line plot of $\Eox$ MAE (eV, log scale, $\sim$0.15 to 0.3) versus $\Ntrain$ (log scale) comparing the standard FJD and the ``linear'' FJD variant, each with and without $\Delta$-ML (four curves). The standard and linear curves track each other closely; the two $\Delta$-ML curves reach the lowest errors ($\sim$0.15 eV) while the non-$\Delta$-ML curves are higher at small $\Ntrain$. A black dashed line marks experimental uncertainty at 0.2 eV and a grey dash-dot line the GFN2-xTB baseline; points carry error bars.}
\end{figure*}

Figure~\ref{fig:FreeSolv_mixed_sorf} compares versions of MSORF protocols using the ``mixed-extensive sorf'' $\MIXEDEXTENSIVE$ layer to their normal counterparts for FreeSolv. As mentioned in Subsection~IV\,D of the main text, in this setup using $\MIXEDEXTENSIVE$ did not significantly affect prediction accuracies. Figure~\ref{fig:LES_FJD_linear} and Table~\ref{tab:LES_FJD_linear} compares results for predicting $\Eox$ with ``linear'' version of FJD defined $\WEIGHTEDSUM\sconfs\WEIGHTEDSUM\sorbs\SORF\NORM\WEIGHTEDSUM\SORF$ to the FJD results presented in Subsection~IV\,C of the main text. The learning curves look very similar, with differences only observed for scaffold train/test split, where optimizing quantity shift noticeably improves performance of FJD without $\Delta$-ML, which agrees with the point made in Section~III of the main text about the shift being physically meaningful in this scenario. However, as was the case in Subsecs.~IV\,C and~IV\,D, combining the representations with $\Delta$-ML largely mitigated the difference between them.

\begin{table}
\begin{tabular}{lccccc}
\toprule
 \multirow{2}{*}{method} & \multicolumn{2}{c}{random} & \phantom{.} & \multicolumn{2}{c}{scaffold (skeleton)} \\
\cline{2-3}\cline{5-6}  & MAE & RMSE &  & MAE & RMSE \\
\midrule \multicolumn{6}{c}{\emph{mean shift}} \\
 FJD & $0.176\pm0.015$ & $0.246\pm0.026$ &  & $0.258\pm0.015$ & $0.368\pm0.022$ \\
 FJD $+\,\Delta$-ML & \underline{$0.144\pm0.004$} & \underline{$0.207\pm0.017$} &  & $0.212\pm0.010$ & $0.314\pm0.011$ \\
 FJD (linear) & $0.177\pm0.008$ & $0.248\pm0.025$ &  & $0.287\pm0.017$ & $0.408\pm0.032$ \\
 FJD (linear) $+\,\Delta$-ML & $0.145\pm0.009$ & $0.212\pm0.033$ &  & \underline{$0.200\pm0.012$} & $0.310\pm0.012$ \\
\midrule \multicolumn{6}{c}{\emph{opt. shift}} \\
 FJD (linear) & $0.172\pm0.005$ & $0.240\pm0.020$ &  & $0.249\pm0.007$ & $0.350\pm0.007$ \\
 FJD (linear) $+\,\Delta$-ML & $0.146\pm0.008$ & $0.209\pm0.032$ &  & $0.202\pm0.011$ & \underline{$0.309\pm0.007$} \\
\bottomrule
\end{tabular}
\caption{Summary of results for machine learning $\Eox$ from Ref.~\onlinecite{Lee_Lilienfeld:2024} dataset with standard FJD protocol used throughout Subsection~IV\,C of the main text and its linear version as defined in Section~\ref{sec:additional_protocols}; for the latter we show results both for shifting $\Eox$ by training set mean (``mean shift'') and for optimizing it w.r.t. cross-validation errors (``opt. shift'').}
\label{tab:LES_FJD_linear}
\par\medskip{\footnotesize\textit{Alt text:} A table of LES prediction accuracies comparing the standard FJD protocol with its ``linear'' variant; for the linear variant results are shown both with a training-set-mean shift and with a cross-validation-optimized shift. Rows list the variants, columns give the random and scaffold split error metrics.}
\end{table}

\section{Comparing $\Delta$-ML with augmenting geometric representations with orbital information}

\begin{figure*}
\center
\includegraphics[width=1.\textwidth]{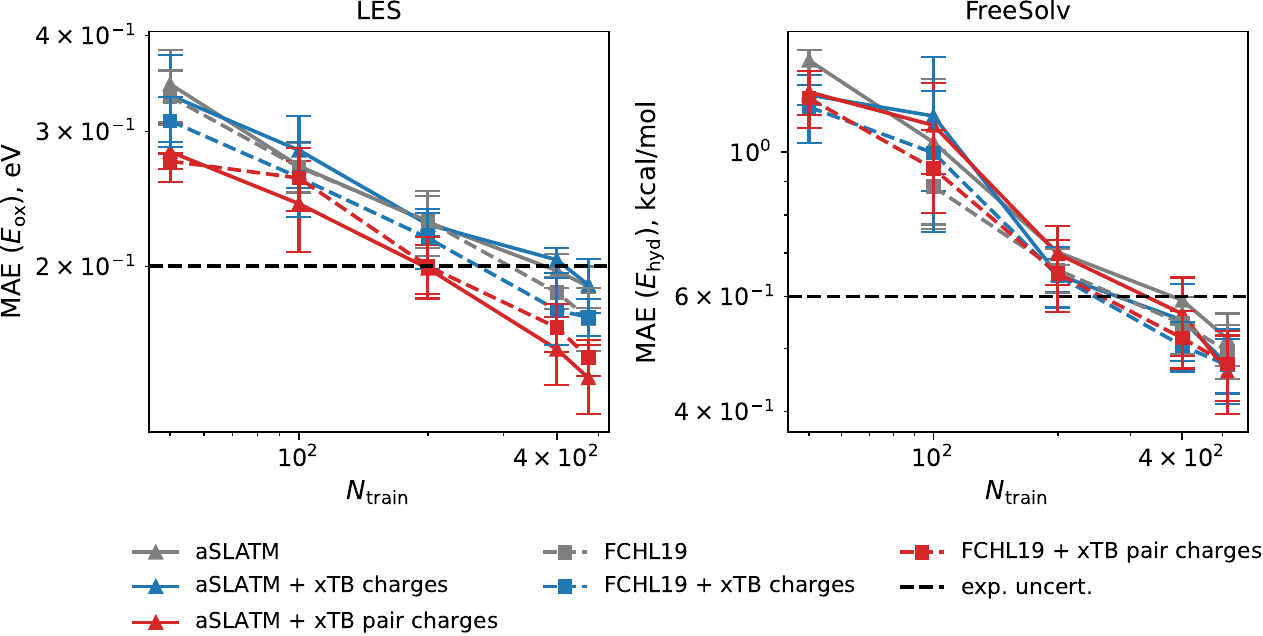}
    \caption{Mean Absolute Error (MAE) of predictions of $\Eox$ in the LES dataset and $\Ehyd$ in the FreeSolv dataset obtained by combining MSORF with aSLATM and FCHL19 representations augmented with GFN2-xTB atomic charges; ``xTB charges'' means augmenting with charges for neutral molecule for LES and solvated molecule for FreeSolv, ``xTB pair charges'' means additionally augmenting with charges for charged molecule for LES and molecule in vacuum for FreeSolv. Assumed experimental uncertainties (``exp. uncert.'') are added for comparison.}
    \label{fig:charge_aug_dml_off} 
\par\medskip{\footnotesize\textit{Alt text:} A two-panel line plot (LES left, $\Eox$ in eV; FreeSolv right, $\Ehyd$ in kcal/mol; both log axes versus $\Ntrain$ on a log scale) showing aSLATM and FCHL19 each plain, with xTB charges, and with xTB pair charges, all without $\Delta$-ML. The ``xTB pair charges'' curves (red) reach the lowest errors and the plain representations (grey) the highest, showing that pair-charge augmentation improves accuracy when $\Delta$-ML is not used. A black dashed line marks experimental uncertainty; points carry error bars.}
\end{figure*}

\begin{figure*}
\center
\includegraphics[width=1.\textwidth]{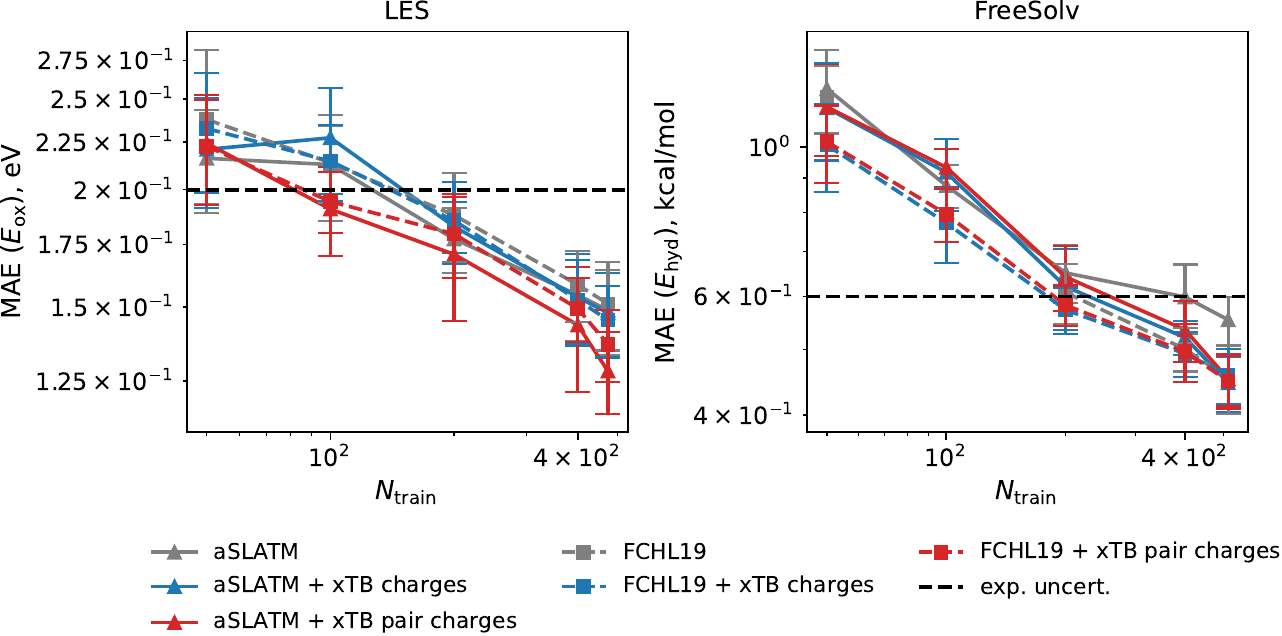}
    \caption{Mean Absolute Error (MAE) of predictions of $\Eox$ in the LES dataset and $\Ehyd$ in the FreeSolv dataset obtained by combining MSORF and $\Delta$-ML with aSLATM and FCHL19 representations augmented with GFN2-xTB atomic charges; ``xTB charges'' means augmenting with charges for neutral molecule for LES and solvated molecule for FreeSolv, ``xTB pair charges'' means additionally augmenting with charges for charged molecule for LES and molecule in vacuum for FreeSolv. Assumed experimental uncertainties (``exp. uncert.'') are added for comparison.}
    \label{fig:charge_aug_dml_on} 
\par\medskip{\footnotesize\textit{Alt text:} A two-panel line plot (LES left, $\Eox$ in eV; FreeSolv right, $\Ehyd$ in kcal/mol; log axes versus $\Ntrain$ on a log scale) of the same aSLATM and FCHL19 charge-augmentation variants as the previous figure, now combined with $\Delta$-ML. The plain, xTB-charges and xTB-pair-charges curves bunch together much more closely than without $\Delta$-ML, indicating that $\Delta$-ML largely removes the benefit of charge augmentation. A black dashed line marks experimental uncertainty; points carry error bars.}
\end{figure*}

Learning curves for MSORF combinations with FCHL19 and aSLATM representations augmented with GFN2-xTB atomic charge information as discussed in Subsection~IV\,F of the main text are presented in Figures~\ref{fig:charge_aug_dml_off} and~\ref{fig:charge_aug_dml_on} for cases when $\Delta$-ML was and was not used. We can observe good learning rates in all cases.

\section{Detailed properties and implementation of ``self-consistent'' loss functions}

In this section we will detail definition and properties of two loss functions that we considered as surrogates of MAE with continuous derivatives in the hyperparameter optimization algorithm described in~Appendix~B of the main text, ``self-consistent Huber'' $\LSCH$ and ``self-consistent LogCosh'' $\LSCLC$ loss functions. The algebra presented in this section is largely a less condensed, differently parametrized re-derivation of the Orlicz-norm regression estimators of Ref.~\onlinecite{Andoni2018Orlicz}, specialized to the losses used here and kept self-contained for the reader's convenience; the differences between our approach and that of Ref.~\onlinecite{Andoni2018Orlicz} will be summarized at the end of this section. We start with a general statement about regression loss functions operating on error vectors of dimensionality $N$.

\begin{theorem}
Suppose $L(\vec{x},\abssmooth)$ is a function that
\begin{enumerate}
\item for a fixed $\abssmooth>0$ is convex and first-order smooth w.r.t. $\vec{x}$, as well as strictly positive for any $\vec{x}\neq\vec{0}$, while for $\vec{x}=\vec{0}$ it equals zero and its gradient w.r.t. $\vec{x}$ equals $\vec{0}$;
\item is positively homogenous of order 1 w.r.t its arguments, \emph{i.e.} for all $\vec{x}\in\vecspace$ and $\posconst>0$
\begin{equation}
L(\posconst\vec{x},\abssmooth\posconst)=\posconst^{m}L(\vec{x},\abssmooth)
\end{equation}
with $m=1$;
\item and satisfies the limiting behavior
\begin{equation}
\lim_{|\vec{x}|/\abssmooth\rightarrow\infty}\frac{L(\vec{x},\abssmooth)-\MAE(\vec{x})}{\abssmooth}=-1.
\label{eq:boundary_point}
\end{equation}
\end{enumerate}

In this case we can define $\LSC$ as the unique solution to the equation
\begin{equation}
L(\vec{x},\gamma \LSC)-\LSC=0,
\label{eq:sc_loss_definition}
\end{equation}
where $\gamma$ is a positive dimensionless parameter. $\LSC$ is a convex function of $\vec{x}$ that is zero at $\vec{x}=\vec{0}$, strictly positive everywhere else, and is related to $\MAE$ by
\begin{equation}
1\leq\frac{\MAE(\vec{x})}{\LSC(\vec{x})}\leq 1+\gamma.
\label{eq:relative_error_bounds}
\end{equation}

\label{theorem:loss_functions}
\end{theorem}

Positive homogeneity of order 1 of $L(\vec{x},\beta)$ allows us to write
\begin{align}
L(\vec{x},\beta)&\phantom{:}=\beta\lperspective(\vec{x}/\beta),\\
\lperspective(\vec{x})&:=L(\vec{x},1),
\end{align}
which defines $L$ as the \emph{perspective}\cite{Boyd_Vandenberghe:2004} of $\lperspective$. Since the latter operation preserves convexity for $\beta>0$ and $\lperspective$ is convex, $L(\vec{x},\beta)$ is jointly convex in $(\vec{x},\beta)$ for $\beta>0$. We can also see that for fixed $\vec{x}$ $L(\vec{x},\beta)$ is a monotonically decreasing function of $\beta$, as can be seen by applying Euler's theorem~\cite{Lewis:1969} to $L$

\begin{equation}
\frac{\partial L(\vec{x},\abssmooth)}{\partial\abssmooth}=\frac{1}{\abssmooth}\left[L(\vec{x},\abssmooth)-\vec{x}\cdot\frac{\partial L(\vec{x},\abssmooth)}{\partial\vec{x}}\right],\label{eq:L_derivative_wrt_delta}
\end{equation}
and using convexity of $L$ to write for $\beta > 0$
\begin{equation}
\frac{\partial L(\vec{x},\abssmooth)}{\partial\abssmooth}<\frac{1}{\abssmooth}\left[L(\vec{0},\abssmooth)+\vec{x}\cdot\frac{\partial L(\vec{x},\abssmooth)}{\partial\vec{x}}-\vec{x}\cdot\frac{\partial L(\vec{x},\abssmooth)}{\partial\vec{x}}\right]=0.
\label{eq:L_beta_negativity}
\end{equation}

We now define for $l>0$
\begin{equation}
\FSC(\vec{x},l):=L(\vec{x},\gamma l)-l
\label{eq:FSC_def}
\end{equation}
and rewrite Equation~\eqref{eq:sc_loss_definition} as 
\begin{equation}
\FSC(\vec{x},\LSC)=0.
\label{eq:LSC_redef}
\end{equation}
Since $L(\vec{x},\beta)$ is jointly convex in $(\vec{x},\beta)$ for $\beta>0$ and $\beta=\gamma l$ is linear in $l$, $\FSC(\vec{x},l)=L(\vec{x},\gamma l)-l$ is jointly convex in $(\vec{x},l)$ as well. Additionally, Equation~\eqref{eq:L_beta_negativity} implies that the partial derivative of $\FSC(\vec{x},l)$ w.r.t. $l$ never exceeds $-1$, while the limit~\eqref{eq:boundary_point} means that $\FSC(\vec{x},l)\rightarrow\MAE(\vec{x})$ as $l\rightarrow0$; thus Equation~\eqref{eq:LSC_redef} has a single solution and $\LSC$~\eqref{eq:sc_loss_definition} is well-defined.

To show $\LSC$ is convex we examine its \emph{epigraph}, i.e. the set $\{(\vec{x},\tau):\tau\ge\LSC(\vec{x})\}$ of points lying on or above the function, which is convex exactly when $\LSC$ is\cite{Boyd_Vandenberghe:2004}. We already demonstrated that $\FSC(\vec{x},l)$ decreases through zero at $l=\LSC(\vec{x})$, so for $\tau >0$ $\LSC(\vec{x})\leq\tau\iff\FSC\leq0$. Hence the epigraph of $\LSC$ (restricted to $\tau>0$) is the \emph{sublevel set}\cite{Boyd_Vandenberghe:2004} $\{\FSC\leq0\}$, which in turn is convex because $\FSC$ is convex, making $\LSC$ convex too.

To prove Equation~\eqref{eq:relative_error_bounds} we first apply first-order convexity condition\cite{Boyd_Vandenberghe:2004} to $L(\vec{x},\beta)$ with fixed $\vec{x}$ and $\beta$ changed to zero, which together with Equation~\eqref{eq:boundary_point} imply
\begin{equation}
L(\vec{x},\abssmooth)\geq\MAE(\vec{x})-\abssmooth.
\label{eq:L_lower_bound}
\end{equation}
Since $L(\vec{x},\beta)$ decreases with $\beta$ and approaches $\MAE(\vec{x})$ for $\beta\rightarrow0^{+}$ we also have
\begin{equation}
L(\vec{x},\abssmooth)\leq\MAE(\vec{x}).
\label{eq:L_upper_bound}
\end{equation}
The inequalities~\eqref{eq:L_lower_bound} and~\eqref{eq:L_upper_bound} together with Equation~\eqref{eq:sc_loss_definition} prove the inequalities in~\eqref{eq:relative_error_bounds}.

There are two closing remarks on Theorem~\ref{theorem:loss_functions}. Firstly, since $\LSC(\vec{x})$ is order 1 positively homogeneous w.r.t. $\vec{x}$, its $n$-th derivative is positively homogeneous of order $1-n$, implying that at $\vec{x}=\vec{0}$ the first and second derivatives of $\LSC(\vec{x})$ have a discontinuity and a singularity. This is the reason we instead use $\LSC^{2}(\vec{x})$ ($\LSC$ squared), as it is order 2 positively homogeneous, making its first derivative continuous for all $\vec{x}\in\vecspace$ and its second derivative bound, though discontinuous at $\vec{x}=\vec{0}$. Secondly, Equation~\eqref{eq:relative_error_bounds} means that using $\LSC(\vec{x})$ or $\LSC^{2}(\vec{x})$ in hyperparameter optimization yields a minimum whose $\MAE$ does not differ from the global minimal $\MAE$ by more than $\gamma$ in relative terms.

To derive $\fulld\LSC(\vec{x})/\fulld\vec{x}$ we differentiate Equation~\eqref{eq:sc_loss_definition} w.r.t. $\vec{x}$ and use Equation~\eqref{eq:L_derivative_wrt_delta} to write
\begin{equation}
\frac{\fulld \LSC(\vec{x})}{\fulld\vec{x}}=\LSC(\vec{x})\frac{\partial L(\vec{x},\gamma\LSC)/\partial\vec{x}}{\vec{x}\cdot\partial L(\vec{x},\gamma\LSC)/\partial\vec{x}}.
\end{equation}
This expression was used in steepest descent optimization as described in~Appendix~B of the main text.

Though it is not, in general, possible to analytically solve equation~\eqref{eq:LSC_redef}, the latter's solvability with Newton method \cite{Boyd_Vandenberghe:2004} can be guaranteed by requiring second derivative of $L(\vec{x},\abssmooth)$ w.r.t. $\abssmooth$ to be continuous and non-zero for $\abssmooth>0$ and fixed $\vec{x}$ (we note in advance that this will be the case for $\LSCLC$, but not $\LSCH$, which, however, can be solved analytically). Combined with convexity of $\FSC(\vec{x},l)$ w.r.t. $l$ for fixed $\vec{x}$, this makes second derivative of $\FSC(\vec{x},l)$ for fixed $\vec{x}$ strictly positive and thus having positive lower and upper bounds for $0<l\leq\LSC$. Therefore, a Newton method run starting at $\LSC=0$ will only consider points in $0<l\leq\LSC$ where second derivative of the r.h.s. of~\eqref{eq:LSC_redef} has positive lower and upper bounds, giving well-defined upper bound on the number of steps needed to converge the solution to a given accuracy~\cite{Boyd_Vandenberghe:2004}.  The derivative required by the Newton method can be obtained by differentiating $\FSC$~\eqref{eq:FSC_def} and applying Euler's theorem to the case of $l>0$

\begin{equation}
\frac{\partial\FSC(\vec{x},l)}{\partial l}=\frac{1}{l}\left[\FSC(\vec{x},l)-\vec{x}\cdot\frac{\partial L(\vec{x},\abssmooth)}{\partial\vec{x}}\right]
\label{eq:Newton_der}
\end{equation}

We now consider two loss functions satisfying the conditions of Theorem~\ref{theorem:loss_functions}. The first one is Huber loss function~\cite{Huber:1964}, which we rescale to satisfy the limit~\eqref{eq:boundary_point}
\begin{align}
\LH(\vec{x},\abssmooth):=&\frac{1}{N}\sum_{i=1}^{N}\fH(x_{i},\abssmooth)\label{eq:Huber_main_definition}\\
\fH(t,\beta):=&\begin{cases}
t^{2}/(4\abssmooth) & |t|<2\abssmooth,\\
|t|-\abssmooth& |t|\geq2\abssmooth.
\end{cases}
\end{align}
Substitution of Equation~\eqref{eq:Huber_main_definition} into Equation~\eqref{eq:sc_loss_definition} yields the equation for self consistent Huber loss function $\LSCH$
\begin{equation}
(N+\gamma N_{\geq})\LSCH^{2}-\LSCH\sum_{\geq}|x_{i}|-\frac{1}{4\gamma}\sum_{<}x_{i}^{2}=0,
\label{eq:LSCH_quadratic_equation}
\end{equation}
where $\sum_{\geq}$ and $\sum_{<}$ are shorthands for summation over $i$ such that $|x_{i}|\geq2\gamma\LSCH$ and $|x_{i}|<2\gamma\LSCH$, $N_{\geq}$ is the number of $|x_{i}|$ values such that $|x_{i}|\geq2\gamma\LSCH$. The sets of $i$ for both summation signs can be found by sorting $|x_{i}|$ values and finding the interval between two consecutive $|x_{i}|$ where r.h.s of~\eqref{eq:LSCH_quadratic_equation} as a function of $\LSCH$ changes sign. Finding the only positive root of~\eqref{eq:LSCH_quadratic_equation} yields the final expression
\begin{equation}
\LSCH=\frac{1}{2(N+N_{\geq}\gamma)}\left[\sum_{\geq}|x_{i}|+\sqrt{\left(\sum_{\geq}|x_{i}|\right)^{2}+\frac{(N+\gamma N_{\geq})}{\gamma}\sum_{<}x_{i}^{2}}\right].
\label{eq:LSCH_analytic_solution}
\end{equation}
An alternative way to find $\LSCH$ without sorting $|x_{i}|$ is derived by considering the $\FSC$ arising in this case as a function defined as equaling  $\pFSC^{(j)}$ on each interval of the form $(|\sortedx_{j}|/(2\gamma),|\sortedx_{j+1}|/(2\gamma))$, where $\sortedvecx$ consists of values of $\vec{x}$ sorted by increasing absolute value and augmented with $\sortedx_{0}=0$, and $\pFSC^{(j)}$ reads
\begin{equation}
\pFSC^{(j)}(\vec{x},l)=\MAE(\vec{x})-(\gamma+1)l+\frac{1}{N}\sum_{j^{\prime}=1}^{j}\mathrm{max}\left[0,\frac{(\sortedx_{j^{\prime}})^{2}}{4\gamma l}-|\sortedx_{j^{\prime}}|+\gamma l\right].
\end{equation}
For fixed $\vec{x}$ each $\pFSC^{(j)}$ is a convex monotonically decreasing function, with $\pFSC^{(j+1)}$ differing from $\pFSC^{(j)}$ by a nonnegative function. This means that each $\pFSC^{(j)}$ has only one positive root w.r.t. $l$, which is also smaller than the positive root of $\pFSC^{(j+1)}$. Uniqueness of $\LSCH$ also implies that of all the roots of $\pFSC^{(j)}$, only one actually lies in the corresponding interval $(|\sortedx_{j}|/(2\gamma),|\sortedx_{j+1}|/(2\gamma))$. That means $\LSCH$ can be evaluated similarly to the Newton method, by starting with the initial guess $\LSCH=0$ and calculating each successive guess with Equation~\eqref{eq:LSCH_analytic_solution} with $\sum_{\geq}$ and $\sum_{<}$ evaluated based on the previous guess. The procedure will yield increasing estimates of $\LSCH$ until the exact value is reached, as signified by $\sum_{\geq}$ and $\sum_{<}$ staying the same as in the previous iteration.

Secondly, we considered the LogCosh loss function \cite{Wang_Tian:2022} which, when rescaled to satisfy~\eqref{eq:boundary_point}, reads
\begin{equation}
\LLC(\vec{x},\abssmooth):=\frac{1}{N}\sum_{i=1}^{N}\frac{\abssmooth}{\ln 2}\ln\left(\cosh\frac{x_{i}\ln2}{\abssmooth}\right).
\label{eq:logcosh_definition}
\end{equation}
It is straightforward to check that for fixed $\vec{x}$ the second derivative of $\LLC$ w.r.t. $\abssmooth$ is never zero, allowing to calculate the self-consistent version of $\LLC$, dubbed $\LSCLC$, numerically via the Newton method as discussed previously.

We implemented both $\LSCH^{2}$ and $\LSCLC^{2}$ in the QML2 code. Though $\LSCLC^{2}$ is significantly more expensive to calculate than $\LSCH^{2}$, we chose the former for our work since its cost is small relative to the cost of finding the SVD of the $\mat{Z}$ matrix and its derivatives are infinitely differentiable for all $\vec{x}\neq\vec{0}$ [as can be verified by differentiating~\eqref{eq:sc_loss_definition}], with $\vec{x}=\vec{0}$ extremely unlikely to be encountered during hyperparameter optimization. We  underline that Equation~\eqref{eq:sc_loss_definition} implies that the $\gamma$ operates similarly to the way $\abssmooth$ operates in the definitions of $\LH$~\eqref{eq:Huber_main_definition} and $\LLC$~\eqref{eq:logcosh_definition}. Since the $\abssmooth$ parameter in this context is sometimes described as an absolute error value beyond which a point is considered an outlier, $\gamma$ parameter can be described as how large an error is compared to $\LSC$ (\emph{i.e.}, to the ``average smoothed error'') to be considered an outlier. As mentioned in~Appendix~B of the main text, we found $\gamma$ to be a more convenient parameter to regulate in our applications.

As mentioned previously, $\LSCH$ and $\LSCLC$ are closely related to Orlicz-norm regression estimators of Ref.~\onlinecite{Andoni2018Orlicz}, and the scale-free Orlicz-norm estimator $G$ proposed in that work can be defined, up to a constant normalization factor, as the solution w.r.t. $l$ of the equation
\begin{equation}
N\LH(\vec{x},l)-l=0.
\label{eq:orlicz}
\end{equation}
Besides the inability to tune the $\gamma$ factor and using sum rather than mean of $\fH$ contributions in Equation~\eqref{eq:orlicz}, another application difference is that Ref.~\onlinecite{Andoni2018Orlicz} avoided explicit evaluation of $G$ altogether in their algorithm for non-regularized regression, while our aim was to evaluate a differentiable loss function from cross-validation errors. We also note that Ref.~\onlinecite{Andoni2018Orlicz} effectively demonstrates that $\LSCH$ and (less directly) $\LSCLC$ can be used in non-regularized regression as well, giving rise to alternative expressions for $\vec{\alpha}$ regression coefficients used in this work; however, at least at present, switching to such expressions would prevent KRR or SORF protocols from capitalizing on efficient procedures for regularized regression and evaluating cross-validation errors used extensively in this work. We also note that the ``self-consistent'' loss expressions remove the extra degree of freedom associated with concomitant scale estimation\cite{CombettesMuller:2018}, where the perspective scale $\abssmooth$ is a free variable fitted jointly with the model through an added penalty (equivalently, Huber's Proposal~2\cite{Huber:1964}).

As a closing remark, we note that $\LSCH^{2}$ and $\LSCLC^{2}$ cannot be straightforwardly used with Stochastic Gradient Descent~\cite{blondel2025elementsdifferentiableprogramming} algorithms since these loss functions are not sums of contributions from different elements of $\vec{x}$. While it should be possible to bypass the issue by using $ \LH$ or $\LSC$ and enforcing Equation~(\ref{eq:sc_loss_definition}) with a Lagrange multiplier, investigating feasibility and usefulness of such a procedure was beyond the scope of this work.

%aipnum4-2.bst 2019-01-14 (MD) hand-edited version of apsrev4-1.bst